\documentclass[12pt,preprint]{aastex}
\usepackage{url}
\usepackage[usenames]{color}
\usepackage{amsmath}
\newcommand{\logg} {\log \textsl{\textrm{g}}}

\newcommand{\Te} {T_{\rm eff}}

\newcommand{\msun} {$M_\odot$}

\newcommand\gta{\lower 0.5ex\hbox{$\buildrel > \over \sim\ $}} 
\newcommand\lta{\lower 0.5ex\hbox{$\buildrel < \over \sim\ $}} 
\newcommand{\nh} {N({\rm H})/N({\rm He})}

\newcommand{\ha} {$\rm{H}{\alpha}$}
\newcommand{\hb} {$\rm{H}{\beta}$}

\newcommand{\rsun} {$R_{\odot}$}

\begin{document}

\title{MEASUREMENTS OF PHYSICAL PARAMETERS OF WHITE DWARFS: A TEST OF THE 
MASS-RADIUS RELATION}

\author{A. B\'edard, P. Bergeron, \& G. Fontaine}
\affil{D\'epartement de Physique, Universit\'e de Montr\'eal,
  C.P.~6128, Succ.~Centre-Ville, Montr\'eal, Qu\'ebec H3C 3J7, Canada}
\email{bedard@astro.umontreal.ca, bergeron@astro.umontreal.ca, 
fontaine@astro.umontreal.ca}

\begin{abstract}

We present a detailed spectroscopic and photometric analysis of 219 DA
and DB white dwarfs for which trigonometric parallax measurements are
available. Our aim is to compare the physical parameters derived from
the spectroscopic and photometric techniques, and then to test the
theoretical mass-radius relation for white dwarfs using these
results. The agreement between spectroscopic and photometric
parameters is found to be excellent, especially for effective
temperatures, showing that our model atmospheres and fitting
procedures provide an accurate, internally consistent analysis. Values
of surface gravity and solid angle, obtained respectively from
spectroscopy and photometry, are combined with parallax measurements
in various ways to study the validity of the mass-radius relation from
an empirical point of view. After a thorough examination of our
results, we find that 73\% and 92\% of the white dwarfs are consistent
within 1 and 2$\sigma$ confidence levels, respectively, with the
predictions of the mass-radius relation, thus providing strong support
to the theory of stellar degeneracy. Our analysis also allows us to
identify 15 stars that are better interpreted in terms of unresolved
double degenerate binaries.  Atmospheric parameters for both
components in these binary systems are obtained using a novel
approach. We further identify a few white dwarfs that are possibly
composed of an iron core rather than a carbon/oxygen core, since they
are consistent with Fe-core evolutionary models.

\end{abstract}


\section{INTRODUCTION} \label{intro}

Our understanding of the nature and the evolution of white dwarf stars
begins with the determination of their physical parameters, which
include effective temperature, mass, radius, surface gravity, chemical
composition, luminosity, cooling age, just to name a few. Some of
these parameters are of course related to each other, but the methods
used to measure these quantities, directly or indirectly, may differ
quite significantly. There are indeed several methods that can be
applied to individual stars, or to statistical ensembles, to estimate
these parameters. The most commonly used method for stars with strong
enough absorption features is the spectroscopic technique, first
applied to a large sample of DA white dwarfs by \citet{BSL92}, where
the predictions from model atmospheres are compared with high
signal-to-noise optical spectra.  This technique has been applied in
several analyses of relatively bright DA stars
\citep{finley97,vennes97,LBH05,koester09,gianninas11} and in
particular to the much fainter DA stars identified in the large
spectroscopic SDSS sample (see, e.g., \citealt{tremblay11},
\citealt{kepler15}), and also to other spectral types as well,
including DB stars \citep{voss07,bergeron11,koester15}, and DO stars
\citep{reindl14}. With the spectroscopic technique, the atmospheric
parameters measured are the effective temperature, the surface
gravity, and the atmospheric composition, because these correspond to
the input parameters of model atmosphere calculations. White dwarf
masses and other relevant quantities can then be determined through
the use of mass-radius relations obtained from evolutionary models
appropriate for these stars. The spectroscopic technique can thus be
successfully applied with excellent precision to large spectroscopic
data (less than 2\% in $\Te$ and about 0.04 dex in $\logg$), given
that the spectra have sufficiently high signal-to-noise ratio (${\rm
  S/N}\gtrsim50$) and that they are properly flux calibrated --- which
is not always the case.

For cool white dwarfs with weak, or no spectroscopic features --- the
DC stars, one must rely on the so-called photometric method (see,
e.g., \citealt{BRL97}) where the spectral energy distribution of a
star, built from as many photometric data points as possible, is
compared with the predictions of model atmospheres --- synthetic
photometry in this case. With this method, magnitudes are converted
into average fluxes, using appropriate photometric zero points, and
compared with model fluxes averaged over the same filter
bandpasses. Here both $\Te$ and the solid angle, $\pi(R/D)^2$, are
considered free parameters for an assumed atmospheric composition,
most commonly pure hydrogen or pure helium.  If the distance $D$ to
the star is known --- from trigonometric parallax measurements for
instance --- its radius $R$ can be obtained directly.  The stellar
radius can then be converted into mass, and thus $\logg$, using
mass-radius relations from evolutionary models. In the absence of
distance estimates, one generally assumes an average value of
$\logg=8$, from which an estimate of the {\it photometric distance}
can be derived. Since trigonometric parallaxes are currently available
for a few hundred white dwarfs (around 350 according to the electronic
version of the Catalog of Spectroscopically Identified White
Dwarfs\footnote{http://www.astronomy.villanova.edu/WDCatalog/index.html}
of \citealt{WDC}), the photometric technique can only be applied to a
restricted sample of objects.  This situation will of course change
dramatically when the {\it Gaia} mission is completed (see
\citealt{tremblay17} for preliminary results). A hybrid photometric
and spectroscopic approach can also be used for fitting DQ and DZ
stars, where the effective temperatures and radii are measured from
the spectral energy distributions using the photometric method, while
the chemical composition is determined from fits to optical spectra,
in an iterative fashion \citep{dufour05,dufour07}.

Gravitational redshift measurements can also be used to measure white
dwarf masses of individual stars \citep{koester87,bergeron95a,reid96},
or average masses of different spectral types using statistical
ensembles \citep{falcon10,falcon12}. However, for individual
measurements, the white dwarf must be part of a common proper motion
system in order to determine its systemic velocity, and thus this
technique cannot be applied to the majority of field white dwarfs.

While the spectroscopic method is arguably the most precise method to
measure the atmospheric parameters of white dwarfs, it is not
necessarily the most accurate since theoretical line profiles are very
sensitive to the input physics (see \citealt{tb09} for instance). Also
the treatment of convective energy transport at low effective
temperatures has been demonstrated to affect significantly the
predicted line profiles and overall energy distribution
\citep{bergeron92,tremblay13}; for these reasons, most spectroscopic
determinations of the atmospheric parameters of DA stars in the past
have been restrained to $\Te>13,000$~K, where convective energy
transport becomes negligible. While the photometric method is
certainly less sensitive to these physical uncertainties, the issues
related to photometric calibrations, in particular when combining
various photometric systems, will always remain an important problem.
More importantly, no matter which method is used, one always needs to
rely on mass-radius relations to derive the mass either from $\logg$
values when using the spectroscopic method, or from the measured
radius $R$ when using the photometric method (if the distance is
known).

Given the new theoretical developments over the last few years
regarding model atmosphere calculations (Stark profiles, 3D
hydrodynamical calculations, etc. --- see references above), and given
the large amount of spectroscopic, photometric, and trigonometric
parallax data currently available to us, we felt it was appropriate to
revisit one more time, and with a more internally consistent approach,
the precision and in particular the accuracy of both the spectroscopic
and photometric methods using a well-defined white dwarf sample with
measured trigonometric parallaxes.  Such comparisons have already been
performed using surface gravities (see Figure 7 of \citealt{BSL92}),
stellar masses (see Figure 19 of \citealt{BLR01}), absolute magnitudes
(see Figure 29 of \citealt{gianninas11}), or even distances (see
Figure 22 of \citealt{bergeron11}). In all cases, these comparisons
always rely on the mass-radius relation for white dwarfs in order to
bring the various physical quantities on the same footing. Specific
studies aimed at testing this mass-radius relation for white dwarfs
include the analyses of trigonometric parallaxes measured by {\it
  Hipparcos} \citep{schmidt96,provencal98}, the analysis by
\citet{holberg12} based on parallax measurements of 12 DA stars from
different sources, while the most recent investigations by
\citet{tremblay17} and \citet{parsons17} rely, respectively, on
parallax measurements from the {\it Gaia} Data Release 1 and eclipsing
binaries. In a few instances, the mass-radius relation can be tested
for stars with a measured astrometric mass (see, e.g., Procyon B by
\citealt{provencal02}, and Sirius B by \citealt{barstow15} and
\citealt{bond17}), or even from the gravitational deflection of light
around a white dwarf --- as predicted by Einstein's general theory of
relativity --- in the particular case of Stein 2051 B \citep{sahu17}.

In this paper we present a detailed study of a large sample of
photometric, spectroscopic, and trigonometric parallax data aimed at
comparing the results of atmospheric parameters obtained from the
photometric and spectroscopic techniques, which are then used to test
the mass-radius relation for white dwarfs under various
assumptions. In Section \ref{data} we describe all the observational
data used in our spectroscopic and photometric analyses, the results
of which are presented and compared in Section \ref{atmo}. Our test of
the mass-radius relation follows in Section \ref{mr}, where we explain
our approach and present a thorough discussion of our findings. We
conclude in Section \ref{concl}.

\section{OBSERVATIONAL DATA} \label{data}

\subsection{Selection of the Sample} \label{sample}

The objects studied in our analysis were selected on the basis of two
criteria: the availability of trigonometric parallax measurements, and
the ability to constrain the surface gravity from spectroscopy. These
two requirements follow from the method we employ to investigate the
mass-radius relation, which will be described in detail in Section
\ref{method}. Firstly, trigonometric parallaxes are an essential
ingredient for testing the mass-radius relation, as shown in previous
studies \citep{provencal98,holberg12,tremblay17}, since they provide a
way to measure stellar radii directly without invoking the mass-radius
relation. Secondly, the approach we use to infer white dwarf masses
requires spectroscopic determinations of $\logg$, which involves
comparing observed and synthetic spectra to measure the atmospheric
parameters ($\Te$ and $\logg$).  Because the spectroscopic method can
only be applied successfully to hydrogen and helium line spectra, we
restrict our analysis to DA and DB white dwarfs that are hot enough to
have sufficiently strong absorption features, that is, DA stars with
$\Te\gtrsim5500$~K and DB stars with $\Te\gtrsim12,000$~K.

\subsection{Trigonometric Parallax Measurements} \label{pi}

After a careful search through the literature, we have defined a
sample of 206 DA and 13 DB white dwarfs with trigonometric parallax
measurements, for a total of 219 objects. The parallax measurements
used in our study are drawn from several sources. Our sample comprises
29 white dwarfs for which we adopt the remarkably precise parallax
measurements provided by the first data release of the {\it Gaia}
astrometric mission \citep{GAIA}.  We also rely on new results from
the CTIOPI and USNO parallax programs for 46 and 15 objects,
respectively \citep{sub09,sub17}. These new {\it Gaia}, CTIOPI and
USNO parallaxes have typical uncertainties of only a few percent ---
even less than 1\% in many cases --- making these white dwarfs highly
reliable candidates to test the mass-radius relation.

Another good source of parallaxes is the {\it Hipparcos} Space
Astrometry Mission. Our sample contains 16 objects for which the
parallax measurements were taken from the revised version of the {\it
  Hipparcos} catalog \citep{hipparcos07}. Most of these measurements
have uncertainties smaller than 10\%, a level of precision that is
quite satisfactory for the purpose of our analysis.  For 52 stars in
our sample, the parallaxes were drawn from the numerous lists
published by the USNO in the eighties
\citep{harr80,dahn82,dahn88,harr85}.  Although this set of
measurements is older than the more recent one described above, our
level of confidence in the reliability of the USNO data remains
high. The uncertainties are typically between 5 and 20\%, but can
reach nearly 40\% for some objects.  For 1639+153 and 2349$-$031, we
use more recent USNO data taken respectively from \citet{harris13} and
\citet{dahn04}. For 57 white dwarfs in our sample, parallax
measurements were taken from the Yale Parallax Catalog
\citep[][hereafter YPC]{YPC}; most of the uncertainties range from 5
to 25\%, but can reach almost 50\% in some cases. Finally, the
parallaxes for two additional objects, 1733$-$544 and 2351$-$368, were
taken from \citet{ruiz96} and \citet{ducourant07} with uncertainties
of approximately 1 and 10\%, respectively.

In the cases where multiple parallax measurements from different
sources are available for the same object, we adopt the following
order of priority: {\it Gaia}, CTIOPI/USNO-new, {\it Hipparcos},
USNO-old, YPC. This order reflects our level of confidence in the
different sources. In particular, we favor the more recent data, which
typically have smaller uncertainties than the older measurements.
However, we make an exception to this rule: if two parallaxes for the
same object show a large discrepancy, we adopt the value that yields
the best agreement with the mass-radius relation, because we have
reasons to believe that the other measurement is erroneous. The best
example is provided by 1314+293. As discussed by \citet{holberg12},
the {\it Hipparcos} parallax is larger than the YPC parallax by nearly
a factor of 2. If the YPC value is used, the star is consistent with
the mass-radius relation, and it is likely that the {\it Hipparcos}
parallax is simply in error for this object. This reasoning is
confirmed by the fact that the more recent {\it Gaia} measurement
(used in the present study) agrees with the YPC measurement.

\begin{figure}[p]
\centering
\includegraphics[width=\linewidth]{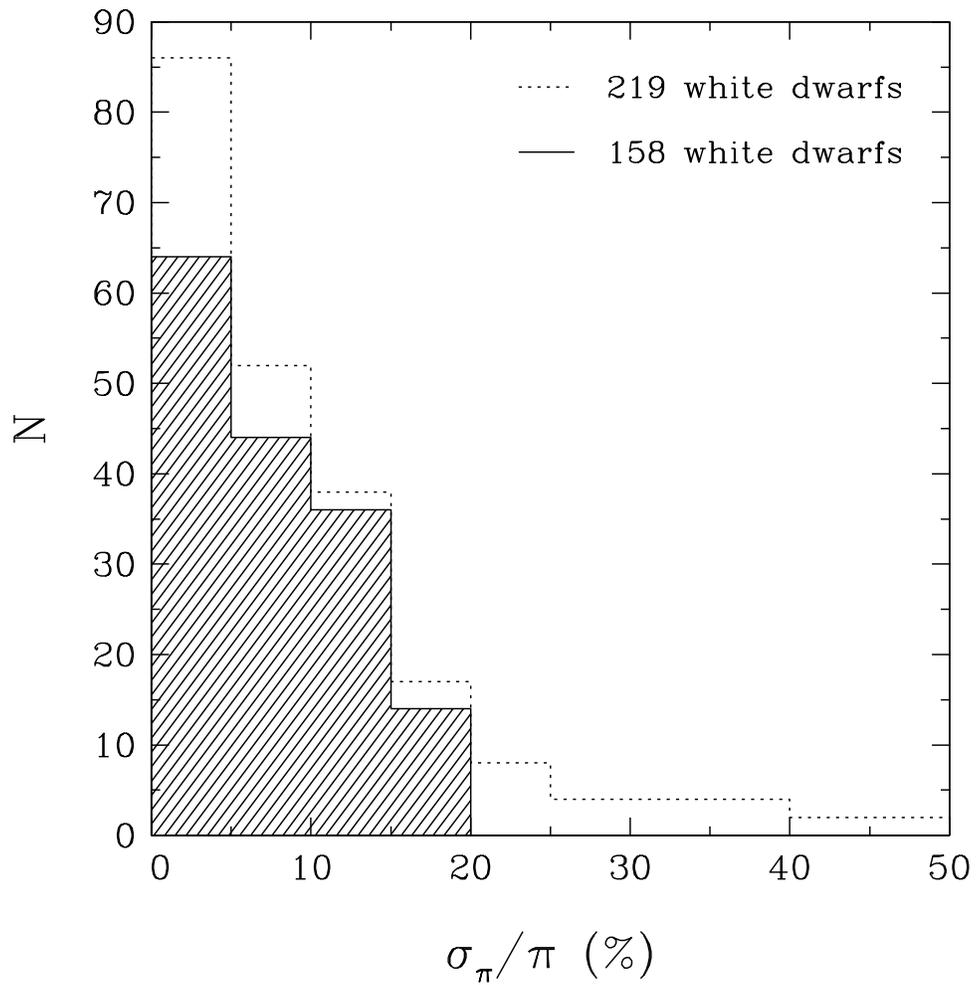}
\vspace*{-4cm}
\caption{Distribution of parallax uncertainties for the 219 white dwarfs in
  our sample. The hatched histogram shows the distribution for the 158 
  objects with reliable trigonometric parallax and spectroscopic data used 
  in our test of the mass-radius relation (see Section \ref{mr}).
  \label{N_Sigpi}}
\end{figure}

\begin{figure}[p]
\centering
\includegraphics[width=\linewidth]{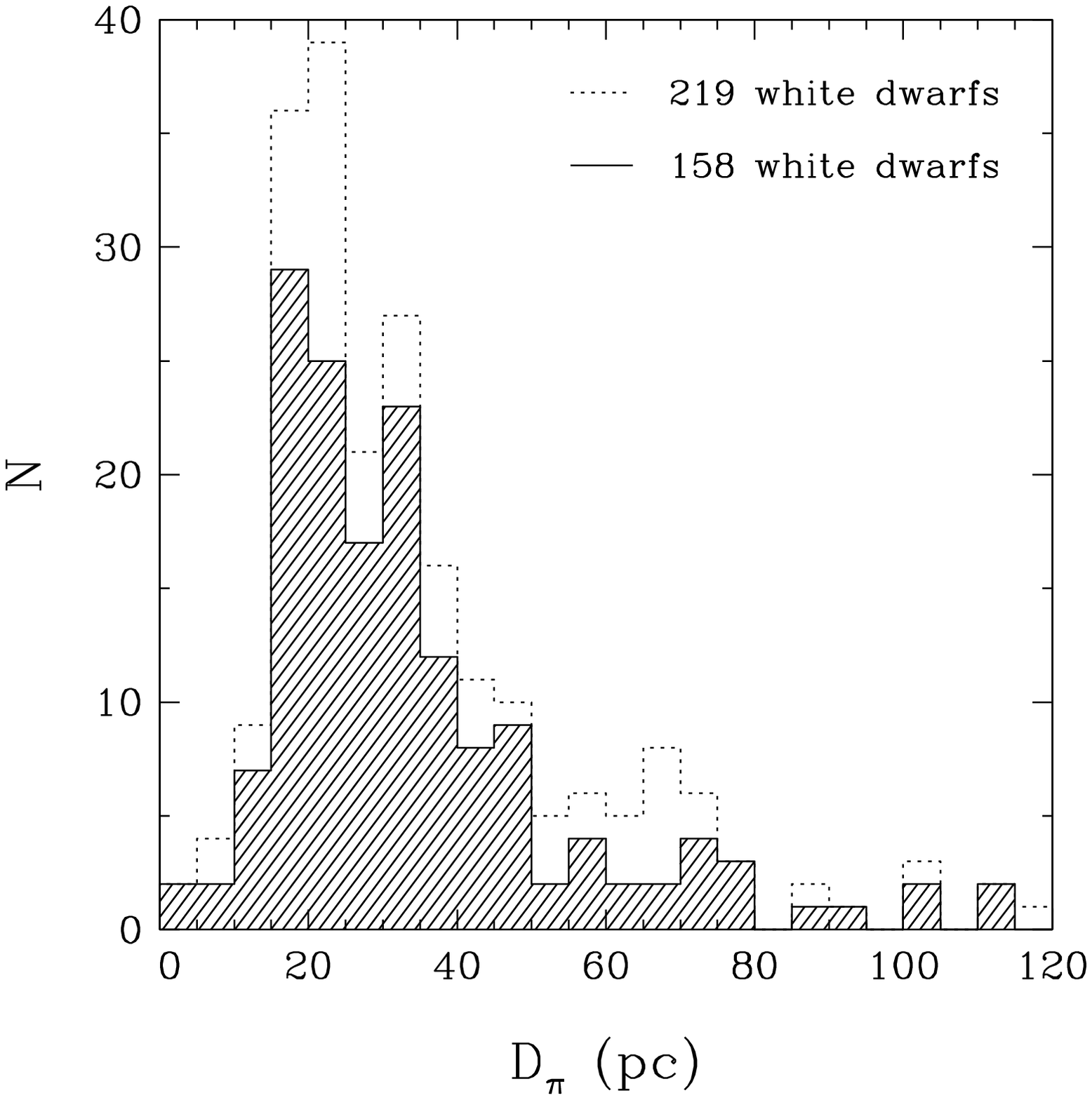}
\vspace*{-4cm}
\caption{Distribution of distances obtained from the parallax measurements 
  for the 219 white dwarfs in our sample. The hatched histogram shows the 
  distribution for the 158 objects with reliable trigonometric parallax and 
  spectroscopic data used in our test of the mass-radius relation (see 
  Section \ref{mr}).\label{N_Dpi}}
\end{figure}

The distribution of parallax uncertainties for the 219 white dwarfs in
our sample is displayed in Figure \ref{N_Sigpi}. Also shown is the
same distribution but for the 158 objects that were kept in our
refined sample used to investigate the mass-radius relation, after our
critical analysis of all the available trigonometric parallax,
spectroscopic, and photometric data, described in Section \ref{atmo}.
Finally, we show in Figure \ref{N_Dpi} the number of white dwarfs in
both samples as a function of distance.

\subsection{Spectroscopic Data} \label{spec_data}

Among the 206 DA white dwarfs included in our sample, four 
are DAZ stars since their spectra show traces of metals, and 
five are members of DA+M dwarf systems. Also, seven out of the 
13 DB white dwarfs have small amounts of hydrogen and 
are thus DBA stars. Five objects in our sample 
are confirmed to be weakly magnetic: 0257+080 \citep{koester09}, 
1953$-$011 \citep{koester98,koester09}, 2047+372 \citep{landstreet16}, 
2105$-$820 \citep{landstreet12}, and 2359$-$434 \citep{landstreet12}, 
but we have intentionally excluded all strongly magnetic white dwarfs, 
which cannot be modeled with sufficient accuracy in the context of our 
theoretical framework.

High signal-to-noise spectroscopic observations were gathered for all
white dwarfs in our sample, particularly in the blue portion of the
spectrum ($\lambda\sim3700-5200$ \AA) to cover the hydrogen Balmer
line series and most of the important neutral helium lines, but we
also secured spectra at \ha\ to constrain the hydrogen abundance in DB
stars. Most of the spectroscopic data for DA and DB stars were
retrieved from our previous white dwarf studies
\citep{BSL92,LBH05,sub07,sub08,sub09,bergeron11,gianninas11,nearby12,limoges15};
the spectral resolution in all cases ranges from 2 \AA\ to 9
\AA\ FWHM.  We also make use of high-resolution ($<1$ \AA\ FWHM)
observations from the ESO SN Ia Progenitor Survey \citep[SPY;][]{SPY}.

\subsection{Photometric Data} \label{phot_data}

For each object in our sample, we searched the literature for several
sets of photometric data, namely Johnson $BVRI$, SDSS $ugriz$, and
Str\"{o}mgren $uvby$ photometry in the optical, as well as Johnson
$JHK$ and 2MASS $JHK_s$ photometry in the infrared. Our purpose was to
constrain as best as possible the energy distribution for each star by
using {\it simultaneously} the available magnitudes in all bandpasses
during the fitting procedure.

The bulk of our Johnson $BVRI$ and $JHK$ photometry was taken from the
studies of cool white dwarfs by \citet[][hereafter BRL97 and BLR01,
  respectively]{BRL97,BLR01}, and from the study of ZZ Ceti stars by
\citet{bergeron09}. For many objects, $VRI$ magnitudes were taken from
the studies of white dwarfs in the solar neighborhood by
\citet{sub07,sub08,sub09,sub17}. We also collected photometric
measurements from additional sources, including $BVRI$ from
\citet{landolt92,landolt09,landolt13}, \citet{holberg08}, and
\citet{koen10}, as well as $BV$ from the USNO parallax lists mentioned
above and from \citet{kidder91}. The $ugriz$ photometry was drawn
directly from the SDSS database, while magnitudes in the Str\"{o}mgren
system were taken from several sources, namely \citet{graham72},
\citet{wegner79,wegner83}, \citet{lacombe81}, \citet{koester82}, and
\citet{fontaine85}. We also make use of photometric data in the
infrared from the Two Micron All-Sky Survey
\citep[2MASS;][]{2MASS}. For one object, 0030+444, we rely on
multichannel $UBGVRI$ photometry from \citet{gr76}.

\section{ATMOSPHERIC PARAMETERS DETERMINATION} \label{atmo}

\subsection{Spectroscopic Analysis} \label{spec_anal}

The spectroscopic analysis of the DA stars in our sample relies on the
so-called spectroscopic technique first described in \citet{BSL92} and
improved by \citet{bergeron95b} and \citet{LBH05}. Briefly, the
observed Balmer line profiles are normalized to a continuum set to
unity, either by using pseudo-Gaussian profiles or model spectra, and
then compared with model spectra convolved with the appropriate
Gaussian instrumental profile. A $\chi^2$ value is defined by
comparing observed and model spectra of all normalized Balmer lines
from \hb\ to H8, which is then minimized using the nonlinear
least-squares method of Levenberg-Marquardt \citep{press86} to obtain
the best fitting parameters, $\Te$ and $\logg$. The model atmospheres
and synthetic spectra we use for the DA stars are similar to those
described in \citet{tb09} and references therein, which are pure
hydrogen models where convective energy transport is treated within
the ML2/$\alpha=0.7$ version of the mixing-length theory, and where
non-LTE effects are taken into account above $\Te=30,000$~K. We also
apply the 3D corrections in both $\Te$ and $\logg$ given in
\citet{tremblay13} to take into account hydrodynamical effects. For
the DAZ stars in our sample, we rely on the fitting technique and
model atmospheres described in \citet{gianninas11}, which include the
opacity from Ca~\textsc{ii} H \& K known to contaminate the
H$\epsilon$ line.

Formal uncertainties in each fitted parameter can be obtained from the
covariance matrix of the fitting algorithm; these internal errors can
become very small if the signal-to-noise ratio of the spectrum is very
high, however, and these have never proven very useful in the context
of the spectroscopic method. Instead, multiple observations of the
same star on different nights, or even different observing runs,
provide a much more realistic estimate of the {\it external error},
which basically gives a measure of the repeatability of the
atmospheric parameter measurements.  By using this approach,
\citet{LBH05} estimated that for DA stars, uncertainties of 1.4\% in
$\Te$ and 0.042 dex in $\logg$ could be achieved\footnote{The
  uncertainties given in Section 2.4 of Liebert et al., 1.2\% in $\Te$
  and 0.038 dex in $\logg$, are erroneous, and those provided here are
  the correct values, as shown in Figure 8 of Liebert et al.}. It is
important to stress that these uncertainties do not only reflect the
precision of the spectroscopic method itself, but also strongly depend
on the quality of the observations.

Because the hydrogen lines reach their maximum strength near
$\Te\sim13,000$~K, two temperature solutions exist when using the
spectroscopic method, one on each side of the peak of the equivalent
widths. In the absence of photometric information, one needs to make
an educated guess at which of the cool or hot solution is the most
appropriate (see, e.g., \citealt{gianninas11}). This ambiguity is not
present in our analysis since we also derive photometric temperatures
for each star in our sample, which can be used to discriminate between
the cool and the hot solutions.

\begin{figure}[p]
\centering
\vspace*{-1cm}
\includegraphics[width=\linewidth]{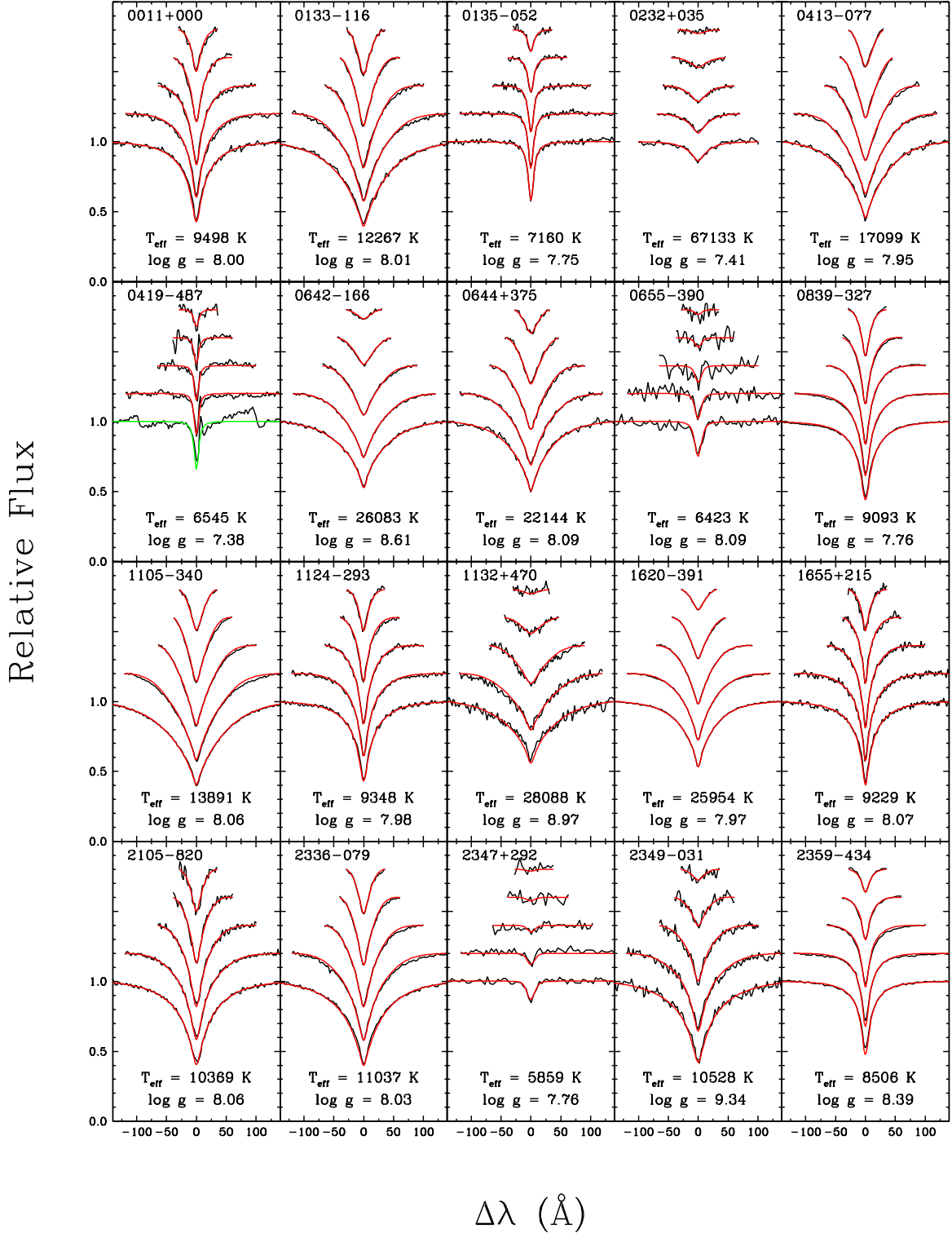}
\vspace*{-1cm}
\caption{Fits to the optical spectra of some DA stars in our
  sample. The lines range from \hb\ (bottom) to H8 (top), each offset
  vertically by a factor of 0.2.  Theoretical line profiles shown in
  green are not used in the fitting procedure. The 3D corrections
  from \citet{tremblay13} have also been applied for both $\Te$ and
  $\logg$.}\label{SampleFitsDA}
\end{figure}

Sample fits for DA stars of particular interest (see Section \ref{mr})
are displayed in Figure \ref{SampleFitsDA}. Since the accuracy of the
spectroscopic technique depends on well-defined absorption line
profiles, the determination of surface gravity for stars with either a
low effective temperature or a noisy spectrum can become somewhat
unreliable. For DA stars in particular, the presence of the high
Balmer lines in the optical spectrum is critical since these lines are
the most sensitive to variations of $\logg$. Thus, some objects
initially included in our sample were flagged as uncertain after a
careful examination of the spectroscopic results. All in all,
35 DA stars in our sample were too cool (32 objects) or had spectra of
too poor quality (3 objects) to allow a significant $\logg$
determination. Note that the $\logg$ values for these objects are not
necessarily inaccurate, but have to be considered uncertain in the
context of our test of the mass-radius relation.

For binary systems composed of a white dwarf and an M dwarf companion,
the optical spectrum of the former is usually contaminated by the
latter, especially at \hb. This is the case for one of the five WD + M
dwarf systems in our sample, 0419$-$487, shown in Figure
\ref{SampleFitsDA}. To circumvent this problem, we simply exclude the
\hb\ line from the fitting procedure for this object, and the
atmospheric parameters are derived from the uncontaminated absorption
lines.

Our sample also contains 13 DB white dwarfs, seven of which are DBA
stars.  The atmospheric parameters for these white dwarfs, including
also the hydrogen abundance $\nh$ --- or upper limits when \ha\ is
absent, are obtained using the same model atmospheres, synthetic
spectra, and fitting procedure as those described at length in
\citet[][see also \citealt{bergeron15}]{bergeron11}, and the details
will not be repeated here. It is worth noting, however, that we
exclude from our reliable subsample two DBA stars, one whose line
profiles are too weak to be fitted with the spectroscopic method
(1917$-$077), and one that is too cool to yield a reliable $\logg$
measurement (2147+280), a problem most likely related to inaccuracies
in the treatment of line broadening at low effective temperatures (see
\citealt{bergeron11} and references therein).

\subsection{Photometric Analysis} \label{phot_anal}

The photometric technique used in our analysis is described at length
in BRL97 (see also \citealt{nearby12} for further
improvements). Briefly, the magnitudes $m$ in all photometric systems
--- described in Section \ref{phot_data} --- are converted into
average fluxes, $f^{m}_{\lambda}$, using the zero points given in
\citet{holberg06}. These are compared with the model Eddington fluxes,
$H^{m}_{\lambda}$, properly averaged over the appropriate filter
bandpass. These two average fluxes are related by the equation

\begin{equation}
f^{m}_{\lambda} = 4\pi(R/D)^2H^{m}_{\lambda} \label{flux}
\end{equation}

\noindent where $R/D$ defines the ratio of the radius of the star to
its distance from Earth. A $\chi^2$ value is defined in terms of the
difference between observed and model fluxes over all bandpasses,
properly weighted by the photometric uncertainties, which is then
minimized using the method of Levenberg-Marquardt mentioned
previously. In the above equation, the Eddington fluxes
$H^{m}_{\lambda}$ depend on $\Te$, $\logg$, and chemical
composition. In our fitting procedure, both $\Te$ and the solid angle
$\pi(R/D)^2$ are considered free parameters, for an assumed chemical
composition, and the uncertainties of each fitted parameter are
obtained directly from the covariance matrix of the fitting
algorithm. Since trigonometric parallaxes are available for all
objects in our sample, the radius $R$ can be obtained directly from
the solid angle and the distance $D$. The value of $\logg$ used to
calculate the model fluxes $H^{m}_{\lambda}$ can be determined from
the radius combined with mass-radius relations, or alternatively, can
be set to the spectroscopic value. In standard photometric analyses
(e.g., BRL97, BLR01), the former approach is used, while the latter
approach is preferred when studying the validity of the mass-radius
relation for white dwarfs (see Section \ref{method}). We note,
however, that the particular choice of $\logg$ for estimating the
model fluxes does not affect the measured photometric temperatures
significantly, as demonstrated by \citet[][see their Figure
  7]{genest14}.

Because the radius measured from the photometric method depends
strongly on the precision and accuracy of the trigonometric parallax
measurements, BLR01 excluded from their analysis all parallaxes with
uncertainties larger than 30\%. Since we want to eventually test the
mass-radius relation for white dwarfs in our analysis described below,
we adopt a more severe constraint and exclude all objects with
uncertainties larger than 20\%, that is, 24 objects out of our initial
sample of 219 white dwarfs with measured parallaxes.

Particular attention must be paid to white dwarfs in DA+M dwarf
systems, since the M dwarf companion contaminates the energy
distribution in the infrared, and even in the optical in some
cases. To prevent this contamination, the infrared measurements for
0232+035, 0419$-$487, 0628$-$020, and 1314+293 were not included in
our photometric analysis. The $R$ and $I$ magnitudes were also omitted
in the cases of 0232+035 and 0419$-$487. For 0518+333, we use the
uncontaminated $BV$ and $JHK$ measurements from \citet{farihi09}.

Finally, a special treatment was required to analyze the energy
distribution of Sirius B (0642$-$166). Due to the overwhelming
luminosity of Sirius A, the only reliable photometric measurement
available is the $V$ magnitude, so the photometric technique cannot be
applied since magnitudes in several bandpasses are needed to constrain
the effective temperature. However, the spectrum we have for Sirius B
is that obtained by the {\it Hubble Space Telescope}
\citep{barstow05}, for which the observed flux is absolutely
calibrated. Hence, the energy distribution can be analyzed by
comparing absolute model fluxes to the observed spectrum at all
wavelength. Thus, in the framework of this spectrophotometric
approach, the spectroscopic data are used in place of the lacking
photometric measurements.

Photometric fits for all 219 objects in our complete sample are
displayed in Section \ref{mr} when we discuss the mass-radius relation
for white dwarfs.

\subsection{Comparison of Atmospheric Parameters} \label{comp}

In this section, we compare the physical parameters obtained from the
spectroscopic and photometric methods described above. For this
comparison, we exclude all objects with trigonometric parallax
uncertainties larger than 20\% (24 objects) or with unreliable
spectroscopic $\logg$ determinations (37 objects), as discussed in
Section \ref{spec_anal}. We thus end up with a subsample of 158 white
dwarfs with reliable trigonometric parallax and spectroscopic data,
displayed as the hatched histograms in Figures \ref{N_Sigpi} and
\ref{N_Dpi} above.

\begin{figure}[p]
\centering
\includegraphics[width=\linewidth]{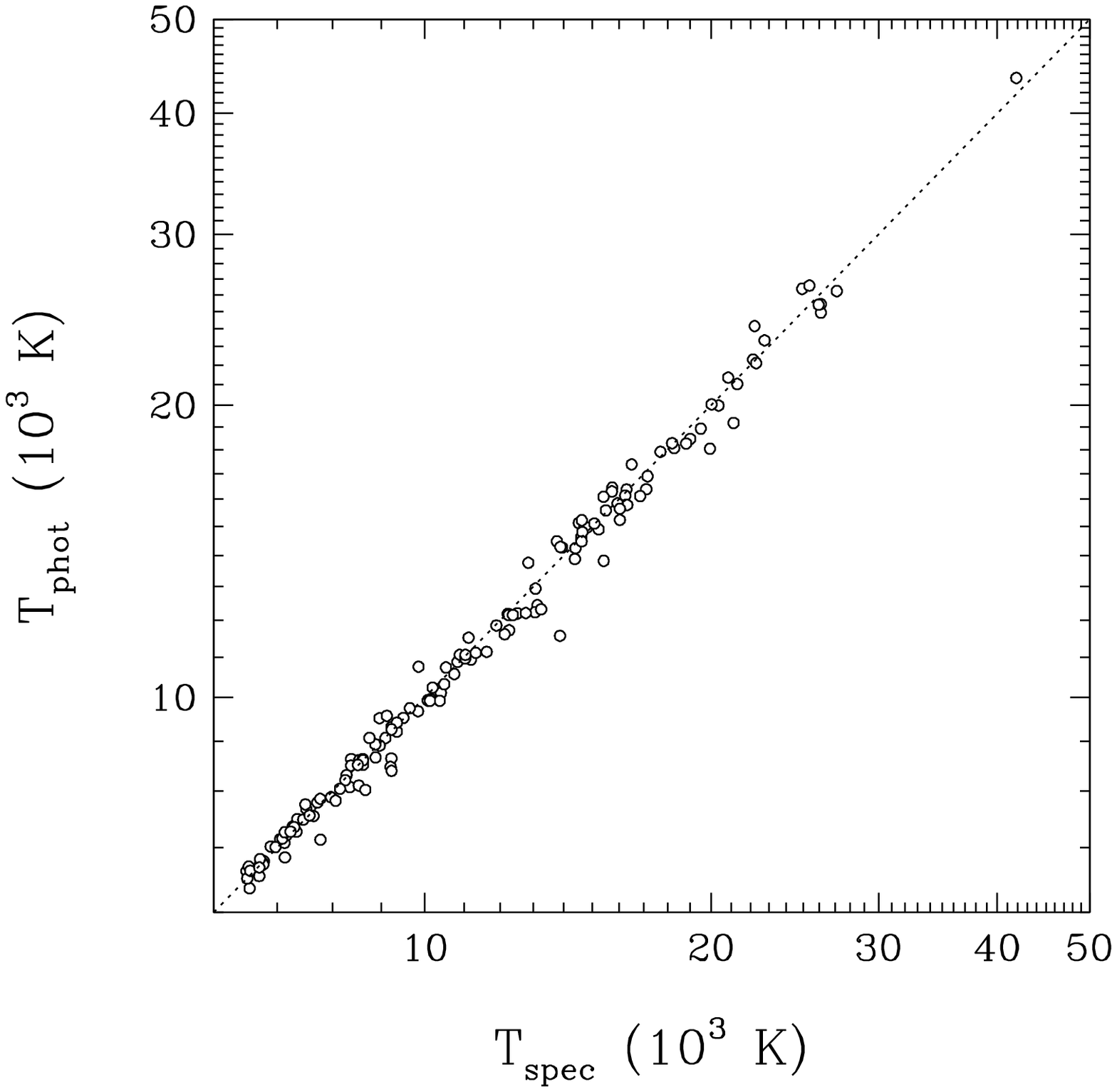}
\vspace*{-4cm}
\caption{Comparison of photometric and spectroscopic effective 
  temperatures for all white dwarfs in our sample with reliable 
  trigonometric parallax and spectroscopic $\logg$ measurements.
  \label{Tphot_Tspec_loglog}}
\end{figure}

\begin{figure}[p]
\centering
\includegraphics[width=\linewidth]{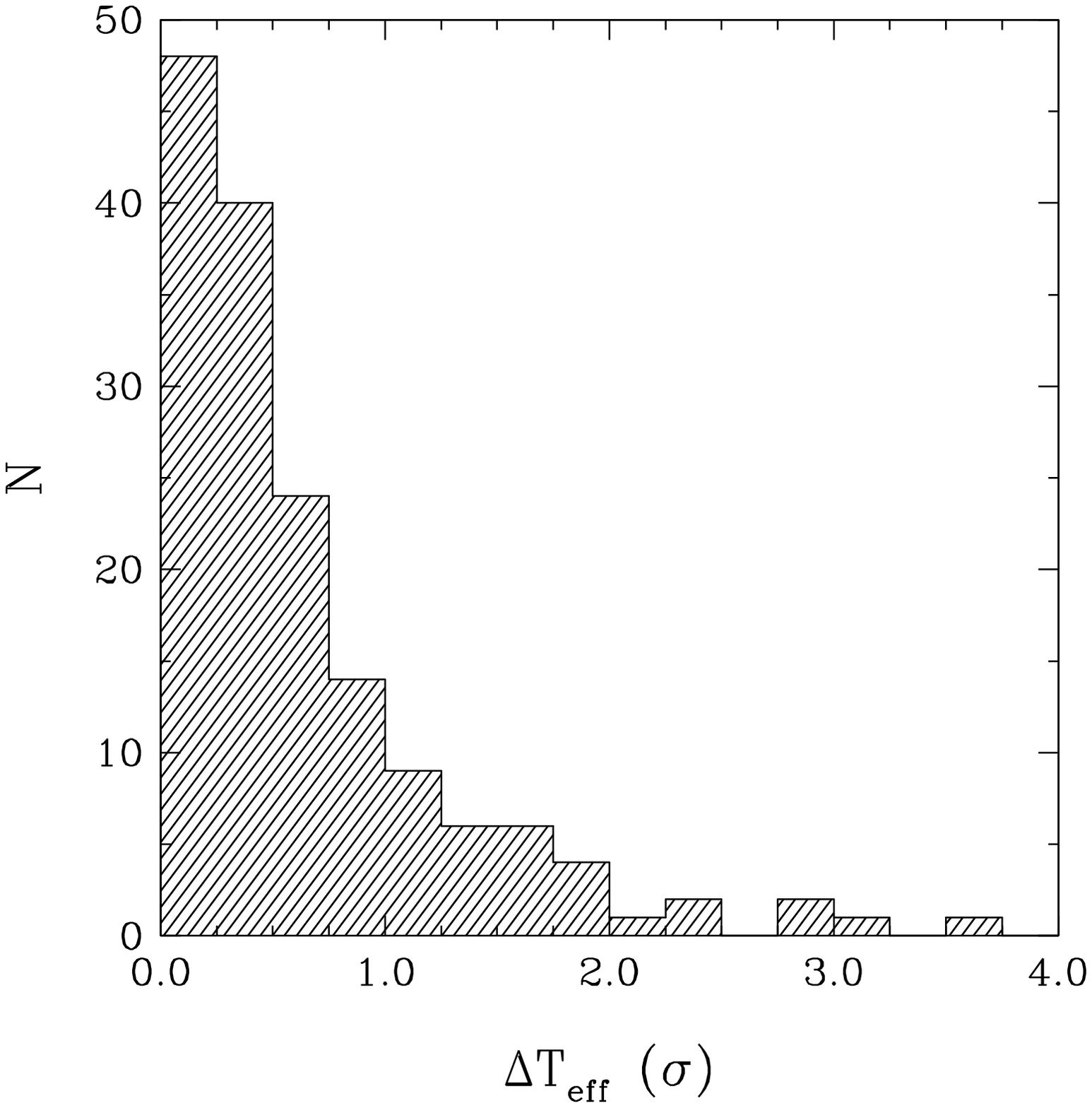}
\vspace*{-4cm}
\caption{Distribution of the (absolute) differences between photometric 
  and spectroscopic effective temperatures, measured in units of 
  $\sigma$, where $\sigma^2\equiv\sigma^2_{T_{\rm phot}}+\sigma^2_{T_{\rm spec}}$, 
  for all stars in our sample with reliable trigonometric parallax 
  and spectroscopic $\logg$ measurements.\label{N_NsigTeff}}
\end{figure}

The comparison of photometric and spectroscopic temperatures for all
white dwarfs in our restricted sample of 158 objects is displayed in
Figure \ref{Tphot_Tspec_loglog}, on a logarithmic scale. The agreement
is excellent through the entire temperature range explored here. The
difference between photometric and spectroscopic temperatures is
explored more quantitatively in Figure \ref{N_NsigTeff} where we show
the distribution of the absolute temperature differences, $\Delta
T_{\rm eff}=|T_{\rm phot}-T_{\rm spec}|$, measured in units of
$\sigma$, which is defined as the combined photometric and
spectroscopic uncertainties, $\sigma^2\equiv\sigma^2_{T_{\rm
    phot}}+\sigma^2_{T_{\rm spec}}$. These results indicate that among
the 158 white dwarfs composing our reliable subsample, the two
temperature measurements agree within 1$\sigma$ for 80\% of the stars,
and within 2$\sigma$ for 96\% of the stars, which is more than
satisfactory given that the values expected from a Gaussian error
distribution (implicitly assumed here) are 68.3 and 95.5\% at 1 and
2$\sigma$ confidence levels, respectively.

Furthermore, for some objects exhibiting differences larger than
1.5$\sigma$, the temperature values can be shown to be problematic due
to various reasons. For instance, several objects for which $\Delta
T_{\rm eff} > 1.5\sigma$ correspond to suspected or confirmed double
degenerate binaries: 0101+048 ($\Delta T_{\rm eff} =1.66\sigma$),
0142+312 ($\Delta T_{\rm eff}=2.92\sigma$), 0326$-$273 ($\Delta T_{\rm
  eff}=3.20\sigma$), 1447$-$190 ($\Delta T_{\rm eff}= 2.35\sigma$),
1606+422 ($\Delta T_{\rm eff}=1.57\sigma$), and 2111+261 ($\Delta
T_{\rm eff}=2.29\sigma$). These objects will be analyzed and discussed
in more detail in Section \ref{DD} following our test of the
mass-radius relation. For the moment, we will only point out that the
derived temperatures for these objects are not meaningful since they
were obtained under the assumption of single stars. For three other
white dwarfs, the photometric temperature is probably inaccurate. The
most striking case is that of 0232+035 ($\Delta T_{\rm
  eff}=3.57\sigma$), the hottest white dwarf in our sample as measured
by the spectroscopic method: $T_{\rm spec}=67,133$~K (not shown in
Figure \ref{Tphot_Tspec_loglog}).  At such high temperature, the
energy distribution in the optical and infrared portions of the
spectrum is almost completely independent of the effective
temperature, so the photometric method yields a value that is far too
low, $T_{\rm phot}=37,329$~K. In the cases of 0518+333 ($\Delta T_{\rm
  eff}=1.89\sigma$) and 2351$-$368 ($\Delta T_{\rm eff}=2.77\sigma$),
we only have two photometric measurements in the optical (in addition
to infrared photometry), so the corresponding photometric temperature
is poorly constrained. Finally, for one object, 2105$-$820 ($\Delta
T_{\rm eff}=1.94\sigma$), the spectroscopic temperature is probably
overestimated. This white dwarf is of particular interest regarding
the results of our investigation of the mass-radius relation, and we
thus defer our discussion of this object to Section \ref{rad}.

\begin{figure}[p]
\centering
\vspace*{-4cm}
\includegraphics[width=\linewidth]{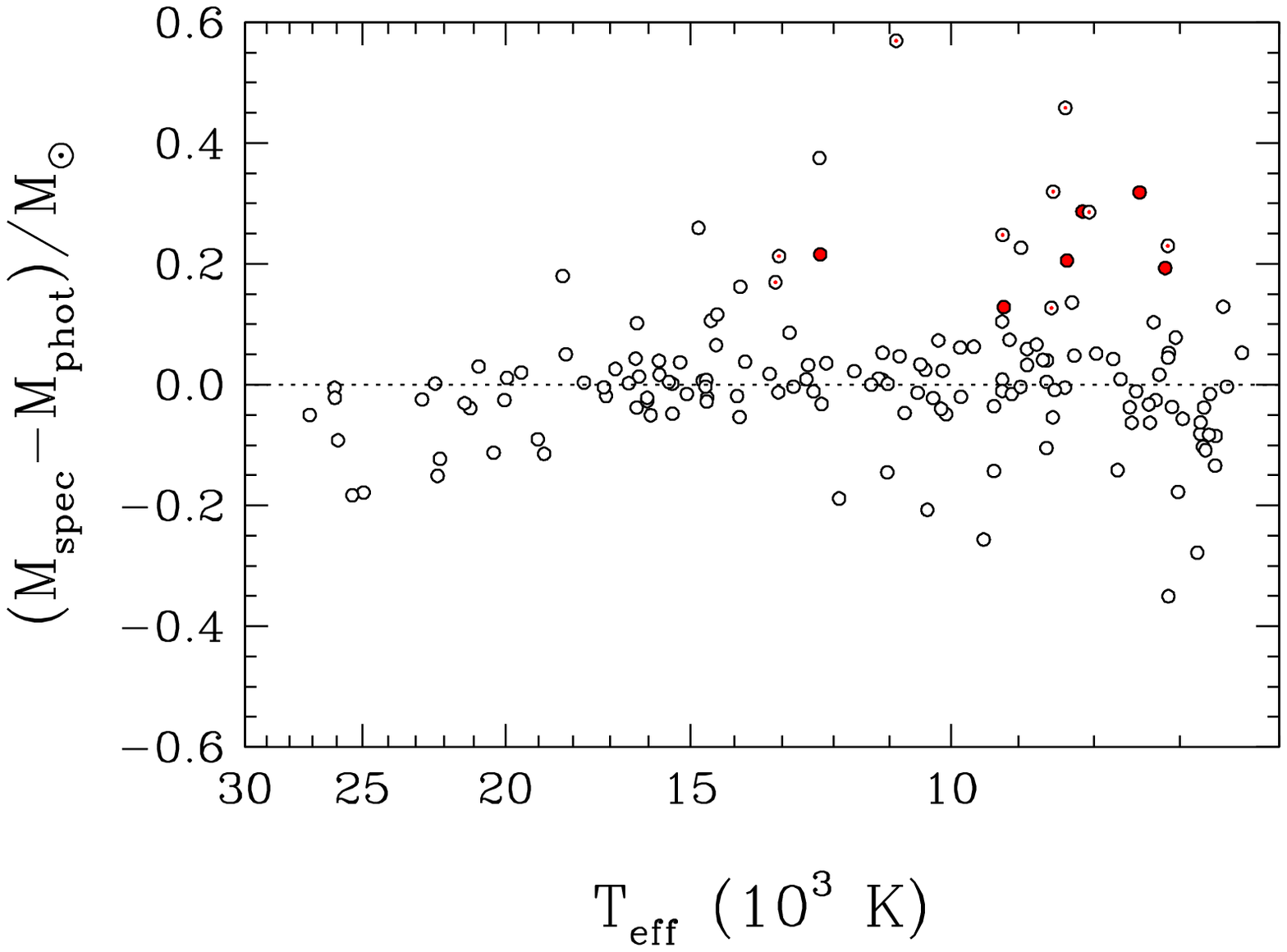}
\vspace*{-6cm}
\caption{Differences in mass obtained from the spectroscopic and
  photometric methods as a function of effective temperature for all
  white dwarfs in our sample with reliable trigonometric parallax
  and spectroscopic $\logg$ measurements. The objects shown as filled 
  red circles and dotted open circles correspond to confirmed and 
  suspected double degenerate binaries, respectively.
  \label{dM_logTeff}}
\end{figure}

The radius and surface gravity values obtained from the photometric
and spectroscopic methods, respectively, can be converted into mass
using the mass-radius relations with C/O cores and thick hydrogen
layers (for hydrogen-rich stars) or thin hydrogen layer (for
helium-rich stars) described in Section \ref{mr_models}. We insist on
the fact that the mass-radius relation is used here only to bring both
measured quantities on an equal footing for comparison, and thus what
follows in the present section is {\it not} part of our test of the
mass-radius relation. The differences between these spectroscopic and
photometric masses are shown as a function of effective
temperature\footnote{In this figure we adopt the spectroscopic or
  photometric effective temperature depending on criteria discussed in
  Section \ref{method}.} in Figure \ref{dM_logTeff} for our 158
reliable objects. On average, the spectroscopic and photometric masses
appear to be consistent, with no obvious systematic trend. Both mass
values agree within 0.1 \msun\ for 70\% of the stars. However, it is
clear that there are significant discrepancies, with differences as
large as $\sim$0.3 \msun\ in both directions, and even more in some
instances. Many of these discrepancies can be explained by the
presence of double degenerate binary systems; in such cases the
objects appear overluminous if analyzed with the photometric technique
under the assumption of a single star, and the radius is thus
overestimated and the photometric mass underestimated. This is
illustrated in Figure \ref{dM_logTeff}, where the known systems are
identified as filled red circles (these correspond to 0101+048,
0135$-$052, 0326$-$273, 1242$-$105, 1639+153, and 1824+040). Most
other objects that have $M_{\rm spec}-M_{\rm phot}>0.1$
\msun\ (including the two most discrepant points for which $M_{\rm
  spec}-M_{\rm phot}>0.4$ \msun) are suspected double degenerates that
will be discussed in Section \ref{DD} (these correspond to 0126+101,
0142+312, 0311$-$649, 1130+189, 1418$-$088, 1447$-$190, 1606+422,
2048+809, and 2111+261). These objects are shown as dotted open
circles in Figure \ref{dM_logTeff}. Some white dwarfs for which
$M_{\rm spec}-M_{\rm phot}<0.1$ \msun, such as 0644+375 and
2105$-$820, will also be investigated on a star by star basis in
further sections of this work, where various scenarios that could
account for the observed discrepancies are explored.

\begin{figure}[p]
\centering
\includegraphics[width=\linewidth]{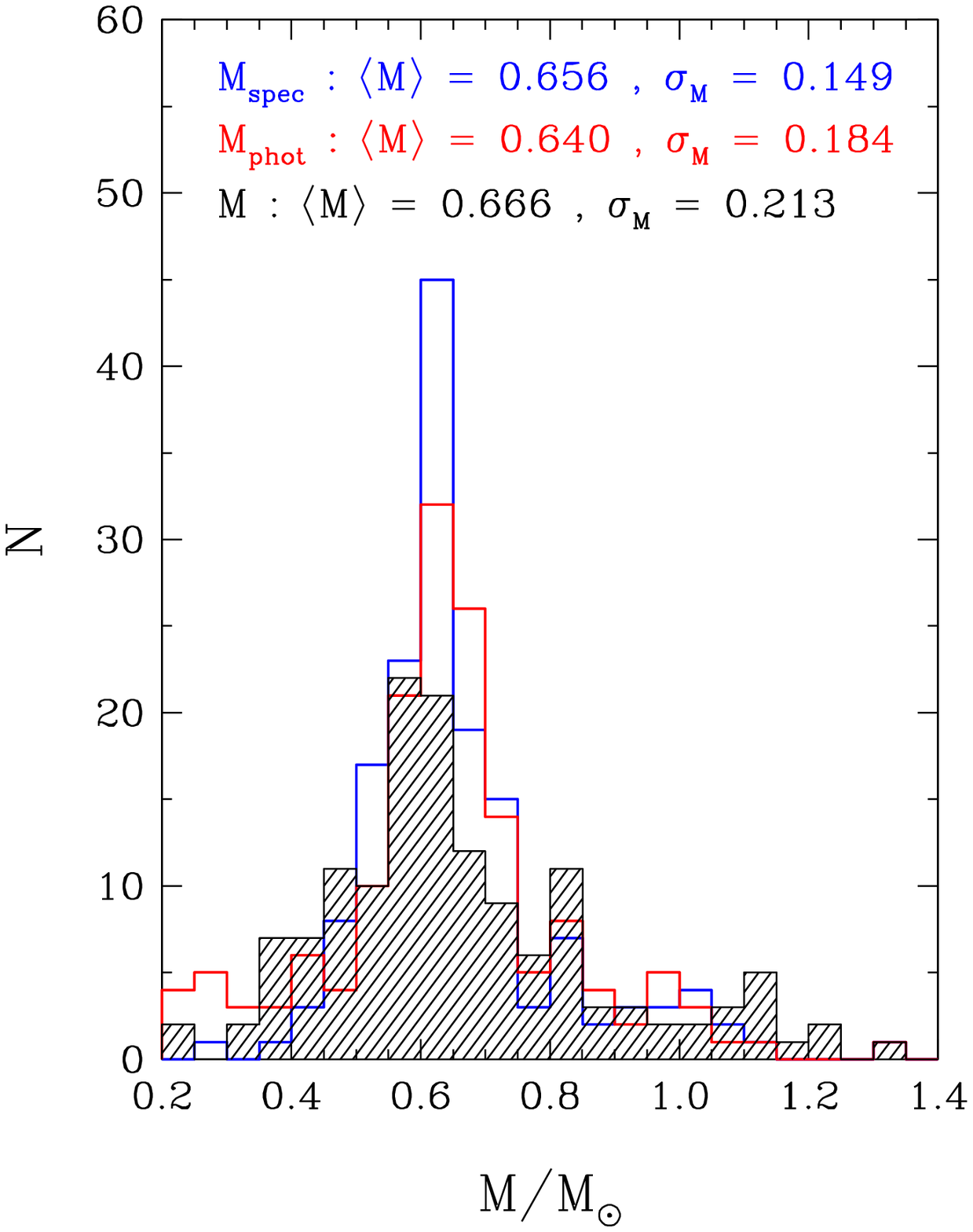}
\vspace*{-4cm}
\caption{Comparison of the mass distributions obtained from the
  spectroscopic (blue) and photometric (red) methods for all white 
  dwarfs in our sample with reliable trigonometric parallax and 
  spectroscopic $\logg$ measurements. The mean masses and standard 
  deviations, in \msun\ units, are given in the figure. Also shown in 
  black is the mass distribution obtained independently of the 
  mass-radius relation (see Sections \ref{method} and 
  \ref{validity}).\label{N_M}}
\end{figure}

The cumulative mass distributions using both spectroscopic and
photometric methods, regardless of the effective temperature, are
displayed in Figure \ref{N_M} in blue and red, respectively. Note that
the mass distributions include here both DA and DB stars. The mean
masses obtained using the two fitting techniques are entirely
consistent, near $\sim$0.65 \msun, and these values are comparable to
other values reported in the literature. We note a more pronounced
extended low-mass tail for the photometric approach, which results
from the presence of the unresolved degenerate binaries discussed in
the previous paragraph. We also note the smaller dispersion and the
sharper peak of the mass distribution obtained from spectroscopy, most
likely because of the higher precision of the spectroscopic technique
when applied to large {\it ensembles} of objects (but not necessarily
for individual objects), and also because of the large trigonometric
parallax uncertainties for some objects in our sample, which translate
into photometric masses of low precision. The third mass distribution
shown in black in the figure is obtained from the method employed to
test the mass-radius relation and will be discussed later, in Section
\ref{validity}.

\section{TEST OF THE MASS-RADIUS RELATION} \label{mr}

\subsection{General Approach} \label{method}

\begin{figure}[p]
\centering
\vspace*{-8mm}
\includegraphics[width=0.85\linewidth]{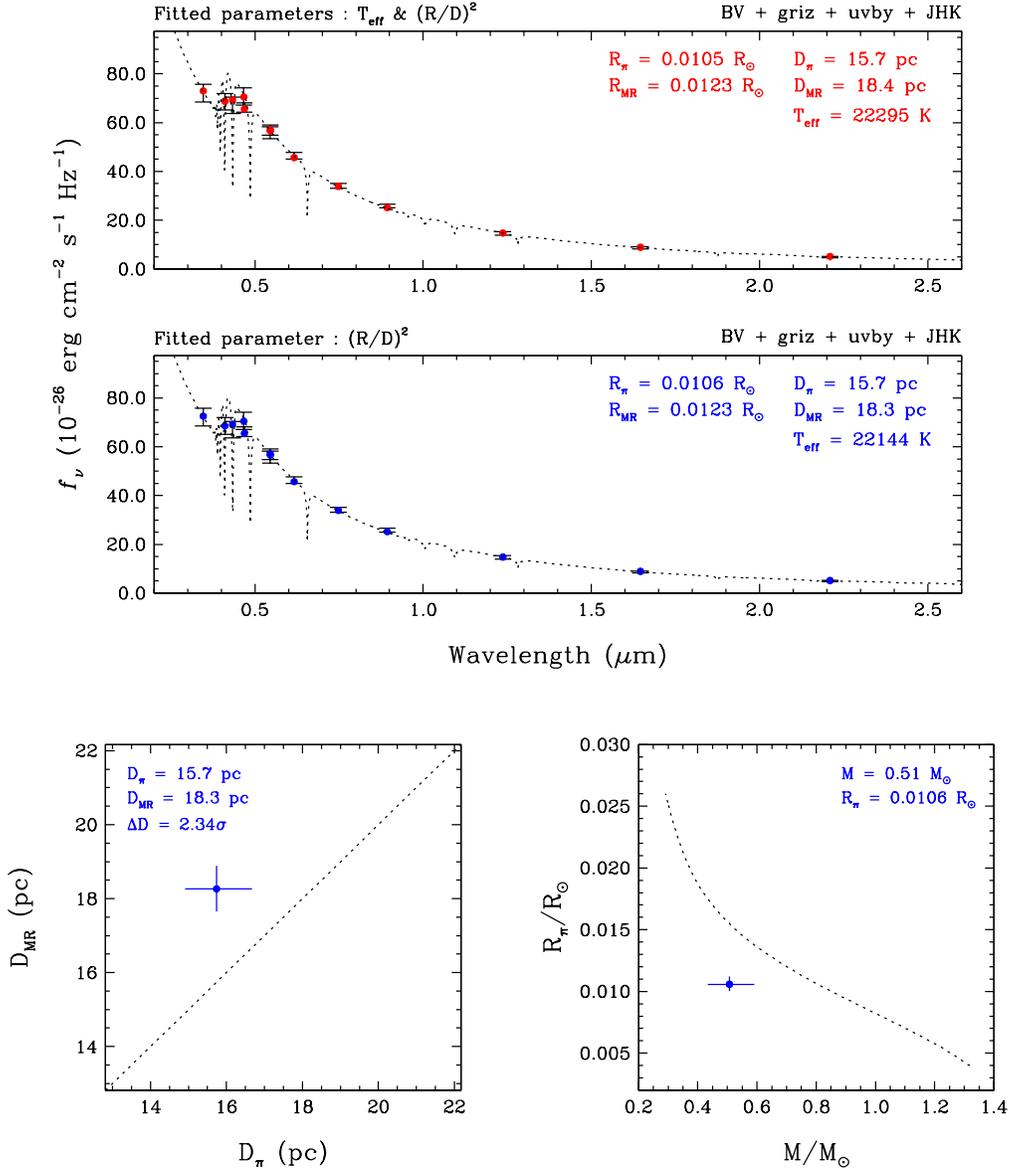}
\vspace*{-6mm}
\caption{{\it Top panels}: Fits to the energy distribution of 0644+375 
  (G87-7), with the effective temperature considered as a free parameter 
  (first panel, in red) or fixed at the spectroscopic value (second panel, 
  in blue). Photometric observations are represented by error bars, while 
  average model fluxes are displayed as filled circles. The monochromatic 
  model flux is also shown as the dotted line. {\it Bottom panels}: 
  Location of 0644$+$375 (G87-7) in the $D_{\rm MR}$ versus $D_\pi$ diagram 
  (left panel) and in the $R_\pi$ versus $M$ diagram (right panel). The 
  dotted lines represent the 1:1 correspondence (left panel) and the 
  mass-radius relation for C/O-core, thick hydrogen envelope models at the 
  derived effective temperature (right panel). The 
  color used in both plots indicates which of the photometric fits is 
  adopted. \label{G87-7}}
\end{figure}

Our empirical test of the mass-radius relation relies on the combined
use of the atmospheric and stellar parameters determined from both the
spectroscopic and photometric techniques. The spectroscopic analysis,
which provides values of $\Te$ and $\logg$, constitutes the first step
in our investigation, since the spectroscopic $\logg$ is needed as
input for the calculation of the model fluxes in the photometric
analysis, as outlined above. Then, the photometric technique is used
to obtain the solid angle $\pi(R/D)^2$; this can be achieved in two
different fashions: one can either consider both the effective
temperature and the solid angle as free parameters during the fitting
procedure (which yields an estimate of the photometric temperature),
or one can set the effective temperature at the spectroscopic value
and consider only the solid angle as a free parameter. This is
illustrated in the top two panels of Figure \ref{G87-7} where we show
our best photometric fits for the DA star 0644+375 (G87-7) as an
example; in the first panel both $\Te$ and $(R/D)^2$ are free
parameters, while in the second panel $\Te$ is forced to its
spectroscopic value and only $(R/D)^2$ is a free parameter. (The full
spectroscopic solution is also indicated as a reference at the top of
the plot together with the source of the trigonometric parallax
measurement.) For the majority of stars in our sample, we choose the
latter solution and hence we adopt the spectroscopic temperature for
the rest of our analysis, given our high level of confidence in the
atmospheric parameters provided by the spectroscopic method. For some
stars whose optical spectrum is noisy or shows weak features, we
prefer to adopt the photometric temperature.  However, it should be
noted that this choice between the two approaches does not affect the
resulting value of $(R/D)^2$ significantly, since the overall
agreement between the spectroscopic and photometric temperatures is
very good, as shown in Section \ref{comp} --- a 0.7\% difference in
the case of G87-7. The color used in the two bottom panels of Figure
\ref{G87-7}, described in the next paragraph, reflects our choice of
photometric (red) or spectroscopic (blue) temperature.

The mass-radius relation can be investigated using a method similar to
that employed by \citet{provencal98} for field white dwarfs. The value
of $(R/D)^2$ is combined with the star's distance from Earth, obtained
directly from the trigonometric parallax, to derive the stellar radius
(exactly as explained in Section \ref{phot_anal}), which we
respectively label $D_\pi$ and $R_\pi$ in Figure \ref{G87-7} and other
following figures. The surface gravity obtained from spectroscopy and
the radius are then used to determine the stellar mass $M$ through the
well-known relation $\textsl{\textrm{g}}=GM/R^2$. Thus, we end up with
radius and mass estimates that are independent of any mass-radius
relation. The derived values can then be compared to the predicted $R$
versus $M$ curve at a given effective temperature and for a specific
core composition and hydrogen layer thickness. This is illustrated in
the bottom right panel of Figure \ref{G87-7} for the particular case
of G87-7, where we use the C/O-core mass-radius relations with thick
hydrogen layers discussed in Section \ref{mr_models} below.

Another way to test the mass-radius relation is to compare the
distance obtained from the parallax, $D_{\pi}$, to another estimate of
the distance that depends on the mass-radius relation, which we denote
$D_{\rm MR}$ in Figure \ref{G87-7} and subsequent figures. This
distance is calculated by first using evolutionary models to convert
the spectroscopic $\logg$ into radius --- labeled $R_{\rm MR}$ in
Figure \ref{G87-7} --- which is then combined with the photometric
value of $(R/D)^2$ to obtain the desired distance $D_{\rm MR}$. This
$D_{\rm MR}$ versus $D_{\pi}$ comparison, illustrated in the bottom
left panel of Figure \ref{G87-7}, has the advantage over the $R_\pi$
versus $M$ comparison that the x-axis (the parallax distance $D_\pi$)
is a well-understood, model-independent quantity, for which the error
bars simply originate from the parallax measurement uncertainty. Also,
by comparing two estimates of the same quantity, the level of
consistency can be easily evaluated in terms of the absolute
difference between the two values expressed in units of the combined
uncertainties $\sigma$, as was done in Section \ref{comp} for
spectroscopic and photometric temperatures. This approach is explained
further in Section \ref{results}.

As mentioned in Section \ref{phot_anal}, the uncertainties on $\Te$
and the solid angle $\pi(R/D)^2$, measured using the photometric
technique, are obtained directly from the covariance matrix of the
Levenberg-Marquardt minimization method. The {\it internal} errors on
$\Te$ and $\logg$ using the spectroscopic technique are obtained in
the same way, but these are combined in quadrature with the {\it
  external} errors, which for the moment we assume to be 1.4\% in
$\Te$ and 0.042 dex in $\logg$, following \citet{LBH05}. The
uncertainties for all other quantities derived from these parameters
--- for instance the radius $R_\pi$ obtained from the solid angle
combined with the measured trigonometric parallax --- are calculated
by propagating in quadrature the appropriate measurement errors
\citep[see also][]{holberg12}. These uncertainties, representing the
1$\sigma$ confidence level, are displayed as error bars in the bottom
panels of Figure \ref{G87-7}. In the particular case of G87-7, a
significant 2.34$\sigma$ difference is observed between the distance
obtained from the trigonometric parallax, $D_\pi=15.7 \pm 0.9$ pc, and
the distance obtained using the mass-radius relation, $D_{\rm MR}=18.3
\pm 0.6$ pc. Accordingly, G87-7 is located quite far from the
mass-radius relation in the $R_\pi$ versus $M$ diagram of Figure
\ref{G87-7}.

\subsection{Mass-Radius Relations} \label{mr_models}

In a few of the results presented so far, and in the remainder of this
analysis, we make extensive use of mass-radius relations using
evolutionary models described at length in \citet{fontaine01} but with
different core compositions and thicknesses of the hydrogen
layers. Our reference stellar models for the DA stars in our sample
have C/O-cores and ``thick'' hydrogen envelopes; that is, a core
consisting of a uniform mixture of carbon and oxygen in equal
proportions ($X_{\rm C} = X_{\rm O} = 0.5$) surrounded by a helium
mantle of $q({\rm He})\equiv M_{\rm He}/M_{\star}=10^{-2}$ and an
outermost hydrogen layer of $q({\rm H})=10^{-4}$. It is of interest to
point out here that the mass-radius relation is not sensitive to the
exact distribution of C and O in the core. For example, a standard 0.6
\msun\ ``thick'' envelope DA model with a uniform (50/50) C/O-core
composition has a total radius at $\Te$ = 15,000 K that is only 0.06\%
larger than the radius of an equivalent model, but with a detailed
nonuniform C/O stratification (with $X_{\rm C}$ = 0.218 and $X_{\rm
  O}$ = 0.782 at the center and becoming increasingly O-poorer in the
above layers) obtained from the calculations of \citet{salaris97},
which incorporate the effects of stellar evolution from the ZAMS. This
difference in radius is a totally negligible effect in the present
context.

We also make use of models with ``thin'' hydrogen envelope, identical
to the thick models described above but with a much thinner hydrogen
layer of only $q({\rm H})=10^{-10}$, which are representative of
hydrogen-atmosphere white dwarfs with thin hydrogen envelopes.  Such a
small layer of hydrogen does not change in any significant way the
mass-radius relation with respect to models with no hydrogen, and from
that point of view, that same mass-radius relation can also be used
for helium-atmosphere white dwarfs. As discussed in BLR01, while it is
possible with our code to compute evolutionary sequences with pure
helium envelopes, the cooling times show an extreme sensitivity to the
presence of even very small traces of heavy elements in the outer
layers, leading to a continuum of ages, hence it is usually preferred
to use models with a thin hydrogen layer for DB stars.

Additional models with a different core composition were calculated 
for the specific purpose of our analysis; these models have pure iron 
cores surrounded by helium and hydrogen layers identical to that of 
our thick models, i.e., $q({\rm He})=10^{-2}$ and $q({\rm H})=10^{-4}$. 
The essential input physics used in this context is the dense-plasma
equation-of-state code developed by \citet{lamb75} as appropriate for
the fully ionized liquid/solid interior of a white dwarf. We used that
code to compute the needed equation-of-state data for pure
Fe. Those data largely specify the mass-radius relation of a model. 
Our Fe-core models will be used and discussed in Section \ref{Fe-core}.

\begin{figure}[p]
\centering
\includegraphics[width=\linewidth]{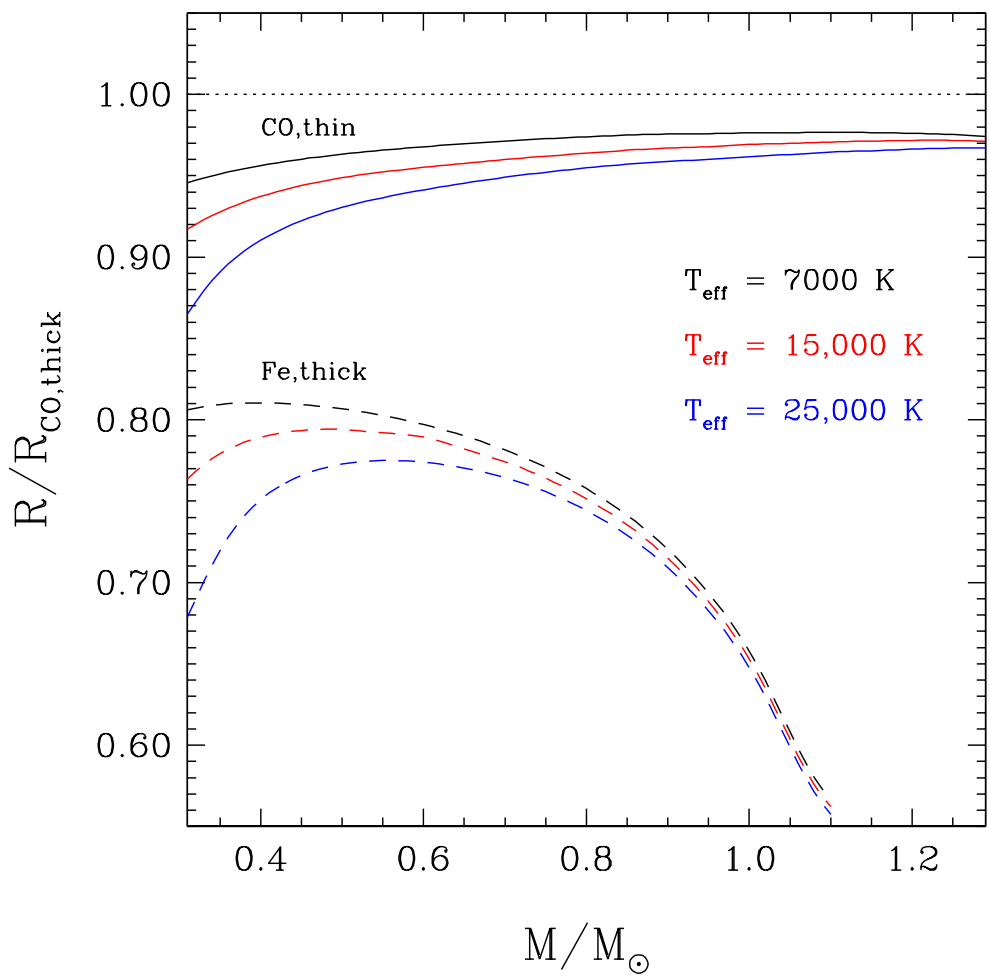}
\vspace*{-4cm}
\caption{Ratio of the radii obtained from thin hydrogen envelope 
  ($q({\rm H})=10^{-10}$) C/O-core models (solid lines) and from 
  thick hydrogen envelope ($q({\rm H})=10^{-4}$) Fe-core models 
  (dashed lines) with respect to our reference thick hydrogen envelope 
  ($q({\rm H})=10^{-4}$) C/O-core models, as a function of mass, for 
  three representative values of effective temperature given in the
  figure.\label{figMR}}
\end{figure}

In Figure \ref{figMR}, inspired from Figure 1 of \citet{tremblay17},
we show the ratio of the radii obtained from our thin and thick
envelope models (with our standard C/O-cores) as a function of mass
for three values of effective temperature, which are representative of
our white dwarf sample. Also shown in Figure \ref{figMR} is the same
ratio but between the radii calculated from our Fe-core and C/O-core
models (both with a thick hydrogen layer). As can be seen from these
results, the thickness of the hydrogen layer has only a moderate
effect ($\lesssim 5\%$) on the expected radius for normal $\sim$0.6
\msun\ white dwarfs, and above. A more important effect is observed
for the core composition, which yields differences as large as 20\% in
radius for normal $\sim$0.6 \msun\ Fe-core white dwarfs with respect
to our reference C/O-core models, and up to 40\% at 1.0 \msun. We also
notice the importance of finite-temperature effects on these
mass-radius relations, in particular for the low-mass models.

\subsection{Global Results} \label{results}

The results for all white dwarfs in our sample --- identical in format
to those shown in Figure \ref{G87-7} for G87-7 --- are provided as
supplementary online figures, available on our Website\footnote{
  http://www.astro.umontreal.ca/$\sim$bergeron/BedardApJ}. If an object
was rejected from our analysis on the basis of unreliable parallax or
spectroscopic $\logg$ measurement, as discussed above, it is flagged
with the label ``R'' placed in the upper right corner of the figure.
Again, the particular color used in the two bottom panels indicates
whether the photometric (red) or spectroscopic (blue) temperature is
adopted in the photometric fit. The corresponding values of the
parameters $D_\pi$, $D_{\rm MR}$, $M$ and $R_\pi$ are given in these two
panels. In the $D_{\rm MR}$ versus $D_\pi$ diagram, we also give, as a
quantitative measure of consistency with the C/O-core mass-radius
relation, the value of the absolute difference between the two distance
estimates, $\Delta D = |D_{\rm MR}-D_\pi|$, expressed in units of the
combined uncertainties $\sigma$, defined as $\sigma^2\equiv\sigma^2_{D_
  {\rm MR}} +\sigma^2_{D_{\pi}}$.

\begin{figure}[p]
\centering
\includegraphics[width=\linewidth]{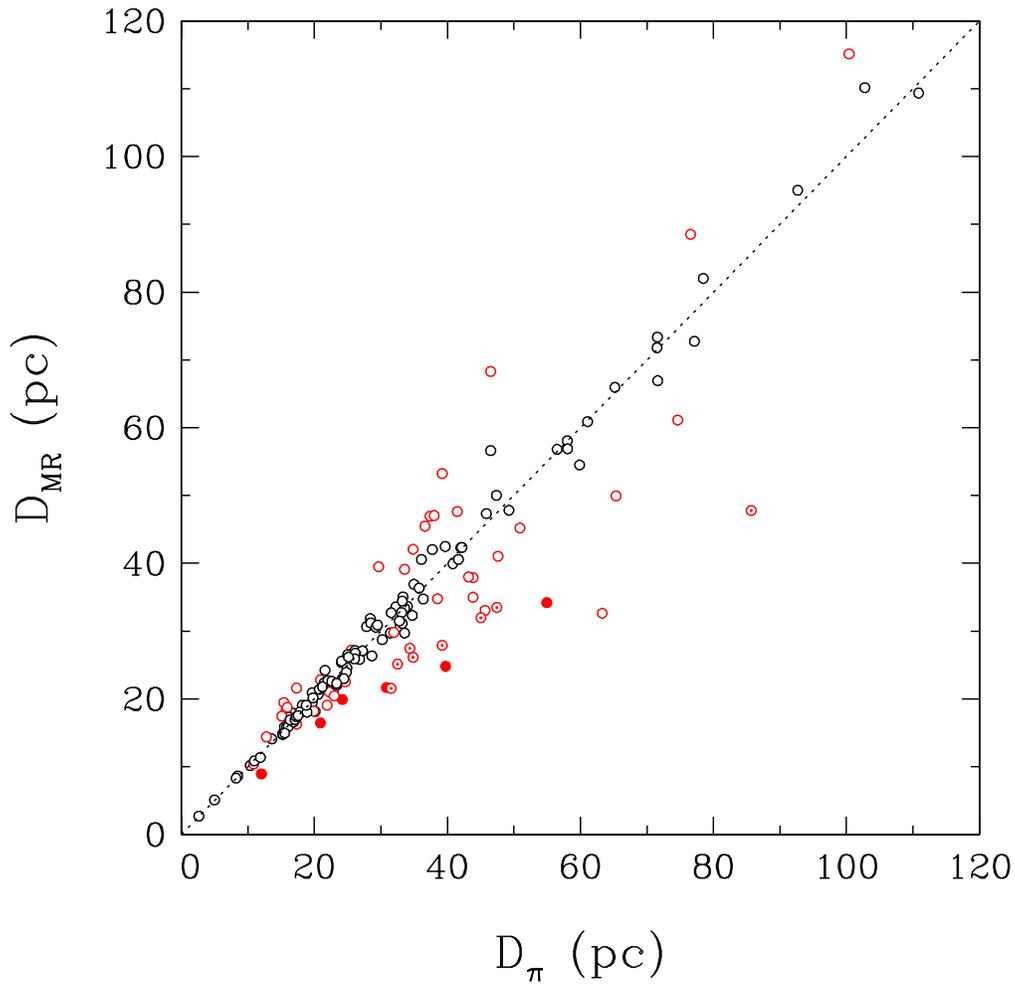}
\vspace*{-4.5cm}
\caption{Comparison of the distances inferred from the mass-radius
  relation, $D_{\rm MR}$, with those obtained directly from the
  trigonometric parallax measurements, $D_\pi$, for all white dwarfs
  in our sample with reliable trigonometric parallax and spectroscopic
  $\logg$ measurements (158 objects). The dotted line indicates the
  1:1 correspondence. The stars shown in red (55 objects) exhibit
  differences larger than a 1$\sigma$ confidence level between the two
  distance estimates. The filled red circles and the dotted open
  circles represent, respectively, known and suspected unresolved
  double degenerate systems.\label{Dmr_Dpi}}
\end{figure}

The comparison of distance estimates is summarized in Figure
\ref{Dmr_Dpi} for the 158 white dwarfs with reliable trigonometric
parallax and spectroscopic $\logg$ measurements in our sample. The 55
objects with distance differences larger than a 1$\sigma$ confidence
level are displayed in red. Also, known unresolved degenerate binaries
are shown as filled red symbols in the same figure; their location in
this diagram suggests that some of the most discrepant results can
probably be explained in terms of unresolved double degenerates, as
noted in Section \ref{comp}. We explore this scenario more
quantitatively in Section \ref{DD} (the double degenerate candidates
that will be identified and discussed in that section are displayed as
dotted open circles in Figure \ref{Dmr_Dpi}).

\begin{figure}[p]
\centering
\includegraphics[width=\linewidth]{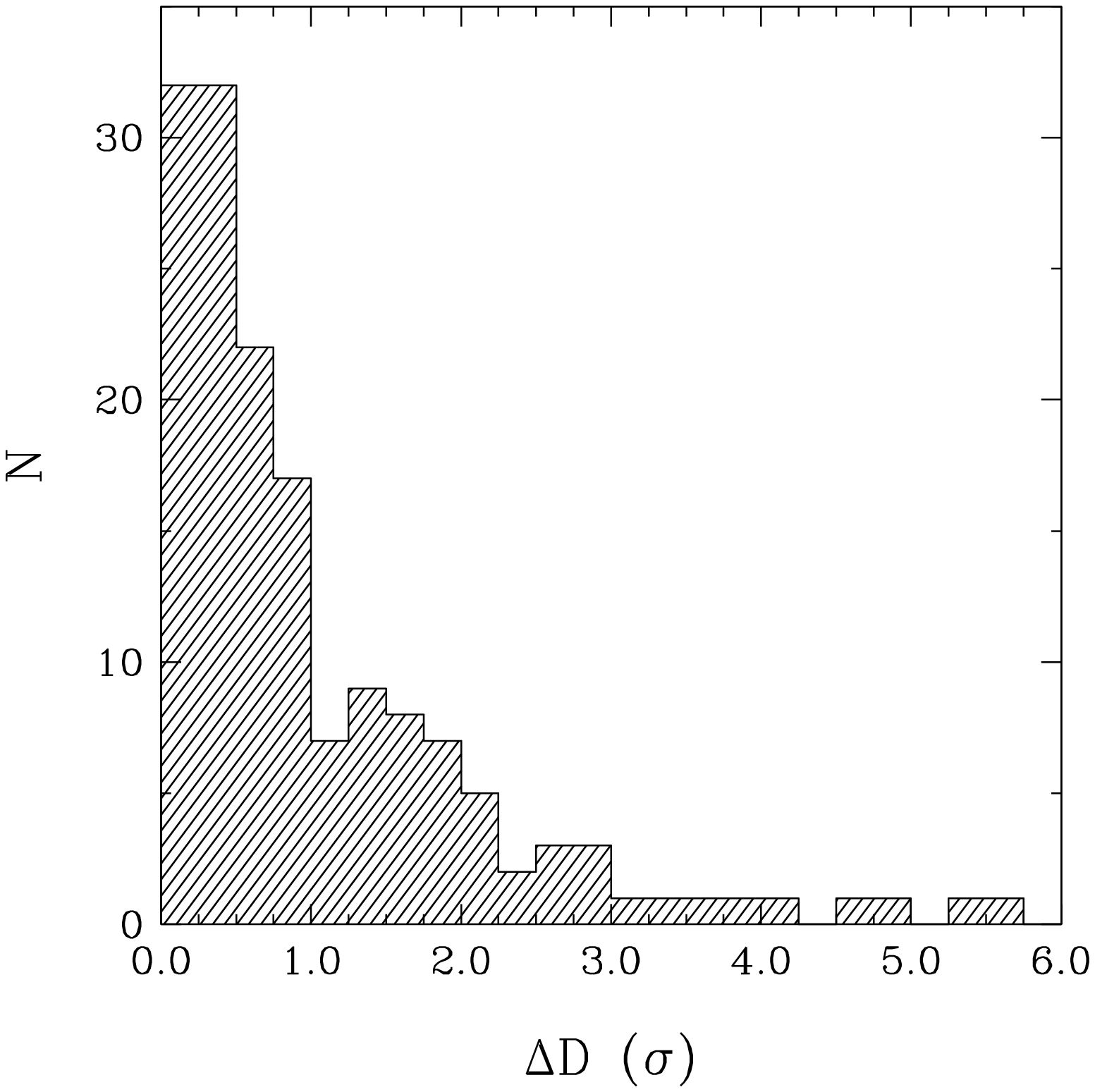}
\vspace*{-4cm}
\caption{Distribution of the (absolute) differences between the
  distances $D_{\rm MR}$ and $D_\pi$, measured in units of $\sigma$, where
  $\sigma^2\equiv\sigma^2_{D_{\rm MR}}+\sigma^2_{D_{\pi}}$, for all white dwarfs 
  in our sample with reliable trigonometric parallax and spectroscopic 
  $\logg$ measurements.
\label{N_NsigD}}
\end{figure}

We show in Figure \ref{N_NsigD}, for the same reliable subsample, the
distribution of the absolute differences in distances measured in
units of $\sigma$, as defined above. We find that the fractions of
white dwarfs having distance estimates within the 1 and 2$\sigma$
confidence levels are 65 and 85\%, respectively. These proportions are
smaller than those expected from a Gaussian distribution, especially
at the 2$\sigma$ level, for which our value departs from the standard
95.5\% value by $\sim$10\%. Furthermore, for 7\% of the stars, the
distance differences are larger than 3$\sigma$, and thus the two
distances are clearly statistically inconsistent. These results
suggest that our sample contains a small number of outliers, which
either have erroneous physical parameters, or simply do not follow the
C/O-core mass-radius relation. We examine and discuss these two
possibilities in the next few sections.

\begin{figure}[p]
\centering
\includegraphics[width=\linewidth]{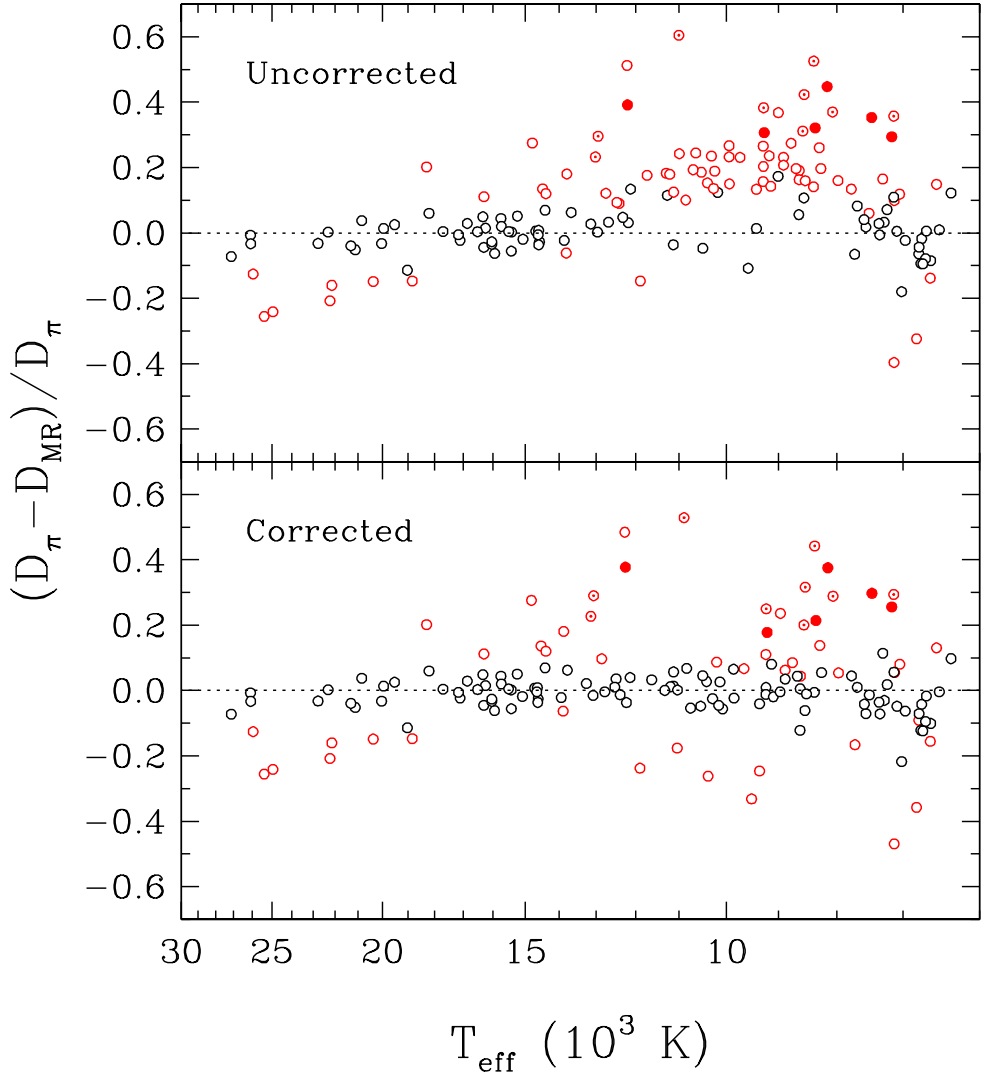}
\vspace*{-4cm}
\caption{Differences between the distances $D_\pi$ and $D_{\rm MR}$, expressed
  as fractions of the parallax distance $D_\pi$, for all white dwarfs in our 
  sample with reliable trigonometric parallax and spectroscopic $\logg$ 
  measurements, as a function of effective temperature; the color coding is 
  identical to that used in Figure \ref{Dmr_Dpi}. The results in the upper 
  panel (uncorrected) do not take into account the hydrodynamical 3D 
  corrections of \citet{tremblay13} for both spectroscopic $\Te$ and 
  $\logg$ determinations.\label{dD_logTeff}}
\end{figure}

We also show, in the bottom panel of Figure \ref{dD_logTeff}, the
differences in distances, $D_\pi-D_{\rm MR}$ (expressed as fractions
of the parallax distances $D_\pi$), but this time as a function of
effective temperature. The results indicate that the general level of
consistency does not vary significantly throughout the temperature
range of our sample. Once again, the confirmed double degenerate
systems occupy a well-defined region in the plot; in all six cases,
$D_\pi$ is greater than $D_{\rm MR}$ by more than 15\%. Interestingly
enough, had we neglected the hydrodynamical 3D corrections of
\citet{tremblay13} applied to both our spectroscopic $\Te$ and $\logg$
determinations for DA stars, severe discrepancies would have been
present at low effective temperatures, where convective energy
transport becomes important ($\Te \lesssim 13,000$~K), as can be seen
in the upper panel of Figure \ref{dD_logTeff}.  The comparison
displayed here thus provides a strong support to the results of
Tremblay et al., and in particular in the interpretation that the
mixing-length theory is the culprit behind the so-called high-$\logg$
problem observed in most spectroscopic analyses of DA stars (see
\citealt{tremblay10} and references therein).

\begin{figure}[p]
\centering
\includegraphics[width=\linewidth]{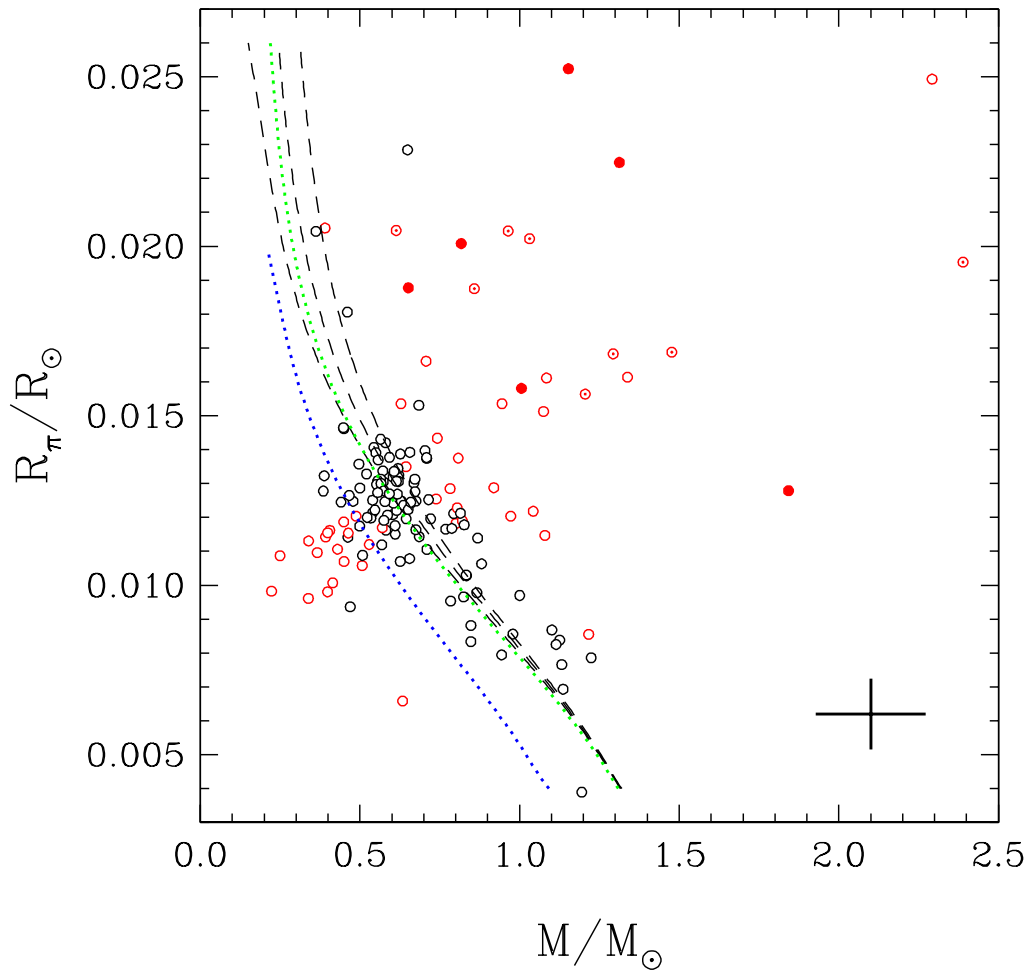}
\vspace*{-4.5cm}
\caption{Location in the $R_\pi$ versus $M$ diagram for all white
  dwarfs in our sample with reliable trigonometric parallax and
  spectroscopic $\logg$ measurements; the color coding is identical to
  that used in Figure \ref{Dmr_Dpi}.  The cross in the lower right
  corner represents the average uncertainties. Also shown are
  mass-radius relations for C/O-core, thick hydrogen envelope models
  at $\Te=7000$, 15,000, and 25,000~K (black dashed lines, from left
  to right), for C/O-core, thin hydrogen envelope models at
  $\Te=15,000$~K (green dotted line), and for Fe-core, thick hydrogen
  envelope models also at $\Te=15,000$~K (blue dotted
  line).\label{R_M}}
\end{figure}

The location of the same white dwarfs in the $R_\pi$ versus $M$
diagram is summarized in Figure \ref{R_M} together with various
theoretical mass-radius relations. We remind the reader that in this
plot, the radius and mass estimates are independent of any mass-radius
relation. The objects shown in red in this figure correspond to the
same objects also shown in red in Figure \ref{Dmr_Dpi}, that is, white
dwarfs that have $\Delta D > 1\sigma$. In particular, the unresolved
double degenerate binaries all appear here on the right of the
theoretical mass-radius relations. The object shown in black near the
top of the figure, 0501+527 (G191-B2B), is within 1$\sigma$ despite
being located far on the right of the predicted curves, because the
parallax measurement has a 17.8\% uncertainty, leading to
uncertainties on the radius and mass that are particularly significant
for this hot star (see online figures).  As can be seen from this
plot, the possibility to ever measure the thickness of the hydrogen
layer in DA white dwarfs using such diagrams is almost hopeless,
because the difference between thin and thick hydrogen envelope models
is way too small. This is particularly true for 40 Eri B (0413$-$077),
which fits better the thin hydrogen envelope C/O-core models, but
which is also consistent with the thick models at the 1$\sigma$ level,
a conclusion also reached by \citet{tremblay17}. The core composition,
however, may be within our reach if a white dwarf possesses a heavy
core made of iron.

A small trend can also be observed in Figure \ref{R_M}, where most
points seem to be roughly distributed along some diagonal pattern,
from the lower left to the upper right.  In this diagram, errors on
the spectroscopic $\logg$ values would result in a horizontal rather
than a diagonal pattern since only the mass $M$ depends on $\logg$.
Instead, the observed feature most likely originates from inaccuracies
on the radius determinations. Indeed, the net effect of a change in
radius is to move a point diagonally since $R_\pi$ appears explicitly
on the y-axis, but is also used in the calculation of $M$ on the
x-axis. Since $R_\pi$ is obtained from the solid angle and the
parallax measurement, and given that the solid angle is a very robust
quantity derived from the energy distribution, the trend observed in
Figure \ref{R_M} can most naturally be explained by the distribution
of errors in the parallax measurements.  The {\it Gaia} mission should
help to reduce the observed scatter in the $R_\pi$ versus $M$ diagram.

Among the stars that were rejected from our reliable subsample (the
objects labeled ``R'' in the online figures), 24 have been excluded on
the basis of parallax uncertainties larger that 20\%, and 37 because
of unreliable spectroscopic $\logg$ measurements.  We note, however,
that despite the high uncertainty associated with these objects, some
of them are still consistent with the expected mass-radius relation;
in some cases, the agreement is even very good.  For instance,
0221+399, 0230$-$144, 0655$-$390, 1630+189, 1710+683, 1811+327B, and
2347+292 are all cool objects ($\Te \lesssim 6500$~K) that were
rejected on the basis of potentially unreliable $\logg$
determinations, but they all accurately match the predictions of the
mass-radius relation (see online figures). This suggests that in some
cases, the spectroscopic technique can provide reliable $\logg$ values
even at such low effective temperatures. The spectroscopic fits for
some of these cool white dwarfs are displayed in Figure
\ref{SampleFitsDA} above. Furthermore, among the stars excluded
because of uncertain parallaxes, 0145$-$174, 0407+179, 1244+149,
1425+540, 1518+636, 1811+327A, 2222+683, and 2329+407 all fall
remarkably close to the expected curves (see bottom panels of the
corresponding online figures), implying that the parallax measurements
for these objects are probably accurate even if they have large
uncertainties. Finally, we have to mention the peculiar case of
1132+470 (G122-31), which was rejected because of a 34.5\% parallax
uncertainty. For this white dwarf, the temperature estimates obtained
from spectroscopy and photometry are radically different, $T_{\rm
  spec}=28,088$~K and $T_{\rm phot}=14,780$~K. This suggests that this
object might actually be an unresolved double degenerate binary, most
likely a DA+DC, an idea further supported by the dubious quality of
our spectroscopic fit (shown in Figure \ref{SampleFitsDA}), especially
at \hb.  \citet{harris13} actually mention in their introduction the
detection of a companion to G122-31, although these results have not
been published yet.

In the remainder of this paper, we investigate the objects showing
distance differences larger than 1.5$\sigma$, with the aim of
determining whether these discrepancies have a physical rather than a
statistical origin. We put forward various physical interpretations
that may explain the measured differences, and thus improve the
agreement with the theoretical mass-radius relation for these stars.

\subsection{Double Degenerate Binaries} \label{DD}

\subsubsection{Confirmed and Suspected Double Degenerate Binaries}

Our sample contains six well-known unresolved double degenerate
binaries, which are indicated by filled red circles in all the
previous plots: 0101+048 \citep[G1-45;][]{maxted00,farihi05},
0135$-$052 \citep[L870-2;][]{saffer88,bergeron89}, 0326$-$273
\citep[L587-77A;][]{zuck03,nelemans05}, 1242$-$105 \citep[LP
  736-4;][]{debes15}, 1639+153 \citep[LHS 3236;][]{harris13}, and
1824+040 \citep[Ross 137;][]{maxted99,morales05,farihi05}.  As
expected, the location of all these binaries in our various diagrams
is at odds with the predictions from the theoretical mass-radius
relations. When analyzed under the assumption of a single star, an
unresolved double degenerate binary will appear overluminous; the
photometric fit will then overestimate the radius $R_\pi$, and also
underestimate the distance $D_{\rm MR}$, as observed in Figures
\ref{Dmr_Dpi}, \ref{dD_logTeff}, and \ref{R_M}. In particular, these
six confirmed binaries show significant discrepancies in their
distance estimates ($\Delta D = 2.52$, 5.36, 4.09, 9.42, 7.69, and
$2.90\sigma$, respectively).

A close inspection of the various results presented in Section
\ref{results} reveals that 12 additional white dwarfs in our sample
exhibit the typical discrepancies expected from unresolved degenerate
binaries. Among these, five have already been identified as unresolved
double degenerate candidates by BLR01: 0126+101 ($\Delta
D=1.81\sigma$), 0142+312 ($\Delta D=1.61\sigma$), 1418$-$088 ($\Delta
D=1.81\sigma$), 1606+422 ($\Delta D=1.78\sigma$), and 2111+261
($\Delta D=2.60\sigma$). Our own results thus reinforce their
conclusion about the likely binary nature of these white
dwarfs. However, we find in our analysis that two additional double
degenerate candidates identified by BLR01 are within the 1$\sigma$
confidence level assuming a single star: 0839$-$327 ($\Delta
D=0.63\sigma$) and 1124$-$293 (named 1124$-$296 in BLR01; $\Delta
D=0.67\sigma$).  In both cases, BLR01 based their conclusion on the
fact that these white dwarfs appeared overluminous, resulting in
extremely low photometric masses of $M=0.44$ \msun\ and $M=0.23$
\msun, respectively. In the case of 0839$-$327, we also obtain a low
mass of $M=0.45$ \msun, but its location right on the expected
mass-radius relation (see online figures) suggests it is a single
low-mass white dwarf, or if there is a companion, its contribution to
the total luminosity of the system is negligible. For 1124$-$293, we
obtain a normal mass of $M=0.54$ \msun, resulting from a more recent
and more precise trigonometric parallax measurement,
$\pi=31.00\pm1.54$ mas \citep{sub17}, a value which is twice as large
as that used by BLR01 in their analysis, $\pi=16.4\pm1.7$ mas
\citep[][]{ruiz96}. Again, the good agreement with the mass-radius
relation shown in the online figures suggests that 1124$-$293 is a
single DA white dwarf.

For two other white dwarfs in our sample, 1130+189 ($\Delta
D=3.24\sigma$) and 2048+809 ($\Delta D=4.77\sigma$), our results agree
with those of \citet{tremblay17}, who found large inconsistencies
between the observed and predicted radii (the equivalent of $R_\pi$
and $R_{\rm MR}$ in our analysis).  Hence, we concur with their
interpretation that these stars are probably unresolved double
degenerate binaries. Also, one object in our sample, 1447$-$190
($\Delta D=2.95\sigma$), was recently identified as a double
degenerate binary by \citet{sub17}, who not only obtained discrepant
distance estimates, but also reported radial velocity variations at
\hb.  Another possible double degenerate is 0311$-$649 ($\Delta
D=4.53\sigma$), also studied by Subasavage et al., who suggested a
binary hypothesis based on their low photometric mass ($M=0.29$
\msun).

The three remaining stars in our sample that could possibly be
explained in terms of unresolved double degenerate binaries are
0133$-$116 ($\Delta D=3.73\sigma$), 0518+333 ($\Delta D=1.58\sigma$),
and 2351$-$368 ($\Delta D=1.54\sigma$). However, we have reasons to
believe that these objects are not double degenerate systems. In the
cases of 0518+333 and 2351$-$368, the energy distributions are poorly
constrained and the parallax uncertainties are quite large (14.4 and
11.2\%, respectively), and hence more reliable data are needed before
any definitive conclusion can be drawn. The puzzling case of
0133$-$116 (Ross 548, ZZ Ceti itself) requires a more extensive
discussion, which we present in the next section.

Thus, besides the six known systems, we identified nine additional
double degenerate candidates based on our analysis, which are shown as
dotted open circles in previous plots.

\subsubsection{Atmospheric Parameters}

To strengthen our double degenerate interpretation for some of the
objects in our sample, we present here a method to derive the
atmospheric parameters of both components in unresolved degenerate
binary systems. Our approach relies on modified versions of the
spectroscopic and photometric techniques, and assumes that the
theoretical mass-radius relation is valid for individual white dwarfs.
Since all the confirmed and suspected binaries identified in our
analysis are hydrogen-rich objects, we assume in the following that
they are composed of two DA stars.

The monochromatic flux received at Earth from an unresolved double
degenerate system is given by the sum of the Eddington flux
originating from each component, properly weighted by their solid
angle (see Equation \ref{flux}):

\begin{equation}
f_{\lambda} = \frac{4\pi}{D^2}\bigg[R_1^2H_{\lambda,1}(T_{\rm{eff},1},\logg_{1})+R_{2}^2H_{\lambda,2}(T_{\rm{eff},2},\logg_{2})\bigg] \label{flux_DD}
\end{equation}

\noindent where we have explicitly written the dependence of the
Eddington flux on the atmospheric parameters. The flux received at
Earth thus seems to depend on seven physical quantities --- the
radius, effective temperature, and surface gravity of each component,
and the distance --- but the mass-radius relation, which relates $R$
to $\logg$, reduces the number of free parameters to only five:
$T_{\rm{eff},1}$, $\logg_{1}$, $T_{\rm{eff},2}$, $\logg_{2}$, and $D$.

With the definition of the combined flux given by Equation
\ref{flux_DD}, the spectroscopic technique can be applied directly as
described in Section \ref{spec_anal}, but this time by considering
$T_{\rm{eff} 1}$, $\logg_{1}$, $T_{\rm{eff} 2}$, and $\logg_{2}$ as
free parameters. Note that the factor $4\pi/D^2$ in Equation
\ref{flux_DD} is irrelevant here since the spectroscopic technique
relies on the normalized Balmer line profiles.  A similar approach can
be used with the photometric technique by averaging the combined
fluxes over the appropriate filter bandpasses.  However, since the
photometric technique is based on absolute fluxes, the factor
$4\pi/D^2$ must be taken into account when comparing to observed
fluxes, with the distance $D$ obtained from the trigonometric
parallax. The fit to the energy distribution thus yields the same four
atmospheric parameters as the spectroscopic technique. Note that even
though the individual energy distributions depend only weakly on
$\logg$, as mentioned in Section \ref{phot_anal}, the combined fluxes
--- as expressed by Equation \ref{flux_DD} --- are strongly dependent
on the stellar radius $R$ of each component of the system, and thus on
the corresponding $\logg$ values through the mass-radius relation. As
such, both $\logg$ values obtained from the photometric fit are
meaningful. Given the large number of fitting parameters, the
application of one technique yields many different solutions,
depending on the number of free and fixed parameters, as well as on
the adopted initial values. However, very few solutions (or in many
cases, only one solution) are consistent with {\it both} the
spectroscopic and the photometric analyses. A consistent solution can
thus be obtained by using the two techniques in an iterative way.

\begin{figure}[p]
\centering
\vspace*{-8cm}
\includegraphics[width=\linewidth]{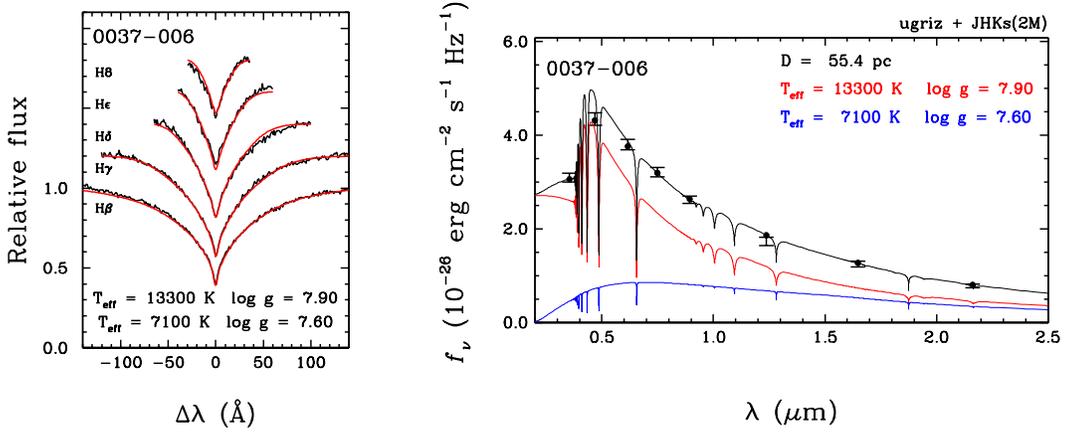}
\vspace*{-9cm}
\caption{Best fit to the optical spectrum (left panel) and energy 
  distribution (right panel) of 0037$-$006, assuming an unresolved double 
  degenerate system composed of two DA stars. In the right panel, the red 
  and blue lines show the contribution of each component to the total 
  model flux, which is displayed as the black line. \label{DDfit_0037-006}}
\end{figure}

\begin{figure}[p]
\centering
\vspace*{-8cm}
\includegraphics[width=\linewidth]{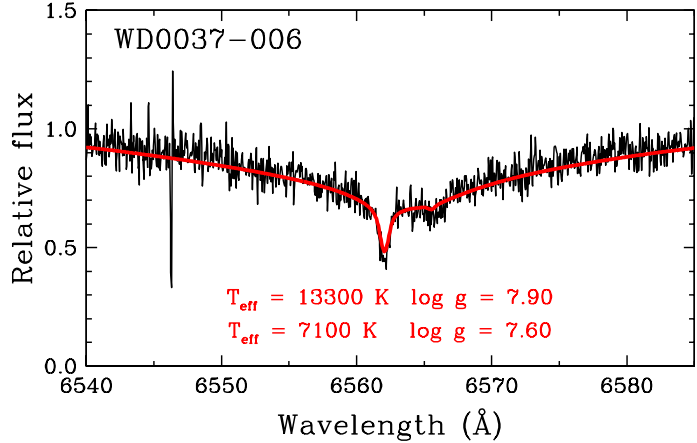}
\vspace*{-9cm}
\caption{Superposition of the observed and model spectra of 0037$-$006 
  at \ha, assuming an unresolved double degenerate system composed of 
  two DA stars whose atmospheric parameters correspond to the best 
  solution shown in Figure \ref{DDfit_0037-006}. 
  \label{Halpha_0037-006}}
\end{figure}

As a test case, we applied our fitting procedure to 0037$-$006, a
known double-lined degenerate binary \citep{koester09}, which is not
included in our sample because no parallax measurement is available
for this star. In such a case, the distance $D$ is considered an
additional free parameter in the photometric fit. This particular
object was selected because high-resolution spectroscopy at \ha\ can
be used to confirm that the model spectrum predicted from the
four-atmospheric parameter solution reproduces reasonably well the
observed double-lined core. Figure \ref{DDfit_0037-006} shows the
solution that best reproduces the spectroscopic and photometric
observations for 0037$-$006. Also shown in the right panel is the
contribution of each component (in red and blue) to the total flux (in
black). The fit to both the Balmer line profiles and the energy
distribution is excellent. We show in Figure \ref{Halpha_0037-006} the
superposition of the observed and model spectra at \ha\ for this
particular solution (a wavelength shift between both components has
been included in our solution to match to observed shift).  We
emphasize that this is not a fit to the \ha\ line profile.  The
double-lined feature is perfectly reproduced, which strengthens our
confidence in our overall procedure to obtain reliable physical
parameters for white dwarfs in unresolved binary systems.

Our fitting technique was applied to the 15 confirmed or suspected
double degenerate binaries identified in our analysis, the results of
which are displayed in Figure \ref{DDfits}. For most of these objects,
values of the atmospheric parameters of both components are reported
for the first time.  Even if our test case example for 0037$-$006
indicates that the atmospheric parameters obtained with our fitting
procedure appear reliable, the quantitative results presented in
Figure \ref{DDfits} must be interpreted with caution. In some
instances, there are more than one acceptable solution, which is
perhaps not surprising given the numerous free parameters. It is also
possible that some of the objects are actually DA+DC systems rather
than DA+DA systems, in which case the true atmospheric parameters
probably differ from those derived here. Still, our method is able to
demonstrate, undeniably, whether or not a white dwarf can be fitted as
a double degenerate, at least in a qualitative way. We finally note
that other estimates obtained from various methods are available in
the literature for five well-known double degenerates (0135$-$052,
0326$-$273, 1242$-$105, 1639+153, and 1824+040), but we are refraining
from making a detailed comparison since our goal here is only to
demonstrate that these are double degenerate systems.

\begin{figure}[p]
\centering
\vspace*{-1cm}
\includegraphics[width=\linewidth]{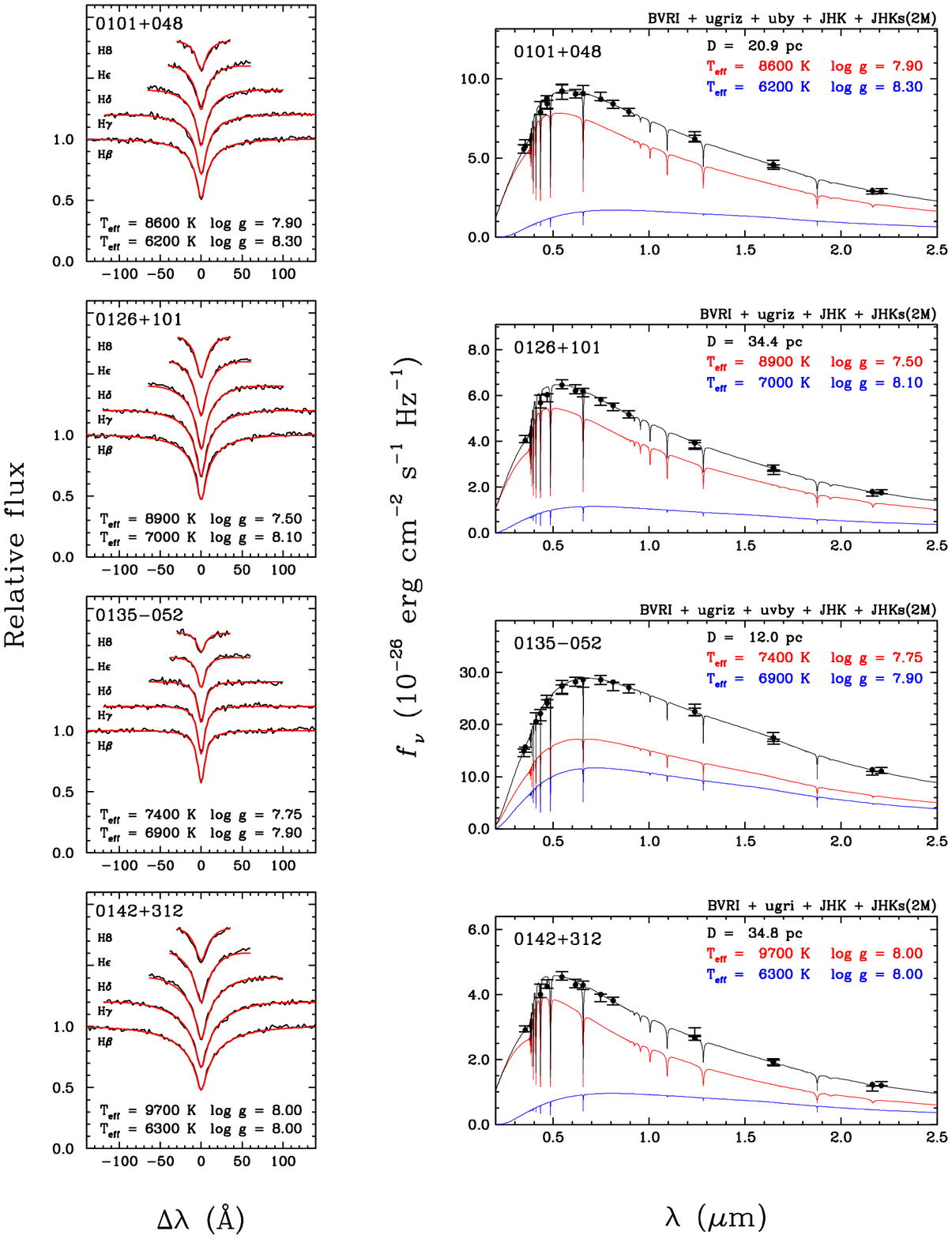}
\vspace*{-2cm}
\caption{Best fit to the optical spectra (left panels) and energy 
  distributions (right panels) of the confirmed and suspected double 
  degenerate systems identified in our analysis, assuming unresolved 
  systems composed of two DA stars. In the right panel, the red and 
  blue lines show the contribution of each component to the total 
  model flux, which is displayed as the black line. \label{DDfits}}
\end{figure}

\begin{figure}[p]
\figurenum{16}
\centering
\vspace*{-1cm}
\includegraphics[width=\linewidth]{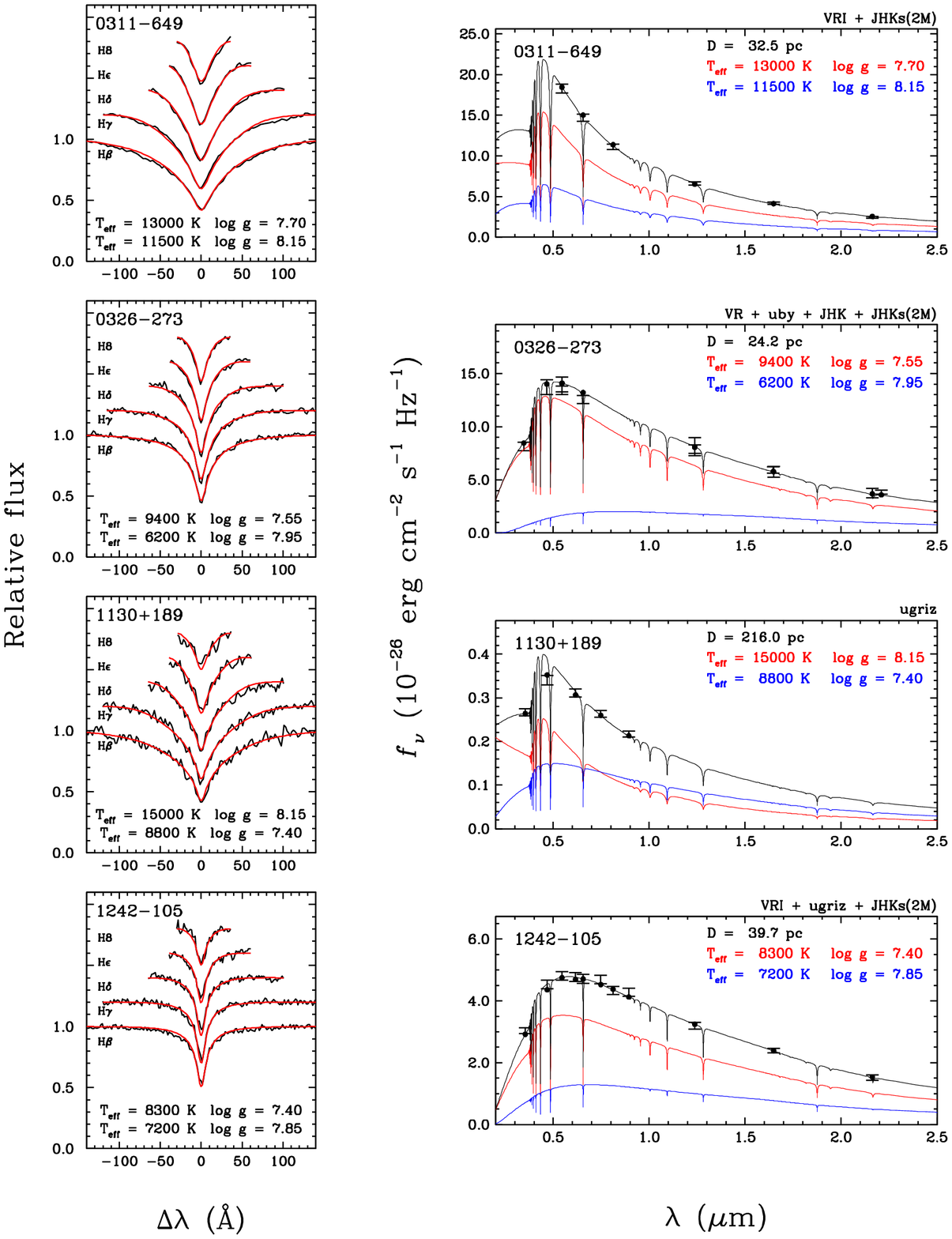}
\vspace*{-2cm}
\caption{(Continued)}
\end{figure}

\begin{figure}[p]
\figurenum{16}
\centering
\vspace*{-1cm}
\includegraphics[width=\linewidth]{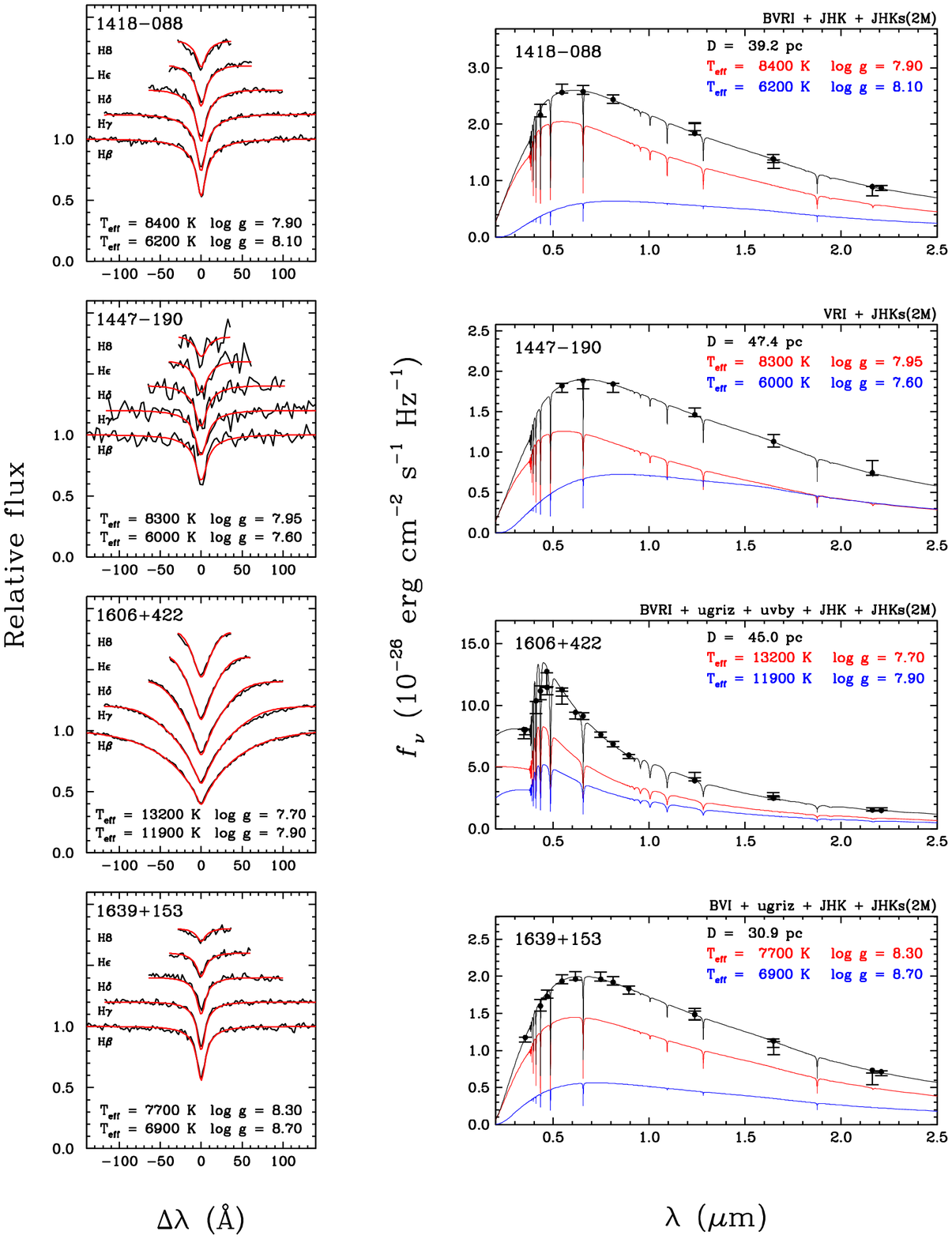}
\vspace*{-2cm}
\caption{(Continued)}
\end{figure}

\begin{figure}[p]
\figurenum{16}
\centering
\includegraphics[width=\linewidth]{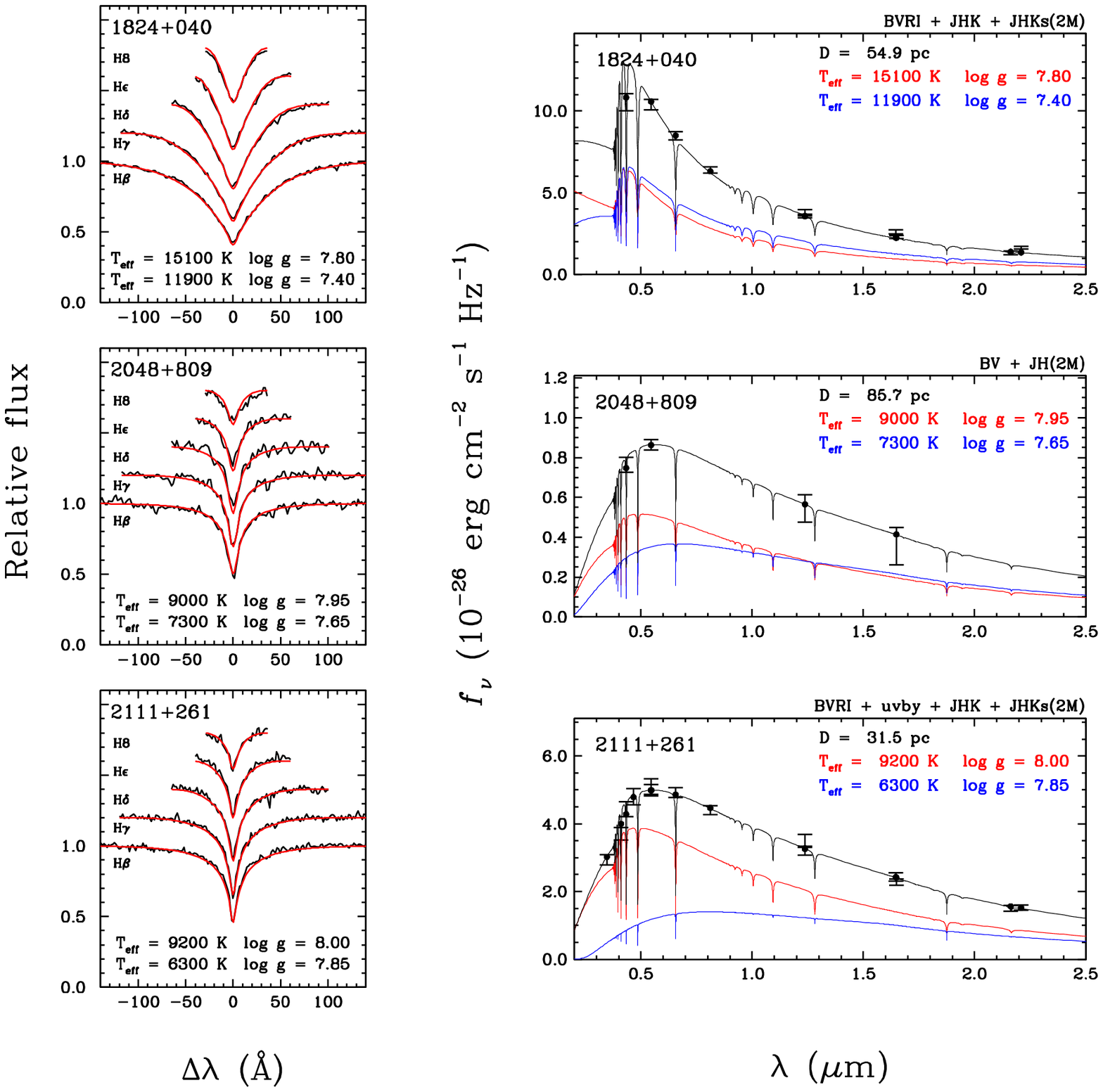}
\vspace*{-6cm}
\caption{(Continued)}
\end{figure}

An excellent counterexample is provided by 0133$-$116 (Ross 548).
Inconsistencies in the parameters of this white dwarf were first
noticed by \citet[][see also \citealt{gianninas11}]{bergeron95b}; our
own analysis of this object presented in the online figures shows a
similar behavior. The parallax measurement gives a distance
$D_{\pi}=63.3 \pm 8.1$ pc, which is twice as large as the distance
computed by invoking the mass-radius relation, $D_{\rm MR}=32.5 \pm
1.1$ pc. Similarly, we obtain a large radius, $R_\pi=0.0250$ \rsun,
which translates into a ridiculously high mass of $M=2.30$ \msun\ for
the measured spectroscopic value of $\logg=8.01$. Given that the
spectroscopic fit is excellent (see Figure \ref{SampleFitsDA}) and
yields a normal $\logg$ value, the inconsistency seems to arise
instead from the photometric fit, as reflected by the high value of
$R_{\pi}$ and the low value of $D_{\rm MR}$. This strongly suggests
that Ross 548 might be a double degenerate system.  However, the
application of our fitting method to this object proved unsuccessful;
there is just no composite model that can fit both the spectroscopy
and the photometry simultaneously. In particular, for all acceptable
sets of effective temperatures, the surface gravities derived from the
Balmer line profiles differ from those obtained from the energy
distribution by more than an order of magnitude. The reason for this
disparity is that the luminosity of Ross 548 appears way too high,
even when considered as an unresolved double degenerate binary. As a
result, the photometric fit yields very low $\logg$ values,
corresponding to large radii, in order to match the extreme inferred
luminosity, in sharp contrast with the spectroscopic solution. There
are two ways to match the high luminosity with physically acceptable
$\logg$ values: either the parallax measurement is faulty and the true
distance is less than 63.3 pc, or there is actually a third component
in the unresolved system.  Therefore, three explanations may be
offered to solve the problem regarding this object: (1) Ross 548 is a
single star and its distance from Earth is about 32 pc (the excess
luminosity is due solely to the parallax being erroneous by a factor
of 2); (2) Ross 548 is a double degenerate system and its distance
from Earth is less than 63.3 pc (the excess luminosity arises from
both an inaccurate parallax and the contribution of two stars); (3)
Ross 548 is a triple white dwarf system and its distance from Earth is
63.3 pc (the parallax is accurate, and the excess luminosity
originates from the contribution of three stars).  The last
possibility is admittedly unlikely but must all the same be
considered. The {\it Gaia} parallax measurement for this object will
hopefully help to settle the issue. We note, in the meantime, that the
independent distance to Ross 548 derived by \citet{noemi16} through
asteroseismological means, $D= 30.0 \pm 0.9$~pc, favors hypothesis
(1).

\subsection{Iron-Core White Dwarfs?} \label{Fe-core}

An examination of the results displayed in Figures
\ref{Dmr_Dpi} and \ref{R_M} shows that another class of objects
occupies the location opposite to that of the double degenerate
binaries in these plots, that is, above the 1:1 correspondence in the
$D_{\rm MR}$ versus $D_{\pi}$ diagram, and on the left of the
theoretical mass-radius relation in the $R_\pi$ versus $M$
diagram. One possible explanation to account for this feature
is that these white dwarfs actually have a core composed of iron
rather than carbon and oxygen. Indeed, Figure \ref{R_M} shows that
these objects fall closer to the predicted curve corresponding to
Fe-core models. We explore this possibility in detail in the present
section.

The most notorious case of a white dwarf that is possibly made of
iron-rich material is 0644+375 (G87-7). The Fe-core hypothesis was
first suggested by \citet{provencal98}, who obtained mass and radius
values consistent with the zero-temperature Fe-core mass-radius
relation of \citet{hamada61}. Further investigations by
\citet{fontaine07} using improved spectroscopic parameters and
finite-temperature evolutionary models with different core
compositions provided additional elements in favor of this
scenario. Since then, there have been several developments regarding
the data and models, namely, the new reduction of the {\it Hipparcos}
data by \citet{hipparcos07}, the upgraded Stark profiles of
\citet{tb09}, and the new Fe-core evolutionary models described in
Section \ref{mr_models} of the present study. Thus, we revisit the
case of G87-7 in light of these new developments, using our standard
approach for testing the mass-radius relation.

To improve the accuracy of our analysis, we first rederive the
spectroscopic parameters of G87-7 by making use of the four
high-quality (S/N $\sim$ 80) optical spectra available to us, as was
done by \citet{fontaine07}. The four individual fits are not displayed
here but are all similar to our original fit shown in Figure
\ref{SampleFitsDA} in terms of quality. We obtain mean values of
$\Te=21,996 \pm 254$~K and $\logg=8.148 \pm 0.043$, where the
uncertainties include both the internal and external errors as usual,
but in the present case we adopt as the external errors the standard
deviations associated with the parameters derived from the four
individual fits, rather than the prescription of \citet{LBH05}.  Our
improved spectroscopic parameters are consistent with our previous
estimates ($\Te=22,143 \pm 337$~K, $\logg=8.094 \pm 0.045$), but we
note that the mean $\logg$ value is slightly larger than our original
value obtained from a single spectrum.

\begin{figure}[p]
\centering
\vspace*{-6cm}
\includegraphics[width=\linewidth]{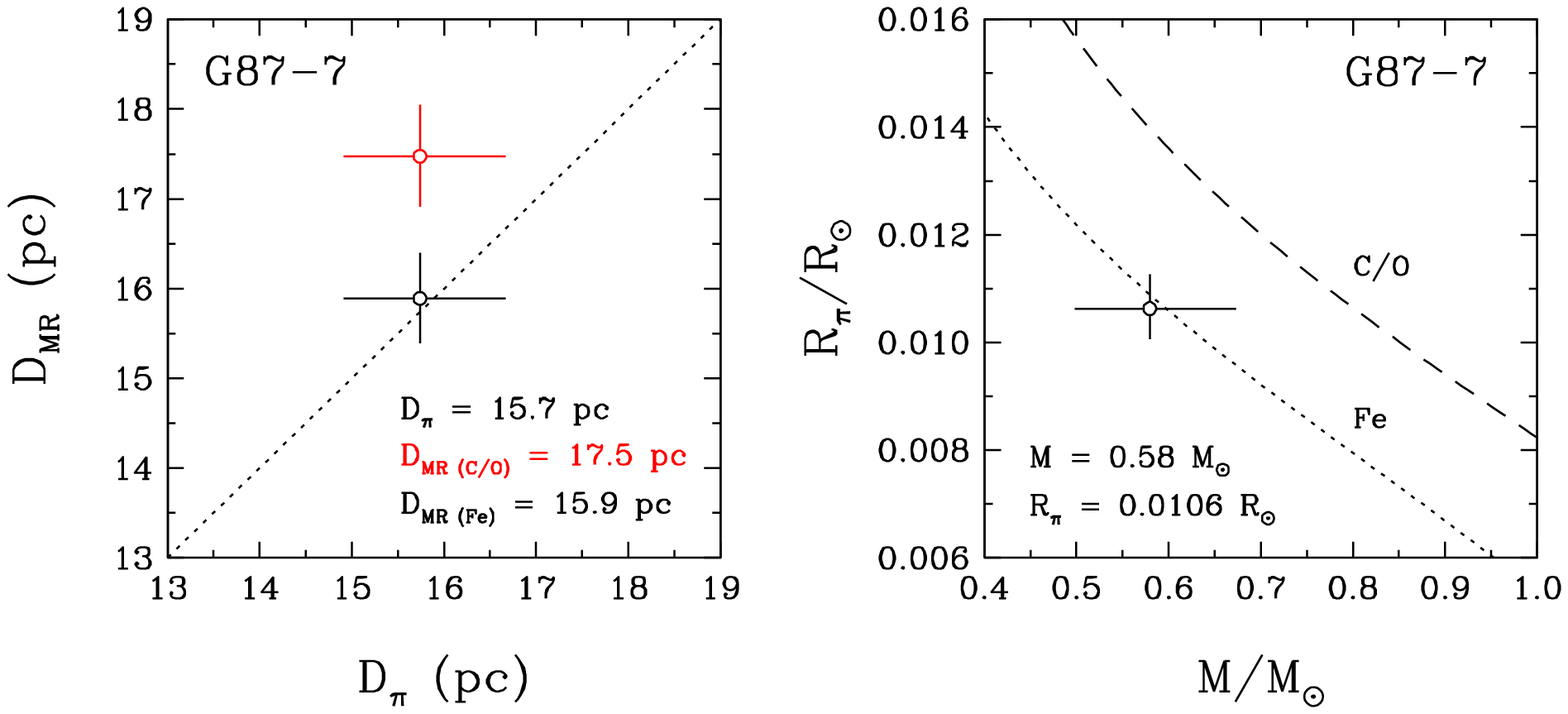}
\vspace*{-7cm}
\caption{Location of 0644$+$375 (G87-7) in the $D_{\rm MR}$ versus
  $D_\pi$ diagram (left panel) assuming C/O-core models (red)
  or Fe-core models (black). Location of the same object
  in the $R_\pi$ versus $M$ diagram (right panel) together with
  mass-radius relations for C/O-core (dashed line) and Fe-core
  (dotted line), thick hydrogen envelope models at the
  effective temperature of G87-7.}\label{MR_Fecore_0644+375}
\end{figure}

We then apply our standard procedure outlined in Section \ref{method},
using successively our C/O-core and Fe-core evolutionary models (still
with thick hydrogen layers). The results are displayed in Figure
\ref{MR_Fecore_0644+375} through our typical $D_{\rm MR}$ versus
$D_{\pi}$ and $R_\pi$ versus $M$ comparisons. Unlike the C/O-core
mass-radius relation, the Fe-core mass-radius relation yields an
almost perfect agreement. More specifically, the difference between
the distance estimates is significantly smaller if we assume an iron
core ($\Delta D=0.15\sigma$) instead of a carbon/oxygen core ($\Delta
D=1.67\sigma$). That being said, $\Delta D=1.67\sigma$ is not a
largely inconsistent result, and it is still plausible that G87-7 is a
normal C/O-core white dwarf. However, we stress that all the physical
quantities for this bright, nearby star are remarkably
well-constrained: the {\it Hipparcos} parallax measurement has a small
5.6\% uncertainty, the spectroscopic parameters are obtained from four
high-quality spectra, and the photometric parameters are also
well-determined since the energy distribution is modeled from
magnitudes in 13 bandpasses in the optical and infrared. Although it
is possible that the {\it Hipparcos} parallax suffers from a
systematic error, as in the case of 1314+293 mentioned in Section
\ref{pi}, this appears highly unlikely since the {\it Hipparcos}
measurement is in good agreement with previous ground-based
measurements. Thus, our analysis provides fairly strong support to
the idea that G87-7 harbors an iron core, although a carbon/oxygen
core cannot be ruled out.

\begin{figure}[p]
\centering
\vspace*{-2cm}
\includegraphics[width=\linewidth]{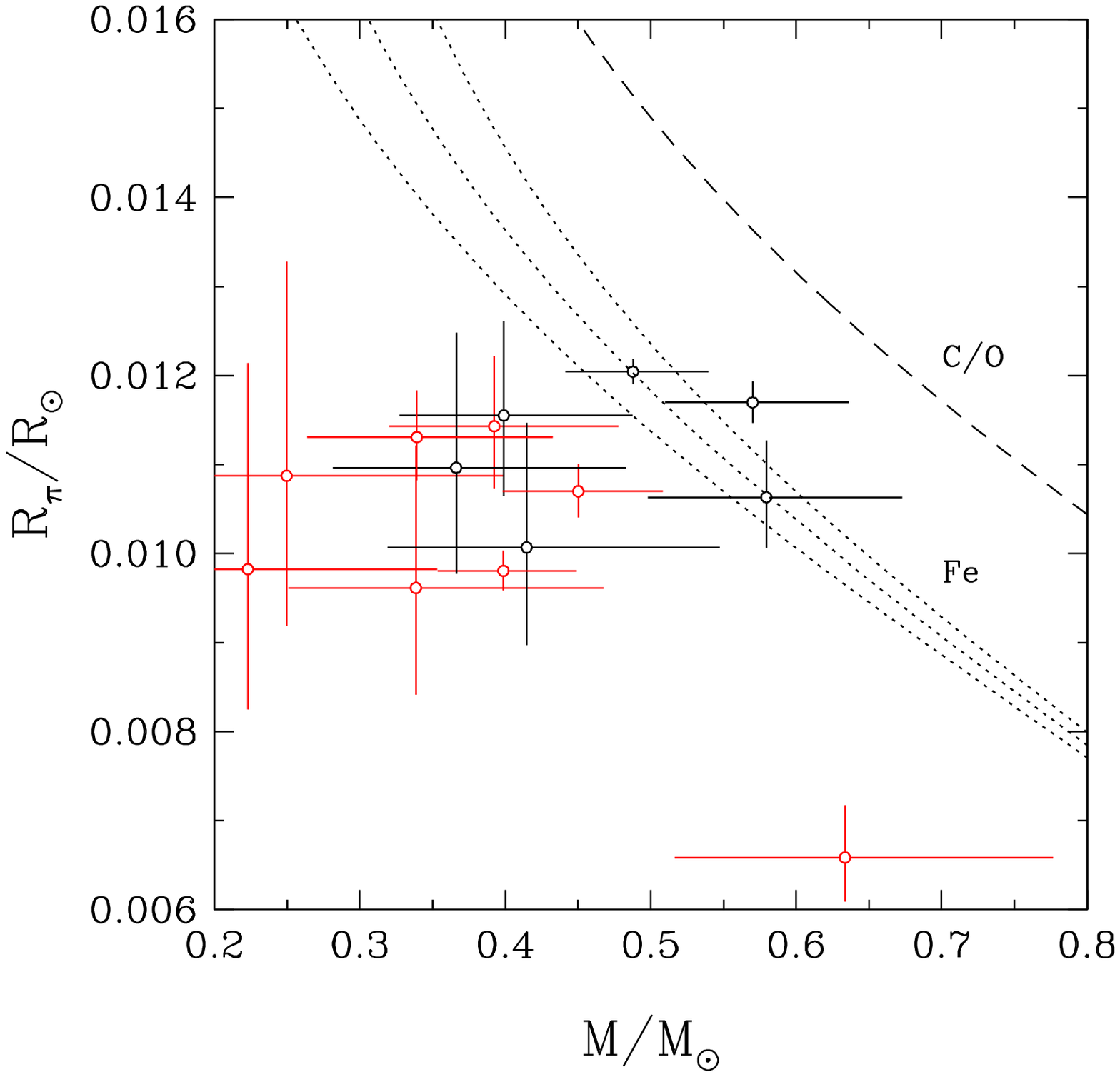}
\vspace*{-4cm}
\caption{Location in the $R_\pi$ versus $M$ diagram of the white dwarfs in 
  our sample that are possibly composed of an iron core.  Also shown are 
  mass-radius relations for Fe-core, thick hydrogen envelope models at 
  $\Te=7000$, 15,000, and 25,000~K (dotted lines, from left to right), 
  and for C/O-core, thick hydrogen envelope models at $\Te=15,000$~K 
  (dashed line). The objects shown in red are not consistent with the Fe-core 
  mass-radius relation within a 1$\sigma$ confidence level.} \label{R_M_Fecore}
\end{figure}

In addition to G87-7, our sample contains 13 other white dwarfs having
distance differences larger than $1.5\sigma$ that might be explained
by an iron-rich core: 0011+000 ($\Delta D=2.17\sigma$), 0842+490
($\Delta D=2.93\sigma$), 0942+236A ($\Delta D=2.16\sigma$), 1105$-$340
($\Delta D=1.64\sigma$), 1609+135 ($\Delta D=2.49\sigma$), 1620$-$391
($\Delta D=3.44\sigma$), 1635+137 ($\Delta D=1.66\sigma$), 1645+325
($\Delta D=1.85\sigma$), 1916$-$362 ($\Delta D=2.62\sigma$), 1936+327
($\Delta D=2.22\sigma$), 1950+250 ($\Delta D=1.82\sigma$), 2105$-$820
($\Delta D=5.54\sigma$), and 2336$-$079 ($\Delta D=3.76 \sigma$). The
location of these objects in the $R_\pi$ versus $M$ diagram is shown
in Figure \ref{R_M_Fecore}, together with three Fe-core theoretical
curves corresponding to $\Te=7000$, 15,000, and 25,000~K. The C/O-core
mass-radius relation at $\Te=15,000$~K is also displayed as a
reference. Six objects (shown in black) are consistent with the
Fe-core mass-radius relation within 1$\sigma$; besides G87-7, these
correspond to 1105$-$340, 1620$-$391\footnote{\citet{holberg12} find
  an excellent agreement with C-core models for 1620$-$391 (CD
  $-$38$^{\rm o}$10980), although their spectroscopic $\logg$ value of
  8.099 is based on an old determination from \citet{bragaglia95},
  which is significantly different from ours ($\logg=7.965$) based on
  a more recent optical spectrum and model atmosphere analysis.},
1645+325, 1936+327, and 1950+250. However, we note from Figure
\ref{R_M_Fecore} that some of these objects are only marginally
consistent with the Fe-core curve, and hence the evidence in favor of
the Fe-core hypothesis is less convincing than for G87-7. Only three
points clearly fall on the predicted curve for an iron core, one of
which corresponds to G87-7. The two other objects are 1105$-$340 and
1620$-$391, which have precise parallax and spectroscopic $\logg$
measurements, as it is the case for G87-7.  These two nearby stars are
thus fairly good candidates for an iron core. The remaining objects
(shown in red) are simply located too far on the left in the diagram
to be considered as possible Fe-core white dwarfs, and we currently
have no alternative physical explanation to account for the observed
discrepancies, except in the case of 2105$-$820, which is discussed in
detail in Section \ref{rad}. Still, we note that 0011+000, 1609+135,
and 2336$-$079 would agree much better with the C/O-core mass-radius
relation had we not applied the 3D correction to the spectroscopic
$\logg$ (these three stars have effective temperatures close to
$\sim$10,000 K, at which the $\logg$ correction reaches its maximum
value). However, there is currently no physical justification to
follow such a procedure. As argued in Section \ref{results}, new
trigonometric parallax measurements might help to improve the
situation.

The major difficulty with the Fe-core hypothesis is that there is no
conventional way in stellar evolution theory to produce iron-core
white dwarfs with, in particular, masses less than the average mass as
shown by the candidates in Figure \ref{R_M_Fecore}. However, we draw
the attention of the reader to the interesting suggestion of
\citet{ouyed11} who argued that, under certain circumstances, Fe-rich
white dwarfs with masses in the range $0.43<M/M_\odot<0.72$ could be
produced through quark-nova explosions in low-mass X-ray binaries. The
predicted mass range is particularly suggestive in view of the results
depicted in Figure \ref{R_M_Fecore}.

\subsection{2105$-$820: A Magnetic White Dwarf with a Radiative Atmosphere?} \label{rad}

\begin{figure}[p]
\centering
\vspace*{-6cm}
\includegraphics[width=\linewidth]{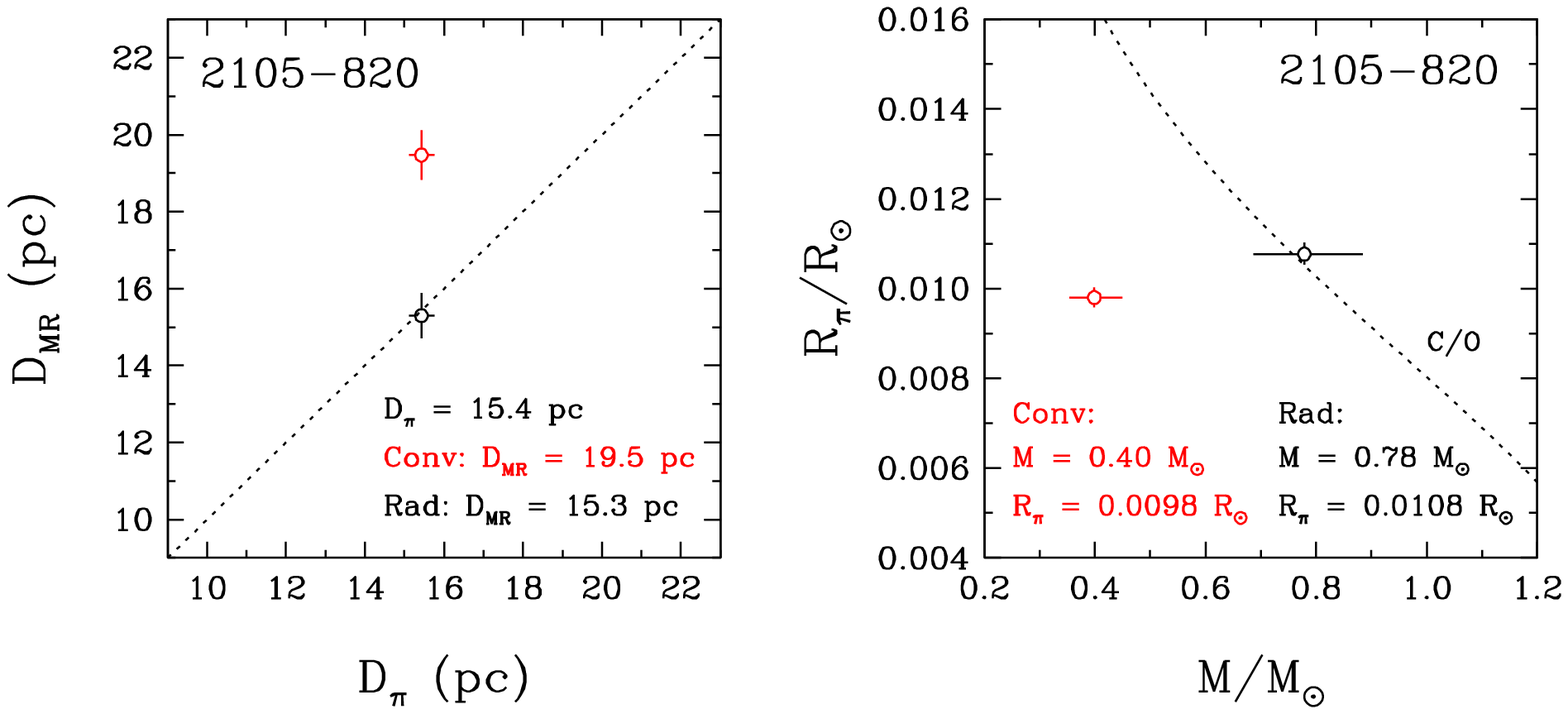}
\vspace*{-7cm}
\caption{Location of 2105$-$820 in the $D_{\rm MR}$ versus
  $D_\pi$ diagram (left panel) and in the $R_\pi$ versus $M$ 
  diagram (right panel) using convective (red) or radiative 
  (black) model atmospheres.} \label{MR_rad_2105-820}
\end{figure}

2105$-$820 (L24-52) is a DA star with a very precise trigonometric
parallax measurement from CTIOPI of $64.81\pm1.39$ mas. Also, because
it is bright ($V\sim13.6$) and hot ($\Te\sim10,000$~K), its
photometric energy distribution and optical spectrum are both
accurately measured, and the corresponding photometric and
spectroscopic solutions are very well constrained. Yet, the distance
inferred from the mass-radius relation, $D_{\rm MR}=19.5$ pc, differs
from the parallactic distance, $D_{\pi}=15.4$ pc, by more than
5.5$\sigma$, and its location in the $R_\pi$ versus $M$ diagram is
largely inconsistent with the mass-radius relations obtained from
C/O-core models, or even from Fe-core models. These results are
summarized in Figure \ref{MR_rad_2105-820} by the red symbols.  A
similar discrepancy, albeit somewhat less significant, is observed
between the spectroscopic temperature, $T_{\rm spec}= 10,369$~K, and
the photometric temperature, $T_{\rm phot}=9921$~K, which corresponds
to $\Delta T_{\rm eff}=1.94\sigma$. Moreover, since 2105$-$820 lies on
the left-hand side of the mass-radius relation, it cannot be
interpreted as an unresolved degenerate binary.

\citet{landstreet12} reported the discovery of an apparently constant
longitudinal magnetic field of $\sim9.5$ kG in 2105$-$820, consistent
with a simple dipolar morphology with a magnetic axis almost parallel
to the rotation axis, and a polar strength of $\sim56$
kG. Interestingly enough, recent studies have shown that magnetic
fields of only a few kG can suppress convective energy transport in
white dwarf atmospheres \citep{valyavin14,tremblay15}. With this idea
in mind, we have reanalyzed all photometric and spectroscopic data
using model atmospheres in which convective energy transport has been
completely suppressed \citep{lecavalier17}. Lecavalier-Hurtubise \&
Bergeron have shown in particular that even though photometric 
temperatures are only marginally affected by the use of purely 
radiative or convective model atmospheres, the spectroscopic 
temperatures measured from both sets of models can differ by more than 
$\sim 2000$~K.

Our results of this experiment for 2105$-$820 are displayed in Figure
\ref{MR_rad_2105-820} by the black symbols. With the use of purely
radiative model atmospheres, the distances are now in perfect
agreement ($\Delta D=0.20\sigma$) and its location in the $R_\pi$
versus $M$ diagram is also entirely consistent with the mass-radius
relations obtained from C/O-core models. Also, the spectroscopic
temperature obtained from purely radiative models, $T_{\rm
  spec}=9797$~K, is now in much better agreement with the
corresponding photometric temperature, $T_{\rm phot}=9957$~K. These
calculations strongly suggest that the correct interpretation for the
discrepant results observed in Figure \ref{MR_rad_2105-820} is that
2105$-$820 does possess a purely radiative atmosphere, and that
convective energy transport has been impeded by the $\sim50$ kG polar
magnetic field present in this star, in agreement with the
calculations of \citet{tremblay15}.

Four other white dwarfs harboring a weak magnetic field are also
included in our sample: 0257+080 \citep{koester09}, 1953$-$011
\citep{koester98,koester09}, 2047+372 \citep{landstreet16}, and
2359$-$434 \citep{landstreet12}. Contrary to the case of 2105$-$820,
all four objects are consistent with the C/O-core mass-radius
relations within 1$\sigma$ using convective model atmospheres.
However, 2047+372 is a rather hot DA white dwarf ($\Te\sim 14,600$~K)
and convection in this star is negligible, while 0257+080 (6620 K),
1953$-$011 (7770 K), and 2359$-$434 (8510 K) are significantly cooler
than 2105$-$820. Since convective energy transport is more important at
lower effective temperatures, our results suggest that it may become
increasingly more difficult for a weak magnetic field to suppress
convection in cooler white dwarfs.

\subsection{Cool White Dwarfs with Mixed He/H Atmospheres?} \label{He/H}

At effective temperatures below $\Te\sim10,000$~K, it is possible to
hide large amounts of helium in a DA white dwarf since helium becomes
spectroscopically invisible. \citet[][see their Figure 16]{bergeron91}
demonstrated that a DA star with an atmosphere enriched with helium
would appear exactly like a pure hydrogen white dwarf but with an
apparent higher $\logg$ value, since both atmospheric parameters have
the similar effect of increasing the atmospheric pressure (see also
Figure 2 of \citealt{tremblay10}). Such helium enrichment --- most
likely resulting from convective mixing --- has been one of the
solutions proposed to explain the longstanding high-$\logg$ problem
observed in all spectroscopic analyses of cool DA stars
($\Te\lesssim13,000$~K; also illustrated in Figure \ref{dD_logTeff}).
However, \citet{tremblay10} ruled out this scenario by determining the
helium abundances in several cool DA white dwarfs using
high-resolution spectra from the Keck 10-m telescope. In all cases
where helium could have been detected, no helium lines were
observed. Tremblay et al.~also explored alternative possibilities to
account for this high-$\logg$ problem, and finally concluded that
convective energy transport treated within the mixing-length formalism
was the most likely origin of this problem, an interpretation later
confirmed by the 3D hydrodynamical calculations of
\citet{tremblay3D11}.

Despite these negative results, we know that helium-rich DA stars with
strong Balmer lines do exist, 1729+371 (GD 362) being the best example
at $\Te=10,540$~K, $\logg=8.24$, and $N(\rm{He})/N({\rm H})\sim15$
\citep{zuckerman07}. The atmosphere of GD 362 is thus actually
dominated by helium.  This DABZ star is also one of the most heavily
polluted white dwarf known, but at medium resolution, the very weak
He~\textsc{i} $\lambda5877$ absorption feature is undetectable. In the
case of GD 362, hydrogen and all heavy elements have most likely been
accreted from a large asteroid, or asteroids, with composition similar
to the Earth-Moon system. As the white dwarf evolves, all the heavy
elements will have diffused at the bottom of the convection zone, but
hydrogen will remain thoroughly mixed within this convection zone, and
the star will have evolved into a helium-dominated DA star. Thus, we
cannot exclude that such objects exist in our sample, although they
may be rare.

The effect of using mixed H/He atmospheres to analyze DA stars is
nicely illustrated in Figure 2 of \citet{tremblay10} in the case of
1655+215 (LHS 3254), also included in our sample. Their figure shows
that a fit to the photometric energy distribution, constrained by the
measured trigonometric parallax, yields a value of $\logg=7.84$ (using
the mass-radius relation) when analyzed with pure hydrogen models, in
sharp contrast with the spectroscopic value of $\logg=8.27$.  It is
also shown that these two surface gravities can be reconciled if model
atmospheres with $N(\rm{He})/N({\rm H})=1$ are used instead (note that
the 3D hydrodynamical $\logg$ corrections were not applied in the
Tremblay et al.~analysis, and the amount of helium required will be
lower if 3D corrections are taken into account). A further examination
of Figure 2 of Tremblay et al.~also reveals that the photometric
method yields $\logg$ values (i.e., radius measurements) that are
completely independent of the assumed atmospheric
composition. Similarly, the photometric and spectroscopic temperatures
are not affected either. Hence the net effect of using mixed H/He
model atmospheres on our own analysis of cool DA stars is to move an
object horizontally to the {\it left} in the mass-radius plot
displayed in Figure \ref{R_M} (when helium-rich models are used, the
measured radius remains unaffected, but the inferred mass is decreased
along the x-axis because of the resulting lower spectroscopic $\logg$
value).

With these conclusions in mind, we thus reexamined all the cool
($\Te\lesssim12,000$~K) DA stars located on the right hand side of the
C/O-core mass-radius relation in Figure \ref{R_M}, excluding of course
the unresolved binaries, or binary candidates, mostly located in the
upper right corner of this figure.  For each of these stars, it is
thus possible to adjust the photospheric helium abundance until the
object moves (left) on top of the C/O-core mass-radius relation for
the appropriate temperature. For instance, a helium abundance between
$N(\rm{He})/N({\rm H})\sim 0.1$ and 1 would bring 1655+215 (see online
figures) precisely on top of the C/O-core mass-radius relation. We
must emphasize, however, that the distance difference for 1655+215
using pure hydrogen models is only a 1.42$\sigma$ result, so the
proposed helium-rich solution may not be significant. In some cases,
the stellar mass inferred from helium-rich models is uncomfortably
low.  For instance, 0518+333 is so far to the right of the mass-radius
relation that the amount of helium required would bring the mass below
0.4 \msun. However, as discussed in Section \ref{DD}, the
trigonometric parallax for this object is most likely unreliable.

All in all, we have identified about two dozen cool DA stars in our
sample for which a modest helium enrichment, around $N(\rm{He})/N({\rm
  H})\sim 0.1$, would improve the agreement with the mass-radius
relation. The best candidates that require an abundance close to
$N(\rm{He})/N({\rm H})\sim 0.1$ are 0148$+$641, 0250$-$007,
2133$-$335, and 2159$-$754 (with distance discrepancies of 1.79, 2.15,
1.95, and 1.61$\sigma$, respectively).  We thus conclude that there is
no compelling evidence in our sample for a significant population of
helium-rich DA stars, at least not at the level observed in GD 362.

\subsection{Validity of the Mass-Radius Relation} \label{validity}

The evidence presented in the last few sections shows that several
discrepant results can be explained by various physical reasons. This
evidence is particularly solid in the cases of the confirmed and
suspected double degenerate binaries, and the peculiar magnetic white
dwarf 2105$-$820. As such, the failure of the mass-radius relation to
accurately predict the measured parameters of these objects is
probably only apparent. Therefore, it is appropriate to review our
test of the mass-radius relation with these considerations in
mind. More specifically, we exclude in the following the 15 double
degenerate systems, and we assume that 2105$-$820 has a purely
radiative atmosphere.

\begin{figure}[p]
\centering
\vspace*{-2cm}
\includegraphics[width=\linewidth]{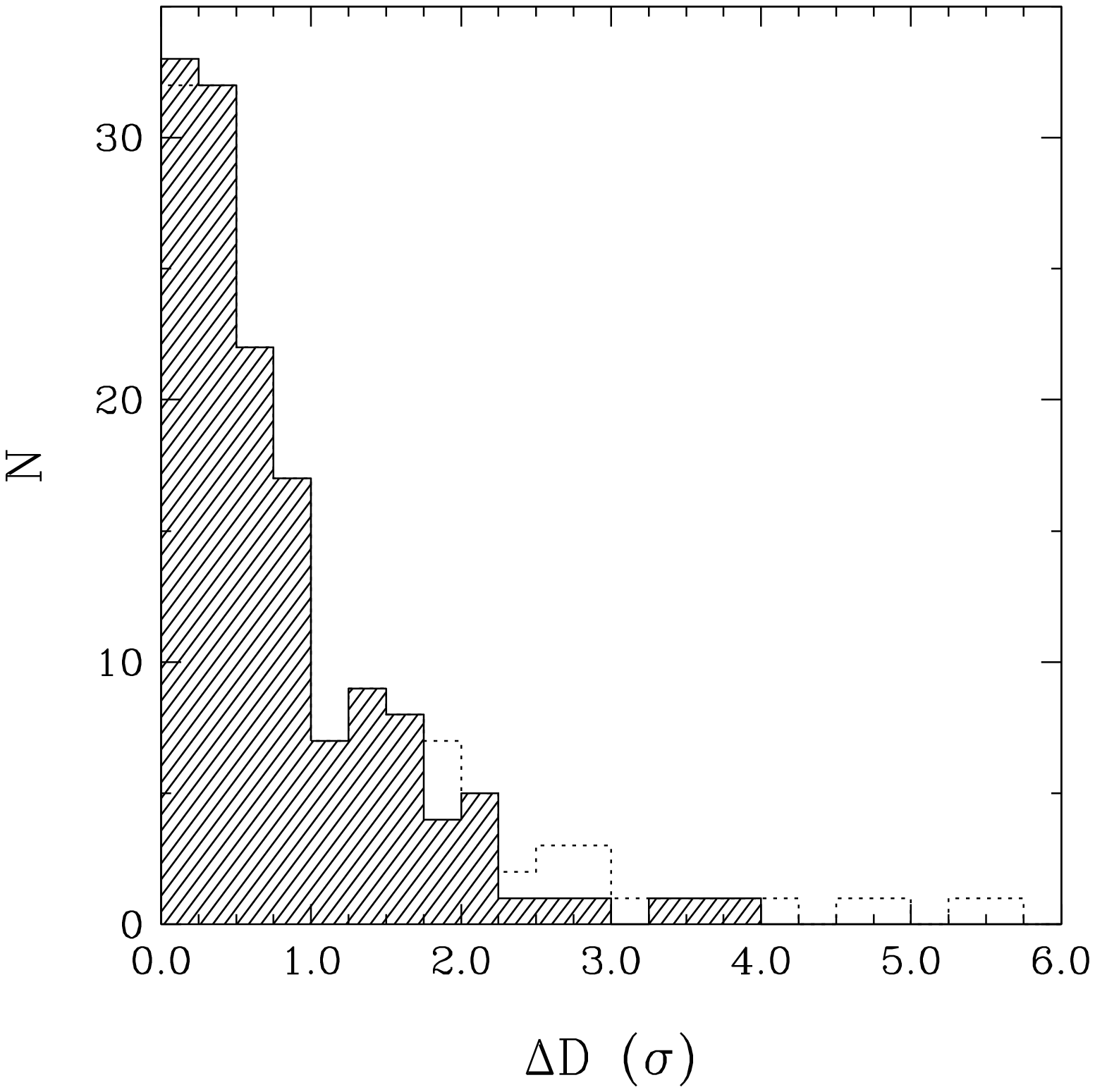}
\vspace*{-4cm}
\caption{Distribution of the (absolute) differences between the
  distances $D_{\rm MR}$ and $D_\pi$, measured in units of $\sigma$,
  where $\sigma^2\equiv\sigma^2_{D_{\rm MR}}+\sigma^2_{D_{\pi}}$, for
  all white dwarfs in our sample with reliable trigonometric parallax
  and spectroscopic $\logg$ measurements, after exclusion of the 15
  double degenerate candidates, and under the assumption that
  2105$-$820 has a purely radiative atmosphere. The initial
  distribution of Figure \ref{N_NsigD} is shown as the dotted line for
  comparison.} \label{N_NsigD_final}
\end{figure}

Figure \ref{N_NsigD_final} displays the distribution of the absolute
differences between the two distance estimates $D_{\rm MR}$ and
$D_\pi$, expressed in units of $\sigma$, under these assumptions. The
initial distribution of Figure \ref{N_NsigD} is also shown as a
reference. Among the 143 objects (the 158 objects in our reliable
subsample minus the 15 double degenerate binaries), the two distances
are now within the 1$\sigma$ confidence level for 73\% of the stars,
and within 2$\sigma$ for 92\% of the stars.  Hence, the distribution
now approaches the expected Gaussian distribution much more
closely. Also, the number of white dwarfs showing large
inconsistencies ($\Delta D \geq 3\sigma$) has dropped to only 3 (these
correspond to 0133$-$116, 1620$-$391, and 2336$-$079, which were
already discussed at length in Sections \ref{DD} and
\ref{Fe-core}). Thus, there is no evidence for a significant
statistical deviation from the theoretical mass-radius relation, and
we can state that our empirical analysis provides strong support to
the current theory of stellar degeneracy.

Furthermore, we can now obtain a mass distribution that is independent
of the mass-radius relation (from the mass values $M$ corresponding to
the x-axis of our $R_\pi$ versus $M$ diagram). To our knowledge, such
a mass distribution is presented for the first time. This mass
distribution is shown as the hatched black histogram in Figure
\ref{N_M}, together with the spectroscopic and photometric mass
distributions discussed previously.  The mean mass, $\langle M \rangle
= 0.666$ \msun, is comparable to those of the two other distributions,
albeit slightly higher. Also, the distribution is somewhat flatter,
which results in a larger standard deviation, $\sigma_M=0.213$
\msun. A careful examination of our results reveals that this larger
standard deviation can be explained in part by the presence of
extended low-mass and high-mass tails, which contain respectively 13
and 14 white dwarfs for which $\Delta D > 1\sigma$. For instance, the
suspiciously high mass of $M=1.34$ \msun\ obtained for 1509+322 must
be considered doubtful since the parallax measurement for this object
has a large 19.2\% uncertainty, close to our confidence limit of
20\%. Similarly, we obtained an aberrant mass of $M=2.30$ \msun\ for
0133$-$116, which was not included in the calculation of the mean
value and standard deviation of the distribution, for the obvious
reason that this mass is undoubtedly largely overestimated, as
discussed in Section \ref{DD}.

We finally summarize in Table \ref{table} the results of our analysis
for the 158 white dwarfs (or white dwarf systems) with reliable data,
where we give for each object the WD number, name, and spectral type,
the parallactic distance ($D_\pi$), the effective temperature ($\Te$)
and the method used to determine this temperature, the spectroscopic
surface gravity determination ($\logg$), the radius ($R_\pi$) derived
from the photometric technique, and the mass ($M$) obtained by
combining the values of $\logg$ and $R_\pi$. We also give, as a
quantitative measure of consistency with the C/O-core mass-radius
relation, the difference between the distances $D_\pi$ and $D_{\rm
  MR}$, in units of $\sigma$. We want to emphasize that since we have
established the validity of the mass-radius relation, the physical
parameters for objects with large $\Delta D$ values (i.e., objects
that deviate from the predictions of the mass-radius relation) should
be regarded with caution. In particular, note that the parameters for
the double degenerate binaries (or binary candidates) are given here
under the assumption of a single star, and the reader should refer to
Figure \ref{DDfits} to obtain the deconvolved parameters. It is
probable that at least some of the remaining discrepant cases will be
resolved by improved astrometric, spectroscopic and/or photometric
data. However, it is also possible that some of these white dwarfs
have yet unknown specific characteristics that still await a correct
interpretation.

\section{CONCLUSION} \label{concl}

We performed a detailed model atmosphere analysis of 219 DA and DB
white dwarfs with measured trigonometric parallaxes, using both the
spectroscopic and photometric techniques. The physical parameters
essential to the characterization of white dwarf stars were obtained
in a homogeneous way from fits to optical spectra and to energy
distributions. After focusing our attention to a subsample of 158
objects with reliable surface gravity determinations and precise
parallax measurements, we showed that the effective temperatures and
masses derived from spectroscopy and photometry are generally in good
agreement.

The physical parameters were then compared to those predicted by the
theoretical mass-radius relation for white dwarfs. Since 92\% of the
stars in our reliable subsample were found to be consistent with the
mass-radius relation within a 2$\sigma$ confidence level, we can
confidently assert that the theory of degenerate stars rests on solid
empirical grounds. We therefore reach the same conclusion as
\citet{holberg12}, \citet{tremblay17}, and \citet{parsons17}, but for
a much larger sample of white dwarfs, thus greatly improving the
observational constraints on this widely used theoretical model.
However, we are on the verge of an even more important improvement of
our ability to investigate the degenerate mass-radius
relation. Indeed, the upcoming data releases of the {\it Gaia} mission
will significantly increase the number and precision of white dwarf
parallax measurements.  However, high-quality spectroscopy and
photometry will be needed as well to obtain accurate physical
quantities, and hence to advance work in this field of research.

It was also demonstrated that the systematic approach used to study
the validity of the mass-radius relation offers the indirect advantage
of unveiling physical peculiarities of individual white dwarfs, which
we fully exploited. Indeed, we presented convincing evidence that 15
objects in our reliable subsample are most certainly unresolved double
degenerate binaries. Based on these results, the proportion of
unresolved double degenerate systems among the white dwarf population
can be estimated to be $\sim$10\%. We pushed our analysis further by
showing that the spectroscopic and photometric observations for all
these objects can be fitted by composite models, and atmospheric
parameters were derived for both components in these binary systems
separately. This opens up a new window to study the physical
properties of white dwarfs in double degenerate systems on a large
scale. The solutions presented here could eventually be confirmed by
comparisons with high-resolution spectroscopy resolving both
components.

While the current level of precision on the white dwarf parameters
does not allow for a significant determination of the hydrogen
envelope thickness, it is nevertheless possible to probe the internal
composition. Our findings indicate that there might be some exceptions
to the widely accepted assumption that stellar evolution produces
white dwarfs having a core composed mostly of carbon and oxygen. More
specifically, three objects in our sample fall directly on the
theoretical mass-radius relation for an iron core.  The most
compelling case is that of 0644+375 (G87-7), for which the parameters
are all firmly pinned down by high-quality observations, and the
possibility of an iron core for this object must be considered
seriously. This interpretation definitely challenges the current
theories of white dwarf formation.

\acknowledgments This work was supported in part by the NSERC Canada
and by the Fund FRQ-NT (Qu\'ebec).

\clearpage
\bibliography{ms}{}
\bibliographystyle{apj}

\clearpage
\clearpage
\begin{deluxetable}{llccccccccc}
\tabletypesize{\scriptsize}
\tablecolumns{12}
\tablewidth{0pt}
\rotate
\tablecaption{Adopted Physical Parameters of White Dwarfs with Reliable Data}
\tablehead{
\colhead{WD} &
\colhead{Name} &
\colhead{Spectral Type} &
\colhead{$D_{\pi}$ (pc)} &
\colhead{$\Te$ (K)} &
\colhead{Method$^a$} &
\colhead{$\logg$} &
\colhead{$R_{\pi}/R_{\odot}$} &
\colhead{$M/M_{\odot}$} &
\colhead{$\Delta D$ ($\sigma$)} &
\colhead{Notes}}
\startdata
       0002+729           & GD 408          & DBA             & 34.7 (5.8)      & 14,406  (235)   & S               & 8.26 (0.10)     & 0.0114 (0.0019) & 0.87 (0.36)     & 0.39            &                 \\
       0008+424           & GD 5            & DA              & 23.2 (0.2)      &  7121   (105)   & S               & 8.11 (0.07)     & 0.0125 (0.0002) & 0.74 (0.13)     & 1.14            &                 \\
       0011+000           & G31-35          & DA              & 29.7 (4.3)      &  9498   (137)   & S               & 8.00 (0.05)     & 0.0096 (0.0014) & 0.34 (0.11)     & 2.17            &                 \\
       0030+444           & G172-4          & DA              & 71.6 (4.1)      & 10,277  (151)   & S               & 8.05 (0.05)     & 0.0121 (0.0007) & 0.60 (0.10)     & 0.36            &                 \\
       0033+016           & G1-7            & DA              & 32.9 (4.4)      & 10,828  (160)   & S               & 8.73 (0.05)     & 0.0079 (0.0011) & 1.22 (0.36)     & 0.49            &                 \\
     0034$-$602           & LP 122-4        & DA              & 24.1 (0.9)      & 15,239  (319)   & S               & 8.64 (0.04)     & 0.0084 (0.0003) & 1.13 (0.15)     & 0.94            &                 \\
     0053$-$117           & L796-10         & DA              & 22.8 (0.2)      &  7045   (106)   & S               & 8.11 (0.08)     & 0.0128 (0.0002) & 0.78 (0.16)     & 1.47            &                 \\
       0101+048           & G1-45           & DA              & 20.9 (1.7)      &  8343   (121)   & S               & 8.04 (0.05)     & 0.0158 (0.0013) & 1.01 (0.21)     & 2.52            & 1               \\
       0126+101           & G2-40           & DA              & 34.4 (3.7)      &  8548   (123)   & S               & 7.60 (0.05)     & 0.0205 (0.0022) & 0.61 (0.15)     & 1.81            & 1               \\
     0133$-$116           & Ross 548        & DA              & 63.3 (8.1)      & 12,267  (186)   & S               & 8.01 (0.05)     & 0.0249 (0.0032) & 2.29 (0.65)     & 3.73            &                 \\
     0135$-$052           & L870-2          & DA              & 12.0 (0.5)      &  7160   (103)   & S               & 7.75 (0.06)     & 0.0201 (0.0008) & 0.82 (0.13)     & 5.36            & 1               \\
       0136+152           & PG 0136+152     & DA              & 22.2 (0.3)      &  7972   (114)   & S               & 8.20 (0.05)     & 0.0118 (0.0002) & 0.80 (0.10)     & 1.46            &                 \\
       0142+312           & G72-31          & DA              & 34.8 (5.3)      &  9223   (134)   & S               & 8.13 (0.05)     & 0.0156 (0.0024) & 1.21 (0.41)     & 1.61            & 1               \\
       0148+467           & GD 279          & DA              & 15.5 (0.8)      & 13,944  (275)   & S               & 8.05 (0.04)     & 0.0122 (0.0007) & 0.61 (0.09)     & 0.35            &                 \\
       0148+641           & G244-36         & DA              & 17.4 (0.2)      &  8879   (127)   & S               & 8.20 (0.05)     & 0.0119 (0.0002) & 0.82 (0.09)     & 1.79            &                 \\
       0150+256           & G94-21          & DA              & 34.0 (2.2)      &  7677   (153)   & P               & 7.94 (0.13)     & 0.0134 (0.0010) & 0.57 (0.19)     & 0.09            &                 \\
       0205+250           & G35-29          & DA              & 33.3 (2.8)      & 21,121  (316)   & S               & 7.91 (0.04)     & 0.0133 (0.0011) & 0.52 (0.10)     & 0.57            &                 \\
       0208+396           & G74-7           & DAZ             & 16.7 (1.0)      &  7332   (109)   & S               & 7.93 (0.08)     & 0.0125 (0.0007) & 0.48 (0.10)     & 0.91            &                 \\
       0213+396           & GD 25           & DA              & 19.7 (0.3)      &  9228   (134)   & S               & 8.39 (0.05)     & 0.0098 (0.0002) & 0.87 (0.11)     & 0.23            &                 \\
       0220+222           & G94-B5B         & DA              & 78.5 (3.4)      & 16,301  (264)   & S               & 8.04 (0.05)     & 0.0121 (0.0005) & 0.59 (0.08)     & 0.80            &                 \\
     0226$-$329           & MCT 0226$-$3255 & DA              & 45.9 (2.7)      & 22,770  (361)   & S               & 7.96 (0.05)     & 0.0131 (0.0008) & 0.57 (0.09)     & 0.46            &                 \\
       0227+050           & Feige 22        & DA              & 26.7 (3.7)      & 19,514  (302)   & S               & 7.95 (0.05)     & 0.0139 (0.0020) & 0.63 (0.19)     & 0.18            &                 \\
       0232+035           & Feige 24        & DA+dM           & 76.6 (6.3)      & 67,133  (1415)  & S               & 7.41 (0.07)     & 0.0205 (0.0017) & 0.39 (0.09)     & 1.42            &                 \\
     0243$-$026           & LHS 1442        & DA              & 20.9 (0.2)      &  6780   (76)    & P               & 8.06 (0.10)     & 0.0112 (0.0002) & 0.53 (0.12)     & 1.25            &                 \\
     0250$-$007           & LP 591-117      & DA              & 47.6 (2.0)      &  8278   (124)   & S               & 8.29 (0.07)     & 0.0122 (0.0006) & 1.04 (0.20)     & 2.15            &                 \\
     0255$-$705           & LHS 1474        & DA              & 24.1 (1.3)      & 10,743  (167)   & S               & 8.07 (0.06)     & 0.0116 (0.0007) & 0.58 (0.10)     & 0.76            &                 \\
       0257+080           & LHS 5064        & DA              & 27.9 (2.7)      &  6619   (130)   & P               & 7.78 (0.14)     & 0.0132 (0.0014) & 0.39 (0.15)     & 0.73            & 2               \\
     0310$-$688           & LB 3303         & DA              & 10.4 (0.1)      & 16,246  (237)   & S               & 8.14 (0.04)     & 0.0120 (0.0002) & 0.72 (0.08)     & 0.45            &                 \\
     0311$-$649           & LEHPM 1-3159    & DA              & 32.5 (1.3)      & 13,135  (290)   & S               & 7.83 (0.06)     & 0.0187 (0.0008) & 0.86 (0.14)     & 4.53            & 1               \\
     0326$-$273           & L587-55A        & DA              & 24.2 (0.8)      &  9207   (133)   & S               & 7.71 (0.06)     & 0.0188 (0.0006) & 0.65 (0.09)     & 4.09            & 1               \\
       0401+250           & G8-8            & DA              & 26.8 (2.0)      & 12,482  (191)   & S               & 8.10 (0.05)     & 0.0125 (0.0010) & 0.71 (0.14)     & 0.43            &                 \\
     0413$-$077           & 40 Eri B        & DA              & 4.98 (0.06)     & 17,099  (256)   & S               & 7.95 (0.04)     & 0.0131 (0.0002) & 0.57 (0.06)     & 0.63            &                 \\
     0415$-$594           & HD 27442B       & DA              & 18.2 (0.2)      & 15,424  (270)   & S               & 7.98 (0.05)     & 0.0132 (0.0010) & 0.60 (0.12)     & 0.03            &                 \\
     0419$-$487           & LHS 1660        & DA+dM           & 20.1 (0.5)      &  6354   (261)   & P               & 7.38 (0.18)     & 0.0204 (0.0022) & 0.36 (0.18)     & 0.70            &                 \\
       0453+418           & GD 64           & DA              & 43.9 (5.1)      & 14,523  (248)   & S               & 7.85 (0.05)     & 0.0166 (0.0019) & 0.71 (0.18)     & 1.15            &                 \\
     0457$-$004           & G84-26          & DA              & 24.9 (0.1)      & 11,121  (166)   & S               & 8.81 (0.05)     & 0.0069 (0.0001) & 1.14 (0.13)     & 0.30            &                 \\
       0501+527           & G191-B2B        & DA              & 59.9 (11.0)     & 60,701  (988)   & S               & 7.53 (0.05)     & 0.0228 (0.0042) & 0.65 (0.26)     & 0.48            &                 \\
       0518+333           & G86-B1B         & DA+dM           & 65.4 (9.6)      &  8963   (129)   & S               & 8.06 (0.05)     & 0.0161 (0.0024) & 1.08 (0.35)     & 1.58            &                 \\
       0612+177           & G104-27         & DA              & 42.0 (5.6)      & 26,097  (380)   & S               & 7.99 (0.04)     & 0.0133 (0.0018) & 0.62 (0.18)     & 0.05            &                 \\
     0615$-$591           & L182-61         & DB              & 36.4 (0.7)      & 15,746  (238)   & S               & 8.04 (0.06)     & 0.0130 (0.0003) & 0.67 (0.10)     & 1.00            &                 \\
     0628$-$020           & LP 600-42       & DA+dM           & 21.5 (0.8)      &  6741   (380)   & P               & 7.92 (0.12)     & 0.0129 (0.0013) & 0.50 (0.17)     & 0.33            &                 \\
     0642$-$166           & Sirius B        & DA              & 2.64 (0.01)     & 26,083  (378)   & S               & 8.61 (0.04)     & 0.0079 (0.0002) & 0.94 (0.11)     & 0.72            &                 \\
     0642$-$285           & LP 895-41       & DA              & 65.2 (2.3)      &  9230   (133)   & S               & 7.95 (0.05)     & 0.0131 (0.0011) & 0.56 (0.11)     & 0.13            &                 \\
       0644+025           & G108-26         & DA              & 18.2 (0.1)      &  7085   (106)   & S               & 8.52 (0.07)     & 0.0083 (0.0001) & 0.85 (0.15)     & 0.76            &                 \\
       0644+375           & G87-7           & DA              & 15.7 (0.9)      & 21,996  (254)   & S               & 8.15 (0.04)     & 0.0106 (0.0006) & 0.58 (0.09)     & 1.67            & 3               \\
     0659$-$063           & LHS 1892        & DA              & 20.6 (0.1)      &  6506   (148)   & P               & 8.03 (0.11)     & 0.0124 (0.0004) & 0.61 (0.16)     & 0.04            &                 \\
       0752+365           & G90-28          & DA              & 33.6 (4.3)      &  7712   (111)   & S               & 7.98 (0.06)     & 0.0111 (0.0014) & 0.43 (0.13)     & 1.21            &                 \\
       0816+387           & G111-71         & DA              & 39.7 (4.0)      &  7544   (123)   & P               & 8.10 (0.09)     & 0.0112 (0.0012) & 0.57 (0.17)     & 0.58            &                 \\
       0827+328           & LHS 2022        & DA              & 22.0 (1.9)      &  7268   (106)   & S               & 8.37 (0.06)     & 0.0095 (0.0008) & 0.78 (0.18)     & 0.31            &                 \\
     0839$-$327           & LHS 253         & DA              & 8.48 (0.08)     &  9093   (130)   & S               & 7.76 (0.05)     & 0.0146 (0.0002) & 0.45 (0.05)     & 0.63            &                 \\
       0842+490           & HD 74389B       & DA              & 111.5 (7.1)     & 41,885  (731)   & S               & 7.92 (0.07)     & 0.0114 (0.0007) & 0.39 (0.08)     & 2.93            &                 \\
       0913+442           & G116-16         & DA              & 28.4 (3.3)      &  8614   (125)   & S               & 8.07 (0.05)     & 0.0109 (0.0013) & 0.51 (0.14)     & 0.97            &                 \\
     0928$-$713           & BPM 5639        & DA              & 26.0 (0.4)      &  8370   (120)   & S               & 8.16 (0.05)     & 0.0114 (0.0002) & 0.69 (0.08)     & 0.13            &                 \\
       0930+294           & G117-25         & DA              & 31.4 (4.6)      &  8250   (120)   & S               & 8.46 (0.06)     & 0.0097 (0.0014) & 1.00 (0.33)     & 0.36            &                 \\
      0942+236A           & LP 370-50       & DA              & 46.5 (9.2)      &  7125   (109)   & S               & 7.80 (0.10)     & 0.0098 (0.0019) & 0.22 (0.10)     & 2.16            &                 \\
      0942+236B           & LP 370-51       & DA              & 46.5 (9.2)      &  7019   (133)   & P               & 8.17 (0.13)     & 0.0094 (0.0019) & 0.47 (0.25)     & 0.94            &                 \\
       0943+441           & G116-52         & DA              & 33.2 (3.3)      & 13,771  (222)   & S               & 7.59 (0.05)     & 0.0181 (0.0018) & 0.46 (0.11)     & 0.60            &                 \\
       0955+247           & G49-33          & DA              & 24.1 (2.6)      &  8529   (123)   & S               & 8.18 (0.05)     & 0.0107 (0.0012) & 0.63 (0.16)     & 0.52            &                 \\
      1012+083A           & G43-38          & DA              & 28.5 (3.3)      &  6689   (81)    & P               & 7.89 (0.12)     & 0.0125 (0.0015) & 0.44 (0.17)     & 0.65            &                 \\
     1016$-$308           & LP 904-3        & DA              & 50.9 (3.2)      & 16,297  (243)   & S               & 8.18 (0.04)     & 0.0129 (0.0008) & 0.92 (0.15)     & 1.60            &                 \\
       1019+637           & LP 62-147       & DA              & 16.2 (1.0)      &  6777   (75)    & P               & 7.90 (0.08)     & 0.0126 (0.0008) & 0.47 (0.11)     & 0.85            &                 \\
       1104+602           & G197-4          & DA              & 43.9 (8.8)      & 18,294  (273)   & S               & 8.11 (0.04)     & 0.0151 (0.0030) & 1.07 (0.46)     & 1.01            &                 \\
     1105$-$340           & SCR J1107$-$342 & DA              & 25.6 (0.4)      & 13,891  (271)   & S               & 8.06 (0.04)     & 0.0117 (0.0002) & 0.57 (0.06)     & 1.64            & 3               \\
       1121+216           & Ross 627        & DA              & 13.6 (0.6)      &  7346   (107)   & S               & 8.10 (0.06)     & 0.0115 (0.0005) & 0.61 (0.11)     & 0.57            &                 \\
     1124$-$293           & ESO 439-80      & DA              & 32.3 (1.6)      &  9348   (134)   & S               & 7.98 (0.05)     & 0.0125 (0.0006) & 0.54 (0.08)     & 0.67            &                 \\
       1130+189           & LP 433-6        & DA              & 216.0 (34.9)    & 10,886  (174)   & S               & 8.40 (0.06)     & 0.0206 (0.0033) & 3.85 (1.39)     & 3.24            & 1               \\
       1134+300           & GD 140          & DA              & 15.8 (0.9)      & 22,313  (339)   & S               & 8.56 (0.05)     & 0.0086 (0.0005) & 0.98 (0.15)     & 0.04            &                 \\
       1143+321           & G148-7          & DA              & 30.2 (3.0)      & 16,332  (245)   & S               & 8.17 (0.04)     & 0.0121 (0.0012) & 0.79 (0.18)     & 0.46            &                 \\
       1147+255           & LP 375-51       & DA              & 47.4 (8.8)      & 10,072  (148)   & S               & 8.00 (0.05)     & 0.0121 (0.0023) & 0.54 (0.22)     & 0.29            &                 \\
     1202$-$232           & LP 852-7        & DA              & 10.9 (0.1)      &  8608   (124)   & S               & 7.99 (0.05)     & 0.0135 (0.0002) & 0.64 (0.08)     & 1.29            &                 \\
       1214+032           & LP 554-63       & DA              & 21.6 (1.6)      &  6751   (151)   & P               & 7.81 (0.14)     & 0.0128 (0.0011) & 0.39 (0.14)     & 0.93            &                 \\
     1223$-$659           & L104-2          & DA              & 16.3 (0.3)      &  7491   (107)   & S               & 7.76 (0.05)     & 0.0147 (0.0003) & 0.45 (0.06)     & 0.35            &                 \\
     1236$-$495           & BPM 37093       & DA              & 15.3 (0.2)      & 11,620  (189)   & S               & 8.72 (0.05)     & 0.0077 (0.0001) & 1.13 (0.14)     & 0.71            &                 \\
     1242$-$105           & LP 736-4        & DA              & 39.7 (1.2)      &  8145   (119)   & S               & 7.85 (0.06)     & 0.0225 (0.0007) & 1.31 (0.21)     & 9.42            & 1               \\
       1257+278           & G149-28         & DAZ             & 33.6 (4.9)      &  8612   (126)   & S               & 8.04 (0.06)     & 0.0125 (0.0018) & 0.63 (0.21)     & 0.03            &                 \\
       1304+227           & LP 378-537      & DA              & 77.2 (3.5)      & 11,119  (173)   & S               & 8.21 (0.05)     & 0.0118 (0.0005) & 0.83 (0.13)     & 0.97            &                 \\
       1314+293           & HZ 43A          & DA+dM           & 58.0 (2.6)      & 55,396  (1170)  & S               & 8.01 (0.07)     & 0.0137 (0.0007) & 0.71 (0.14)     & 0.01            &                 \\
     1314$-$153           & LHS 2712        & DA              & 58.1 (4.3)      & 15,736  (261)   & S               & 7.90 (0.05)     & 0.0142 (0.0011) & 0.58 (0.11)     & 0.25            &                 \\
       1325+581           & G199-71         & DA              & 36.1 (7.5)      &  6725   (83)    & P               & 7.99 (0.21)     & 0.0114 (0.0024) & 0.46 (0.30)     & 0.48            &                 \\
     1327$-$083           & Wolf 485A       & DA              & 16.1 (0.3)      & 14,714  (237)   & S               & 8.00 (0.05)     & 0.0131 (0.0003) & 0.62 (0.07)     & 0.20            &                 \\
       1333+487           & GD 325          & DBA             & 35.0 (4.0)      & 15,420  (242)   & S               & 8.01 (0.09)     & 0.0120 (0.0014) & 0.53 (0.17)     & 0.42            &                 \\
       1337+705           & G238-44         & DA              & 26.1 (2.1)      & 21,311  (328)   & S               & 7.96 (0.05)     & 0.0129 (0.0010) & 0.56 (0.11)     & 0.46            &                 \\
       1344+106           & LHS 2800        & DA              & 19.7 (1.4)      &  6968   (107)   & S               & 8.00 (0.08)     & 0.0120 (0.0009) & 0.52 (0.13)     & 0.68            &                 \\
       1354+340           & G165-B5B        & DA              & 92.7 (5.0)      & 14,608  (315)   & S               & 7.97 (0.06)     & 0.0129 (0.0007) & 0.57 (0.10)     & 0.38            &                 \\
       1408+323           & GD 163          & DA              & 37.7 (4.0)      & 19,015  (287)   & S               & 8.00 (0.04)     & 0.0117 (0.0013) & 0.50 (0.12)     & 0.99            &                 \\
     1418$-$088           & G124-26         & DA              & 39.2 (6.1)      &  8061   (117)   & S               & 8.10 (0.06)     & 0.0168 (0.0026) & 1.29 (0.45)     & 1.81            & 1               \\
       1422+095           & GD 165          & DA              & 31.5 (2.5)      & 12,227  (186)   & S               & 8.08 (0.05)     & 0.0118 (0.0009) & 0.61 (0.12)     & 0.43            &                 \\
     1447$-$190           & LP 801-14       & DA              & 47.4 (1.9)      &  7132   (229)   & P               & 7.80 (0.20)     & 0.0204 (0.0013) & 0.96 (0.47)     & 2.95            & 1               \\
       1455+298           & LHS 3007        & DA              & 33.6 (4.7)      &  7290   (105)   & S               & 7.90 (0.06)     & 0.0153 (0.0021) & 0.68 (0.22)     & 0.79            &                 \\
       1509+322           & GD 178          & DA              & 45.7 (9.1)      & 14,807  (259)   & S               & 8.15 (0.05)     & 0.0161 (0.0032) & 1.34 (0.57)     & 1.38            &                 \\
     1544$-$377           & L481-60         & DA              & 15.3 (0.2)      & 10,402  (150)   & S               & 7.97 (0.05)     & 0.0134 (0.0003) & 0.62 (0.07)     & 0.77            &                 \\
       1554+215           & PG 1554+215     & DA              & 102.8 (7.2)     & 27,123  (410)   & S               & 7.87 (0.05)     & 0.0136 (0.0010) & 0.50 (0.09)     & 0.90            &                 \\
       1559+369           & Ross 808        & DA              & 33.1 (3.7)      & 11,187  (165)   & S               & 7.99 (0.05)     & 0.0131 (0.0014) & 0.61 (0.15)     & 0.10            &                 \\
       1606+422           & Case 2          & DA              & 45.0 (7.3)      & 13,064  (216)   & S               & 7.84 (0.05)     & 0.0202 (0.0033) & 1.03 (0.36)     & 1.78            & 1               \\
       1609+135           & LHS 3163        & DA              & 17.3 (1.4)      &  9349   (136)   & S               & 8.60 (0.05)     & 0.0066 (0.0005) & 0.63 (0.13)     & 2.49            &                 \\
       1619+123           & PG 1619+123     & DA              & 56.5 (1.7)      & 17,156  (259)   & S               & 7.88 (0.05)     & 0.0141 (0.0004) & 0.54 (0.07)     & 0.13            &                 \\
     1620$-$391           & CD $-$38 10980  & DA              & 12.8 (0.1)      & 25,954  (368)   & S               & 7.96 (0.04)     & 0.0120 (0.0001) & 0.49 (0.05)     & 3.44            & 3               \\
       1625+093           & G138-31         & DA              & 23.4 (2.0)      &  7129   (134)   & S               & 8.60 (0.16)     & 0.0087 (0.0008) & 1.10 (0.45)     & 0.37            &                 \\
       1633+433           & G180-63         & DAZ             & 15.1 (0.7)      &  6625   (61)    & P               & 7.92 (0.15)     & 0.0116 (0.0006) & 0.41 (0.14)     & 1.33            &                 \\
       1635+137           & G138-47         & DA              & 39.2 (7.3)      &  6811   (92)    & P               & 7.76 (0.13)     & 0.0109 (0.0020) & 0.25 (0.12)     & 1.66            &                 \\
       1637+335           & G180-65         & DA              & 29.2 (3.3)      & 10,151  (147)   & S               & 8.05 (0.05)     & 0.0119 (0.0013) & 0.57 (0.15)     & 0.39            &                 \\
       1639+153           & LHS 3236        & DA              & 30.9 (0.2)      &  7452   (110)   & S               & 8.49 (0.07)     & 0.0128 (0.0001) & 1.84 (0.29)     & 7.69            & 1               \\
       1645+325           & GD 358          & DB              & 36.6 (4.5)      & 24,937  (1018)  & S               & 7.92 (0.05)     & 0.0110 (0.0014) & 0.37 (0.10)     & 1.85            &                 \\
       1647+591           & G226-29         & DA              & 11.0 (0.1)      & 12,517  (195)   & S               & 8.34 (0.05)     & 0.0103 (0.0001) & 0.83 (0.09)     & 0.30            &                 \\
       1655+215           & LHS 3254        & DA              & 23.0 (1.6)      &  9229   (133)   & S               & 8.07 (0.05)     & 0.0137 (0.0010) & 0.81 (0.15)     & 1.42            &                 \\
     1659$-$531           & L268-92         & DA              & 27.2 (0.5)      & 15,507  (230)   & S               & 8.07 (0.04)     & 0.0124 (0.0003) & 0.66 (0.07)     & 0.14            &                 \\
       1706+332           & G181-B5B        & DA              & 71.5 (2.7)      & 12,774  (211)   & S               & 8.06 (0.05)     & 0.0124 (0.0005) & 0.64 (0.09)     & 0.07            &                 \\
       1713+695           & G240-51         & DA              & 25.0 (2.8)      & 15,952  (243)   & S               & 8.00 (0.05)     & 0.0122 (0.0014) & 0.55 (0.14)     & 0.53            &                 \\
       1716+020           & G19-20          & DA              & 40.8 (7.2)      & 13,255  (268)   & S               & 7.88 (0.05)     & 0.0143 (0.0025) & 0.56 (0.22)     & 0.12            &                 \\
     1733$-$544           & L270-137        & DA              & 21.9 (0.4)      &  6542   (85)    & P               & 8.35 (0.11)     & 0.0115 (0.0003) & 1.08 (0.29)     & 1.69            &                 \\
       1736+052           & G140-2          & DA              & 42.2 (7.1)      &  8969   (132)   & S               & 8.14 (0.06)     & 0.0116 (0.0020) & 0.68 (0.26)     & 0.02            &                 \\
     1743$-$132           & G154-B5B        & DA              & 38.5 (1.1)      & 12,850  (210)   & S               & 8.00 (0.05)     & 0.0143 (0.0005) & 0.74 (0.10)     & 2.13            &                 \\
       1756+827           & LHS 56          & DA              & 16.9 (1.4)      &  7227   (106)   & S               & 7.89 (0.07)     & 0.0139 (0.0011) & 0.55 (0.13)     & 0.20            &                 \\
       1824+040           & Ross 137        & DA              & 54.9 (7.1)      & 12,252  (190)   & S               & 7.70 (0.05)     & 0.0252 (0.0032) & 1.15 (0.33)     & 2.90            & 1               \\
     1826$-$045           & G21-16          & DA              & 28.7 (3.2)      &  9125   (131)   & S               & 8.00 (0.05)     & 0.0140 (0.0015) & 0.70 (0.18)     & 0.70            &                 \\
       1840+042           & GD 215          & DA              & 24.8 (2.1)      &  8877   (128)   & S               & 8.19 (0.05)     & 0.0117 (0.0010) & 0.77 (0.16)     & 0.39            &                 \\
       1855+338           & G207-9          & DA              & 32.8 (4.8)      & 12,137  (184)   & S               & 8.33 (0.05)     & 0.0106 (0.0016) & 0.88 (0.28)     & 0.26            &                 \\
     1916$-$362           & SCR J1920$-$361 & DB              & 37.4 (1.6)      & 25,383  (3225)  & S               & 7.86 (0.10)     & 0.0113 (0.0005) & 0.34 (0.08)     & 2.62            &                 \\
       1919+145           & GD 219          & DA              & 19.8 (2.2)      & 15,080  (258)   & S               & 8.20 (0.05)     & 0.0111 (0.0012) & 0.71 (0.18)     & 0.15            &                 \\
       1935+276           & G185-32         & DA              & 17.9 (0.9)      & 12,381  (186)   & S               & 8.09 (0.05)     & 0.0120 (0.0006) & 0.64 (0.10)     & 0.21            &                 \\
       1936+327           & GD 222          & DA              & 34.8 (2.9)      & 22,228  (339)   & S               & 7.91 (0.05)     & 0.0116 (0.0010) & 0.40 (0.08)     & 2.22            &                 \\
       1940+374           & L1573-31        & DB              & 49.3 (6.9)      & 16,851  (267)   & S               & 8.07 (0.10)     & 0.0125 (0.0018) & 0.68 (0.25)     & 0.19            &                 \\
       1943+163           & G142-50         & DA              & 41.5 (4.7)      & 20,372  (310)   & S               & 7.94 (0.05)     & 0.0119 (0.0014) & 0.45 (0.11)     & 1.24            &                 \\
       1950+250           & GD 385          & DA              & 38.0 (4.7)      & 11,897  (181)   & S               & 8.05 (0.05)     & 0.0101 (0.0012) & 0.41 (0.11)     & 1.82            &                 \\
     1953$-$011           & LHS 3501        & DA              & 11.9 (0.5)      &  7765   (113)   & S               & 8.20 (0.06)     & 0.0117 (0.0005) & 0.79 (0.13)     & 0.75            & 2               \\
     2007$-$219           & L710-30         & DA              & 26.2 (0.6)      &  9842   (144)   & S               & 8.01 (0.05)     & 0.0125 (0.0003) & 0.58 (0.08)     & 0.50            &                 \\
     2007$-$303           & L565-18         & DA              & 16.4 (1.2)      & 16,041  (233)   & S               & 7.95 (0.04)     & 0.0130 (0.0010) & 0.55 (0.10)     & 0.41            &                 \\
       2032+248           & Wolf 1346       & DA              & 15.5 (0.6)      & 20,851  (303)   & S               & 7.97 (0.04)     & 0.0139 (0.0006) & 0.66 (0.09)     & 0.74            &                 \\
     2035$-$369           & L495-42         & DA              & 31.9 (1.0)      &  9647   (139)   & S               & 8.16 (0.05)     & 0.0123 (0.0004) & 0.80 (0.10)     & 1.45            &                 \\
     2039$-$202           & L711-10         & DA              & 20.7 (1.6)      & 20,031  (295)   & S               & 7.96 (0.04)     & 0.0130 (0.0010) & 0.57 (0.11)     & 0.38            &                 \\
     2040$-$392           & L495-82         & DA              & 22.6 (0.5)      & 11,029  (160)   & S               & 8.08 (0.05)     & 0.0122 (0.0003) & 0.65 (0.08)     & 0.05            &                 \\
       2047+372           & G210-36         & DA              & 17.1 (0.1)      & 14,628  (279)   & S               & 8.33 (0.04)     & 0.0103 (0.0001) & 0.83 (0.09)     & 0.25            & 2               \\
       2048+809           & LP 25-436       & DA              & 85.7 (7.6)      &  8365   (124)   & S               & 8.23 (0.07)     & 0.0195 (0.0018) & 2.39 (0.58)     & 4.77            & 1               \\
     2105$-$820           & L24-52          & DA              & 15.4 (0.3)      &  9797   (141)   & S               & 8.27 (0.05)     & 0.0108 (0.0002) & 0.78 (0.10)     & 0.20            & 2               \\
       2109+011           & PM J21117+0120  & DA              & 61.1 (3.7)      & 16,506  (256)   & S               & 8.07 (0.05)     & 0.0124 (0.0009) & 0.66 (0.11)     & 0.04            &                 \\
       2111+261           & G187-32         & DA              & 31.5 (3.7)      &  8525   (124)   & S               & 8.15 (0.06)     & 0.0169 (0.0020) & 1.48 (0.41)     & 2.60            & 1               \\
       2117+539           & G231-40         & DA              & 17.3 (0.3)      & 14,648  (237)   & S               & 7.91 (0.05)     & 0.0137 (0.0003) & 0.56 (0.06)     & 0.10            &                 \\
       2124+550           & Ross 198        & DA              & 33.2 (5.0)      & 14,620  (383)   & S               & 8.48 (0.05)     & 0.0088 (0.0013) & 0.85 (0.28)     & 0.23            &                 \\
       2126+734           & G261-43         & DA              & 21.2 (1.1)      & 16,040  (236)   & S               & 7.99 (0.04)     & 0.0128 (0.0007) & 0.58 (0.08)     & 0.43            &                 \\
       2129+000           & G26-10          & DB              & 43.2 (1.0)      & 14,383  (231)   & S               & 8.27 (0.12)     & 0.0120 (0.0003) & 0.97 (0.28)     & 1.49            &                 \\
     2133$-$135           & Ross 203        & DA              & 24.7 (0.8)      & 10,193  (149)   & S               & 7.86 (0.05)     & 0.0154 (0.0005) & 0.63 (0.09)     & 1.95            &                 \\
       2136+229           & G126-18         & DA              & 41.7 (5.5)      & 10,127  (148)   & S               & 7.93 (0.05)     & 0.0138 (0.0018) & 0.59 (0.17)     & 0.19            &                 \\
       2136+828           & G261-45         & DA              & 26.0 (3.1)      & 17,698  (263)   & S               & 7.97 (0.04)     & 0.0134 (0.0016) & 0.61 (0.16)     & 0.04            &                 \\
       2149+021           & G93-48          & DA              & 24.4 (1.5)      & 18,202  (265)   & S               & 8.01 (0.04)     & 0.0138 (0.0008) & 0.71 (0.11)     & 0.89            &                 \\
     2159$-$754           & LHS 3752        & DA              & 19.9 (0.5)      &  8748   (129)   & S               & 8.66 (0.06)     & 0.0086 (0.0002) & 1.22 (0.18)     & 1.61            &                 \\
       2207+142           & G18-34          & DA              & 25.1 (2.8)      &  7567   (110)   & S               & 8.19 (0.06)     & 0.0108 (0.0012) & 0.66 (0.17)     & 0.35            &                 \\
       2229+235           & HS 2229+2335    & DA              & 110.9 (10.5)    & 19,952  (360)   & S               & 8.05 (0.06)     & 0.0128 (0.0012) & 0.67 (0.16)     & 0.13            &                 \\
       2246+223           & G67-23          & DA              & 18.9 (1.5)      & 10,484  (153)   & S               & 8.65 (0.05)     & 0.0083 (0.0006) & 1.11 (0.21)     & 0.52            &                 \\
     2253$-$081           & G156-64         & DA              & 35.8 (0.7)      &  6678   (128)   & P               & 7.97 (0.18)     & 0.0127 (0.0004) & 0.56 (0.24)     & 0.14            &                 \\
       2258+406           & G216-B14B       & DA              & 71.6 (3.8)      &  9852   (146)   & S               & 8.18 (0.06)     & 0.0121 (0.0012) & 0.81 (0.19)     & 0.66            &                 \\
       2326+049           & G29-38          & DA              & 17.5 (0.1)      & 11,315  (180)   & S               & 8.02 (0.06)     & 0.0127 (0.0001) & 0.62 (0.08)     & 0.01            &                 \\
     2336$-$079           & GD 1212         & DA              & 15.9 (0.4)      & 11,037  (162)   & S               & 8.03 (0.05)     & 0.0107 (0.0003) & 0.45 (0.06)     & 3.76            &                 \\
       2341+322           & G130-5          & DA              & 18.8 (0.1)      & 13,076  (218)   & S               & 8.00 (0.05)     & 0.0127 (0.0001) & 0.59 (0.07)     & 0.41            &                 \\
     2349$-$031           & LHS 4033        & DA              & 29.5 (0.5)      & 10,528  (162)   & S               & 9.34 (0.06)     & 0.0039 (0.0001) & 1.19 (0.16)     & 0.76            &                 \\
     2350$-$083           & G273-B1B        & DA              & 100.4 (11.5)    & 18,832  (353)   & S               & 7.98 (0.06)     & 0.0115 (0.0013) & 0.46 (0.13)     & 1.18            &                 \\
     2351$-$335           & LHS 4040        & DA              & 23.4 (1.3)      &  8663   (125)   & S               & 8.03 (0.06)     & 0.0131 (0.0008) & 0.67 (0.12)     & 0.66            &                 \\
     2351$-$368           & LHS 4041        & DA              & 74.6 (8.5)      & 13,875  (351)   & S               & 8.04 (0.05)     & 0.0154 (0.0018) & 0.95 (0.24)     & 1.54            &                 \\
     2359$-$434           & LHS 1005        & DA              & 8.23 (0.06)     &  8506   (122)   & S               & 8.38 (0.05)     & 0.0097 (0.0001) & 0.82 (0.09)     & 0.26            & 2               \\
\enddata
\tablecomments{
$^a$ S: spectroscopic effective temperature;  P: photometric effective temperature. (1) Unresolved double degenerate system.  The parameters derived under the  assumption of a single star are meaningless.   Atmospheric parameters of both components are given  in Figure 16. (2) Weakly magnetic white dwarf.  (3) Possible iron-core white dwarf.
}
\label{table}
\end{deluxetable}

\end{document}